%% file: ms.tex
\documentclass[journal, comsoc, twoside]{IEEEtran}
\usepackage[T1]{fontenc}

\usepackage{amsmath}
\usepackage[cmintegrals]{newtxmath}
\usepackage{bm}

\usepackage{tikz}
\usepackage{pgfplots}
\pgfplotsset{compat=newest}
\usetikzlibrary{pgfplots.groupplots}
\usetikzlibrary{calc,shapes.geometric,shapes.misc,arrows,matrix}
\input{commands}
\usepackage{aligned-overset}

\usepackage{scalerel}
\usetikzlibrary{svg.path}
\definecolor{orcidlogocol}{HTML}{A6CE39}
\tikzset{
	orcidlogo/.pic={
		\fill[orcidlogocol] svg{M256,128c0,70.7-57.3,128-128,128C57.3,256,0,198.7,0,128C0,57.3,57.3,0,128,0C198.7,0,256,57.3,256,128z};
		\fill[white] svg{M86.3,186.2H70.9V79.1h15.4v48.4V186.2z}
		svg{M108.9,79.1h41.6c39.6,0,57,28.3,57,53.6c0,27.5-21.5,53.6-56.8,53.6h-41.8V79.1z M124.3,172.4h24.5c34.9,0,42.9-26.5,42.9-39.7c0-21.5-13.7-39.7-43.7-39.7h-23.7V172.4z}
		svg{M88.7,56.8c0,5.5-4.5,10.1-10.1,10.1c-5.6,0-10.1-4.6-10.1-10.1c0-5.6,4.5-10.1,10.1-10.1C84.2,46.7,88.7,51.3,88.7,56.8z};
	}
}

\newcommand\orcidicon[1]{\href{https://orcid.org/#1}{\mbox{\scalerel*{
				\begin{tikzpicture}[yscale=-1,transform shape]
				\pic{orcidlogo};
				\end{tikzpicture}
			}{U}}}} 
\usepackage{hyperref}

\definecolor{darkgreen}{rgb}{0,0.498039,0}
\definecolor{darkyellow}{rgb}{0.75,0.75,0}
\definecolor{magenta}{rgb}{0.75,0,0.75}
\newcommand{\clsd}{darkgray} 
\newcommand{\clmmse}{darkgray} 
\newcommand{\clmosic}{purple} 
\newcommand{\clamp}{darkgreen} 
\newcommand{\clsdr}{orange} 
\newcommand{\clmmnet}{darkyellow} 
\newcommand{\cldetnet}{magenta} 
\newcommand{\cloamp}{blue} 
\newcommand{\clcmd}{red} 
\newcommand{\clawgn}{green!50!black} 

\pgfplotsset{ptsd/.style={color=\clsd, mark=*, mark size=1.4
			}, 
			ptmmse/.style={color=\clmmse, mark=triangle*, dashed, mark options={solid}, mark size=1.4
			}, 
			ptmosic/.style={color=\clmosic, mark=triangle*, mark size=1.4
			}, 
			ptamp/.style={color=\clamp, mark=diamond*, mark size=1.4
			}, 
			ptsdr/.style={color=\clsdr, mark=*, dashed, mark options={solid}, mark size=1.4
			}, 
			ptdetnet/.style={color=\cldetnet, mark=square*, dashed, mark options={solid}, mark size=1.4
			}, 
			ptmmnet/.style={color=\clmmnet, mark=diamond*, dashed, mark options={solid}, mark size=1.4
			}, 
			ptoamp/.style={color = \cloamp, mark=x, mark size=4*0.7, mark options={solid}
			}, 
			ptcmd/.style={color=\clcmd, mark=square*, mark size=1.4
			}, 
			ptawgn/.style={color=\clawgn, semithick
			}, 
}

\begin{document}
\title{\cmdnet: \\Learning a Probabilistic Relaxation of Discrete Variables for Soft Detection with Low Complexity}
\author{Edgar~Beck$^{\orcidicon{0000-0003-2213-9727}}$,~\IEEEmembership{Graduate~Student~Member,~IEEE,}
        Carsten~Bockelmann$^{\orcidicon{0000-0002-8501-7324}}$,~\IEEEmembership{Member,~IEEE,}
        and~Armin~Dekorsy$^{\orcidicon{0000-0002-5790-1470}}$,~\IEEEmembership{Senior~Member,~IEEE}
\thanks{This work was partly funded by the German Ministry of Education and Research (BMBF) under grant 16KIS1028 (MOMENTUM).}
\thanks{The authors are with the Department of Communications Engineering, University of Bremen, 28359 Bremen, Germany (e-mail: \{beck, bockelmann, dekorsy\}@ant.uni-bremen.de).}}

\maketitle

\begin{abstract}
Following the great success of Machine Learning (ML), especially Deep Neural Networks (DNNs), in many research domains in 2010s, several ML-based approaches were proposed for detection in large inverse linear problems, e.g., massive MIMO systems. The main motivation behind is that the complexity of Maximum A-Posteriori (MAP) detection grows exponentially with system dimensions. Instead of using DNNs, essentially being a black-box, we take a slightly different approach and introduce a probabilistic Continuous relaxation of disCrete variables to MAP detection. Enabling close approximation and continuous optimization, we derive an iterative detection algorithm: \cmdfull\ (\cmd). Furthermore, extending \cmd\ by the idea of deep unfolding into \cmdnet, we allow for (online) optimization of a small number of parameters to different working points while limiting complexity. In contrast to recent DNN-based approaches, we select the optimization criterion and output of \cmdnet\ based on information theory and are thus able to learn approximate probabilities of the individual optimal detector. This is crucial for soft decoding in today's communication systems. Numerical simulation results in MIMO systems reveal \cmdnet\ to feature a promising accuracy complexity trade-off compared to State of the Art. Notably, we demonstrate \cmdnet's soft outputs to be reliable for decoders.
\end{abstract}

\IEEEpeerreviewmaketitle

\section{Introduction}
\label{sec:1}

\IEEEPARstart{C}{ommunications} is a long standing engineering discipline whose theoretical foundation was laid by Claude Shannon with his landmark paper "A Mathematical Theory of Communication" in $1948$ \cite{shannon_mathematical_1948}. Since then, the theory has evolved into an own field known as information theory today and found its way into many other research areas where data or information is processed including artificial intelligence and especially its subdomain Machine Learning (ML). Information theory relies heavily on description with probabilistic models playing a significant role for design of new generations of cellular communication systems from 2-6G with respective increases in data rate. Probabilistic models have shown to be advantageous also in the ML research domain. Accordingly, both fields, communications and ML, have touched repeatedly in the past, e.g, \cite{viterbi_error_1967, lin_variational_2009, riegler_merging_2013}.

In the early 2010s, a special class of these models gave rise to several breakthroughs in data-driven ML research: Deep Neural Networks (DNNs). Inspired by the brain, several layers of artificial neurons are stacked on top of each other to create an expressive feed forward DNN able to approximate arbitrarily well \cite{hornik_multilayer_1989} and thus to learn higher levels of abstraction, i.e., features, present in data \cite{simeone2018very}. This is of crucial importance for tasks where there are no well-established models but data to be collected. Previously considered intractable to optimize, dedicated hardware and software, i.e., Graphics Processing Units (GPU) and automatic differentiation frameworks \cite{martin_abadi_tensorflow_2015}, innovation to DNN models \cite{glorot_deep_2011, he_deep_2016} and advancements in training \cite{glorot_deep_2011} have made it possible to build algorithms that equal or even surpass human performance in specific tasks such as pattern recognition \cite{ciresan_multi-column_2012} and playing games \cite{silver_mastering_2016}. The impact included all ML subdomains, e.g., classification \cite{he_deep_2016, ciresan_multi-column_2012} in supervised learning, generative modeling in unsupervised learning \cite{goodfellow_generative_2014} and Q-learning in reinforcement learning \cite{silver_mastering_2016}.

\subsection{ML in Communications}

The great success of DNNs in many domains has stimulated large amount of work in communications just in recent years \cite{simeone2018very}. Especially in problems with a model deficit, e.g., detection in molecular and fiber-optical channels \cite{farsad_neural_2018, karanov_end--end_2018}, or without any known analytical solution, e.g., finding codes for AWGN-channels with feedback \cite{kim_deepcode_2020}, DNNs have already proven to allow for promising application. Notably, the authors of the early work \cite{oshea_introduction_2017} demonstrate a complete communication system design by interpreting transmitter, channel and receiver as an autoencoder which is trained end-to-end similar to one DNN. The resulting encodings are shown to reach the BER performance of handcrafted systems in a simple AWGN scenario. A model-free approach based on reinforcement learning is proposed in \cite{aoudia_model-free_2019}. Using advances in unsupervised learning, also blind channel equalization can be improved \cite{caciularu_unsupervised_2020}.

In contrast to typical ML research areas, a model deficit does not apply to wireless communications. The models, e.g., AWGN, describe reality well and enable development of optimized algorithms. However, those algorithms may be too complex to be implemented. This algorithm deficit applies to the core problem typical for communications: classification in large inverse problems. Therefore, it is crucial to find an approximate solution with an excellent trade-off between detection accuracy and complexity.

\subsection{Related Work}

A prominent example for large inverse problems under current deep investigation and a key enabler for better spectral efficiency in 5G/6G are massive Multiple Input Multiple Output (MIMO) systems \cite{bjornson_massive_2017}. In an uplink scenario, a Base Station (BS) is equipped with a very large number of antennas (around $64$-$256$) and simultaneously serves multiple single-antenna User Equipments (UEs) on the same time-frequency resource. As a first step in receiver design, different tasks such as channel equalization/estimation and decoding are typically split to lower complexity. But still, an algorithm deficit applies to both MIMO detection and decoding of large block-length codes, e.g., LDPC and Polar codes, since Maximum A-Posteriori (MAP) detection has high computational complexity growing exponentially with system or code dimensions. Even its efficient implementation, the Sphere Decoder (SD), remains too complex in such a scenario \cite{jalden_complexity_2005}.

Hence, in communications history, many suboptimal solutions have been proposed to overcome the complexity bottleneck of the optimal detectors. One key approach is to relax the discrete Random Variables (\rv) to be continuous: Remarkable examples include Matched Filter (MF), Zero Forcing and MMSE equalization. But linear equalization with subsequent detection leads to a strong performance degradation compared to SD in symmetric systems.

A heuristic based on the latter is the V-Blast algorithm which first equalizes and then detects one layer with largest Signal-to-Noise Ratio (SNR) successively to reduce interference iteratively. A more efficient and sophisticated implementation, MMSE Ordered Successive Interference Cancellation (\mosic), is based on a sorted QR decomposition of a MMSE extended system matrix with post sorting and offers a good trade-off between complexity and accuracy \cite{wubben2003mmse}.

Pursuing another philosophy of mathematical optimization, the SemiDefinite Relaxation (SDR) technique \cite{luo_semidefinite_2010} treats MIMO detection as a non-convex homogeneous quadratically constrained quadratic problem and relaxes it to be convex by dropping the only non-convex requirement. Proving to be a close approximation, SDR is more complex than MOSIC and solved by interior point methods from convex optimization.

Furthermore, also probabilistic model-based ML techniques were introduced to improve the trade-off and to integrate detection seamlessly with decoding: Mean Field Variational Inference (MFVI) provides a theoretical derivation of soft Successive Interference Cancellation (SIC) and the Bethe approach lays the foundation for loopy belief propagation \cite{simeone2018brief}. Simplifying the latter, Approximate Message Passing (AMP) is derived known to be optimal for large system dimensions in \iid Gaussian channels and computational cheap \cite{jeon2015optimality}. As a further benefit, soft outputs are computed, today a strict requirement to account for subsequent soft decoding. But in practice, the performance of probabilistic approximations like MFVI and AMP suffers if the approximating conditions are not met, i.e., from the full-connected graph structure and finite dimensions in MIMO systems, respectively.

More recent work considers DNNs for application in MIMO systems and focus on the idea of deep unfolding \cite{monga_algorithm_2021, balatsoukas-stimming_deep_2019}. In deep unfolding, the number of iterations of a model-based iterative algorithm is fixed and its parameters untied. Further, it is enriched with additional weights and non-linearities to create a computational efficient DNN being optimized for performance improvements in MIMO detection \cite{samuel2017deep, samuel2019learning}, belief propagation decoding \cite{nachmani_learning_2016, nachmani_deep_2018, gruber_deep_2017} and MMSE channel estimation \cite{neumann_learning_2018}. The former approach DetNet, a generic DNN model with a large number of trainable parameters based on an unfolded projected gradient descent, proves DNNs to allow for a promising trade-off between accuracy and complexity. In \cite{he_model-driven_2018}, unfolding of an extension of AMP to unitarily-invariant channels, the Orthogonal AMP (OAMP), into OAMPNet is proposed adding only $2$ trainable parameters per layer. Offering promising performance, the complexity bottleneck of one matrix inversion per iteration makes this model-driven approach rather unattractive compared to DetNet. Another DNN-like network MMNet is inspired by iterative soft thresholding algorithms and AMP \cite{khani_adaptive_2020}: Striking the balance between expressiveness and complexity, and exploiting spectral and temporal locality, MMNet can be trained online for any realistic channel realization if coherence time is large enough. Since online training is in general wasteful, an efficient implementation non-trivial and requires particularly deep analysis, we focus in this work on offline learning. One major drawback of the latter approaches is that they focus on MIMO detection and do not provide soft outputs.

\subsection{Main Contributions}

The main contributions of this article are manifold: Inspired by recent ML research, we first introduce a CONtinuous relaxation of the probability mass function (pmf) of the disCRETE \rv s by a probability density function (pdf) from \cite{maddison2016concrete, jang2016categorical} to the MAP detection problem. The proposed CONCRETE relaxation offers many favorable properties: On the one hand, the pdf of continuous \rv s converges to the exact pmf in the parameter limit. On the other hand, we notice good algorithmic properties like avoiding marginalization and allowing for differentiation instead. By this means, we replace exhaustive search by computationally cheaper continuous optimization to approximately solve the MAP problem in any probabilistic non-linear model. We name our approach \cmdfull\ (\cmd).

Second, following the idea of Deep Unfolding, we unfold the gradient descent algorithm into a DNN-like model \cmdnet\ with a fixed number of iterations to allow for parameter optimization and to further improve detection accuracy while limiting complexity. By this means, we are able to combine the advantages of DNNs and model-based approaches. As the number of parameters is small, we are able to dynamically adapt them to easily adjust \cmdnet\ to different working points. Further, the resulting structure potentially allows for fast online training of \cmdnet.

Thirdly, we derive the optimization criterion from an information theoretic perspective and are hence able to provide probabilities of detection, i.e., reliable soft outputs. We show that optimization is then equivalent to learning an approximation of the Individual Optimal (IO) detector. This allows us to account for subsequent decoding, e.g., in MIMO systems, in contrast to literature \cite{samuel2019learning, khani_adaptive_2020}.

Finally, we provide numerical simulation results for use of \cmd\ and \cmdnet\ in MIMO systems including a variety of simulation setups, e.g., correlated channels, revealing \cmdnet\ to be a generic and promising approach competitive to State of the Art (SotA). Notably, we show superiority to other recently proposed ML-based approaches and demonstrate with simulations in coded systems \cmdnet's soft outputs to be reliable for decoders as opposed to \cite{samuel2019learning}. Furthermore, by estimating the computational complexity, we prove \cmd\ to feature a promising trade-off between detection accuracy and complexity. Notably, only the Matched Filter has lower complexity.

In the following, we first introduce the concrete relaxation to MAP detection in Section \ref{sec:2} using the example of an inverse linear problem. In Section \ref{sec:3}, we follow a different route and explain how to learn the posterior, i.e., replacing it by some tractable approximation. To yield a suitable model for this approximation, we propose to unfold \cmd\ into \cmdnet\ which we are then able to train by variants of Stochastic Gradient Descent (SGD). Finally in Section \ref{sec:num_res} and \ref{sec:conclusion}, we provide numerical results of the bit error performance in comparison to other SotA approaches using the example of uncoded and coded MIMO systems and summarize the main results, respectively.

\section{Concrete relaxation of MAP problem}
\label{sec:2}

\subsection{System Model and Problem Statement}

To motivate the concrete relaxation, we consider a probabilistic and (possibly) non-linear observation model described by a continuous and differentiable pdf $\prob(\yvec|\xvec)$. Based on this model, the task is to classify/detect the discrete multivariate \rv\ $\xvec$, i.e., $\xvec=\{\xvar_{\indx}\}_{\indx=1}^{\Nt}$ whose \iid elements are from a set $\disset$, given the observation $\yvec\in \complexnum^{\Nr \times 1}$.

To illustrate our findings with an example typically encountered in communications, we focus on a linear complex-valued observation model, e.g., MIMO system, although the following derivations hold without loss of generality for general $\prob(\yvec|\xvec)$. We first exclude coding from our model:
\begin{subequations}
	\begin{align}
		\yvec &= \chmat \xvec + \noisevec \label{eq:sys_model}\\
		\textrm{with} \quad \prob(\yvec|\xvec,\chmat,\noisestd^2)&=\frac{1}{\pi^{\Nr} \noisestd^{2\Nr}} e^{-\frac{1}{\noisestd^2}(\yvec-\chmat\xvec)^\Hm(\yvec-\chmat\xvec)} \label{eq:sys_model_prob} \eqpoint
	\end{align}
	\label{eq:mimomodel}
\end{subequations}
There, a linear channel $\chmat\in \complexnum^{\Nr\times \Nt}$ with statistic $\prob(\chmat)$, e.g., such that taps $\chtap_{\indy\indx}\sim \normdisc(0, 1/\Nr)$ are \iid Gaussian distributed, introduces correlation between the elements $\xvar_{\indx}$ with $\evaltxt[\abstxt[\xvar_{\indx}]^2]=1$ from typical modulation sets $\disset$, e.g., BPSK, $8$-PSK or $16$-QAM. Then, Gaussian noise $\noisevec\sim \normdisc(\zero,\noisestd^2 \eye_{\Nr})$ with variance $\noisestd^2$ distributed according to $\prob(\noisestd^2)$ interferes. The matrix $\eye_{\Nr}$ denotes the identity matrix of dimension $\Nr\times\Nr$. For the following derivations, note that we are able to replace $\yvec$ by one total observation $\tobsvec$ including \rv s $\chmat$ and $\noisestd^2$ without loss of generality since $\xvec$, $\chmat$ and $\noisestd^2$ are statistically independent:
\begin{align}
 \prob(\tobsvec|\xvec)&=\prob(\yvec,\chmat,\noisestd^2|\xvec)=\prob(\yvec|\xvec,\chmat,\noisestd^2) \cdot \prob(\chmat) \cdot \prob(\noisestd^2) \eqpoint \label{eq:tobs}
\end{align}
In this detection problem, there exist two optimal detectors from a probabilistic Bayesian viewpoint: First, we have the likelihood function $\prob(\yvec|\xvec)$ but would like to infer the most likely transmit signal $\xvec$ based on an a-posteriori pdf $\prob(\xvec|\yvec)$. Using Bayes rule, we are able to reform the MAP problem w.r.t. the known likelihood into
\begin{subequations}
	\begin{align}
	\xest &= \argmax[\xvec\in \disset^{\Nt\times 1}] \prob(\xvec|\yvec) \label{eq:map0} \\
	&= \argmax[\xvec\in \disset^{\Nt\times 1}] \prob(\yvec|\xvec) \cdot \prob(\xvec)\\
	&= \argmin[\xvec\in \disset^{\Nt\times 1}] -\ln \prob(\yvec|\xvec) - \ln \prob(\xvec) \label{eq:map}
	\end{align}
\end{subequations}
where $\prob(\xvec)$ is the known a-priori pdf. Since the \rv\ is discrete, i.e., $\xvar_{\indx}\in\disset$, an exhaustive search over all element combinations is required to solve the MAP problem becoming computational intractable for large system dimensions. Note that the Sphere Detector (SD) provides an efficient implementation \cite{jalden_complexity_2005}. Second, we notice that the MAP detector only delivers the most likely received vector $\xvec$. Hence, it minimizes frame error rate and provides hard decisions.

In coded systems with soft decoders usually employed today, delivering soft information is a strict requirement. The Individual Optimal (IO) detector delivers such soft output as probabilities and is optimal in terms of minimizing the Symbol Error Rate (SER) per individual symbol without coding. It is obtained by evaluating the marginal posterior distribution w.r.t. every single $\xvar_{\indx}$:
\begin{align}
\xvarest_{\indx} &= \argmax[\xvar_{\indx}\in \disset] \prob(\xvar_{\indx}|\yvec)= \argmax[\xvar_{\indx}\in \disset] \frac{\summ{\xvec\textbackslash \xvar_{\indx}}{} \prob(\yvec|\xvec)\cdot \prob(\xvec)}{\summ{\xvar_{\indx}} \summ{\xvec\textbackslash \xvar_{\indx}} \prob(\yvec|\xvec)\cdot \prob(\xvec)} \eqpoint
\label{eq:io_detector}
\end{align}
However, it has higher complexity due to required marginalization w.r.t. $\xvec$. Since the MAP detector performance coincides with the IO detector in the high SNR regime and is of lower complexity, we restrict to the MAP detector as a benchmark in simulations without coding.

\subsection{Concrete Distribution}

We now focus on the following question to improve the performance complexity trade-off: How to model the prior information $\prob(\xvec)$ accurately by some approximation $\prob(\xt)$? In \cite{beck_concrete_2020}, we proposed to use ML tricks from \cite{maddison2016concrete, jang2016categorical} to achieve this and to make inference computationally tractable. The idea was recently discovered in the ML community in the context of unsupervised learning of generative models \cite{maddison2016concrete, jang2016categorical}. There, marginalization to compute the objective function, the evidence, becomes intractable. Therefore, the Evidence is replaced by its Lower BOund (ELBO) by means of an auxiliary posterior function. But optimizing w.r.t. the ELBO results in high variance of the gradient estimators. For variance reduction, the so called reparametrization trick is used and leads to an optimization structure similar to an autoencoder known as the variational autoencoder \cite{simeone2018brief}. There, the stochastic node is reparametrized by a continuous \rv, e.g., a Gaussian, and its parameters, e.g., mean and variance. In contrast to continuous \rv s, reparametrization of discrete \rv\ is not possible. Hence, a CONtinuous relaxation of disCRETE \rv s, the CONCRETE distribution, was proposed in \cite{maddison2016concrete, jang2016categorical} independently.

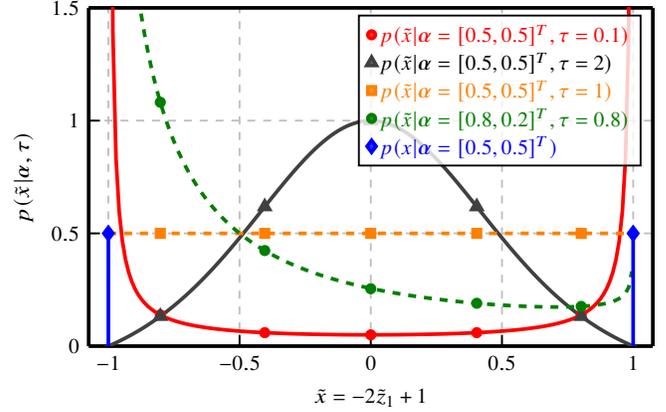
\begin{figure}[!t]
	\centerline{\input{TikZ/pres_concrete_bern_resized2.tikz}}
	\caption{The concrete pdf $p(\xtvar|\priorvec,\softmaxtemp)$ shown for different parameter sets and $\Nclass=2$. It relaxes the Bernoulli pmf $p(\xvar|\priorvec)$ into the interior. Notably, for $\softmaxtemp\leq (\Nclass-1)^{-1}$, it is log-convex and log-concave otherwise. Symmetry results if $\prior_1=\ldots=\prior_{\Nclass}$.}
	\label{fig1:concrete_bern}
\end{figure}

To explain the introduction of this relaxation to MAP detection, let us assume that we have the discrete binary \rv\ $\xvar\in \disset$ with $\disset= \{-1,+1\}$. Further, we define the discrete \rv\ $\oh$ as a one-hot vector where all elements are zero except for one element, i.e., $\oh\in \{0,1\}^{2\times 1}$ with two possible realizations $\oh_1=[1,0]^\trapo$, $\oh_2=[0,1]^\trapo$. In addition, we describe the values of $\disset$ by the representer vector $\dissetvec=[-1,1]^\trapo$. That way, we can write $\xvar=\oh^\trapo \dissetvec$, e.g., $\xvar=[1,0]\cdot [-1,1]^\trapo=-1$. Now, the one-hot vector $\oh\in \{0,1\}^{\Nclass\times 1}$ represents a categorical \rv\ with $\Nclass=\abstxt[\disset]$ classes. Connecting Monte Carlo methods to optimization \cite{maddison2016concrete}, the Gumbel-Max trick states that we are able to generate samples, i.e., classes, of such a categorical \rv\ or pmf $\prob(\xvar)$ by sampling an index $\indiopt$ from $\Nclass$ continuous \iid Gumbel \rv s $\gumbel_{\indi}$ known from extreme value theory:
\begin{equation}
	\indiopt=\argmax[\indi=1,\dots,\Nclass] \ln \prob(\xvar=\dissetel_{\indi}) + \gumbel_{\indi} \eqpoint
\label{eq:argmax0}
\end{equation}
Defining the function $\ohfunc(\indiopt)$ which sets the $\indiopt$-th element in the one-hot vector $\ohvar_{\indiopt}=1$ and $\ohvar_{\indl\neq\indiopt}=0$, the Gumbel-Max trick hence allows to sample one-hot vectors $\oh$. Thus, we are able to reparametrize $\oh$ through a continuous multivariate Gumbel \rv\ $\gumbelvec\in\realnum^{\Nclass\times 1}$ and a vector $\priorvec\in [0,1]^{\Nclass\times 1}$ of class probabilities $\prob(\xvar=\dissetel_{\indc})$ with $\sum_{\indc=1}^{\Nclass} \prior_{\indc}=1$:
\begin{equation}
	\oh = \ohfunc\left(\argmax[\indi=1,\dots,\Nclass] \left[\ln(\priorvec) + \gumbelvec\right] \right) \eqpoint
\label{eq:argmax}
\end{equation}
Note that \eqref{eq:argmax} and equally $\xvar$ are still discrete \rv s, i.e., $\prob(\oh)\hat{=}\prob(\xvar)$, but represented in probabilistic sense by continuous \rv s $\gumbelvec$. To arrive at a continuous \rv\, we now replace the one-hot and $\armax$ computation in \eqref{eq:argmax} by the softmax function \cite{maddison2016concrete, jang2016categorical}:
\begin{equation}
	\oht=\softmax{\gumbelvec}=\frac{e^{(\ln (\priorvec)+\gumbelvec)/\tau}}{\sum_{\indi=1}^{\Nclass}e^{(\ln \prior_\indi+g_\indi)/\tau}} \eqpoint
\label{eq:softmax}
\end{equation}
The resulting \rv\ $\oht\in [0,1]^{\Nclass\times 1}$ is the so called concrete or Gumbel-Softmax \rv\ and now continuous, e.g., $\oht=[0.2, 0.8]^\trapo$. It is controlled by a parameter, the softmax temperature $\tau$. The distribution of $\oht$ in \eqref{eq:softmax} was found to have a closed form density in \cite{maddison2016concrete, jang2016categorical} which gives the definition of the concrete distribution:
\begin{equation}
	\prob(\oht|\priorvec,\tau)=(\Nclass-1)!\ \tau^{\Nclass-1} \prod_{\indc=1}^{\Nclass} \left(\frac{\prior_\indc \ohtvar_\indc^{-\tau-1}}{\sum_{\indi=1}^{\Nclass}\prior_\indi \ohtvar_{\indi}^{-\tau}}\right) \eqpoint \label{eq:conc_dist}
\end{equation}
With $\oht$, we are finally able to relax the discrete \rv\ $\xvar$ into a continuous \rv\ $\xtvar$ by defining $\xtvar=\oht^\trapo \dissetvec$. Now, our derivation of the relaxation is complete. In Fig. \ref{fig1:concrete_bern}, we illustrate the distribution $\prob(\xtvar)$ for the special case $\Nclass=2$ of binary \rv s in comparison to the original categorical pmf $\prob(\xvar)$, i.e., a Bernoulli pmf. It has the following properties \cite{maddison2016concrete}: First, we are able to reparametrize the concrete \rv\ $\oht$ and hence the \rv s $\xtvar$ by Gumbel variables $\gumbelvec$, a direct result from the initial idea \eqref{eq:softmax}. Moreover, the smaller $\tau$, the more $\oht$ approaches a categorical \rv\ and the approximation becomes more accurate. Thus, the statistics of $\xvar$ and $\xtvar$ remain the same for $\tau\rightarrow 0$.

\subsection{Reparametrization}

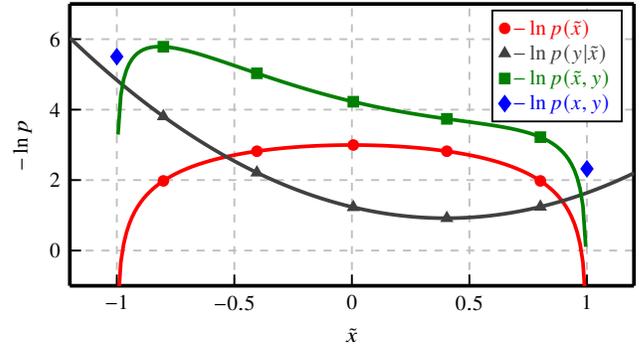
\begin{figure}[!t]
	\centerline{\input{TikZ/pres_concrete_MAP_resized.tikz}}
	\caption{Exemplary plot of the concrete binary MAP cost function (green) for model \eqref{eq:mimomodel} (with $\Nt=1$, $\chmat=1$, $y=0.4$, $\noisestd^2=4$, $\prior_1=0.5$, $\prior_2=0.5$ and $\softmaxtemp=0.1$) and the contribution of conditional (black) and prior pdf (red) to it. The original binary MAP cost function (blue) is shown for comparison.}
	\label{fig2:concreteMAP}
\end{figure}

In \cite{beck_concrete_2020}, our idea is to use the concrete distribution in order to relax the MAP problem \eqref{eq:map} to
\begin{align}
\xest &= \argmin[{\xt\in\left[\min(\disset),\max(\disset)\right]^{\Nt \times 1}}] -\ln \prob(\yvec|\xt) - \ln \prob(\xt) \eqpoint \label{eq:map_relax}
\end{align}
Note that the original MAP problem is included or recovered in the zero temperature limit $\softmaxtemp^{(\indgrad)}\rightarrow 0$. Moreover, the objective function in \eqref{eq:map_relax} may be non-convex as illustrated in Fig. \ref{fig2:concreteMAP} for $\Nclass=2$. The conditional pdf $\prob(\yvec|\xt)$ is log-concave and the prior concrete pdf $\prob(\xt)$ log-convex for $\tau\leq (\Nclass-1)^{-1}$ \cite{maddison2016concrete}, so the negative log joint distribution $\prob(\yvec,\xt)$ forms a non-convex objective function \eqref{eq:map_relax}. The reparametrization of $\oht$ by $\gumbelvec$ helps to rewrite \eqref{eq:map_relax} by expressing each $\xtvar_{\indx}$ in $\xt$ with \eqref{eq:softmax} by the \rv\ $\gumbelvec_{\indx}$, $\indx=1,\dots,\Nt$ of \iid Gumbel \rv s $\gumbel_{\indc\indx}$:
\begin{align}
\xt(\gumbelmat)=\begin{bmatrix}
\xtvar_1 \\
\vdots \\
\xtvar_{\Nt}
\end{bmatrix}&=\begin{bmatrix}
\oht_1^T \\
\vdots \\
\oht_{\Nt}^T
\end{bmatrix} \dissetvec
= \begin{bmatrix}
\softmax{\gumbelvec_1}^T \\
\vdots \\
\softmax{\gumbelvec_{\Nt}}^T
\end{bmatrix} \dissetvec \\
\textrm{with} \quad \gumbelmat &= \begin{bmatrix}
\gumbelvec_1 & \cdots & \gumbelvec_{\Nt} \\
\end{bmatrix} \in \realnum^{\Nclass\times \Nt} \eqpoint
\end{align}
By doing so, we will obtain an unconstrained optimization problem w.r.t. matrix $\gumbelmat$. Now, we reformulate the relaxed MAP problem \eqref{eq:map_relax}: This means, we replace the likelihood $\prob(\yvec|\xt)$ by $\prob(\yvec|\gumbelmat)$ and introduce the Gumbel distribution $\prob(\gumbel_{\indc\indx})=\exp{(-\gumbel_{\indc\indx}-\exp{(-\gumbel_{\indc\indx})})}$ as the new prior distribution:
\begin{subequations}
\begin{align}
\gumbelest =& \argmin[\gumbelmat\in \realnum^{\Nclass\times \Nt}] -\ln \prob(\yvec|\gumbelmat) -\ln \prob(\gumbelmat) \\
=& \argmin[\gumbelmat\in \realnum^{\Nclass\times \Nt}] -\ln \prob(\yvec|\gumbelmat) - \sum \limits_{\indx=1}^{\Nt}\sum \limits_{\indc=1}^{\Nclass}\ln \prob(\gumbel_{\indc\indx}) \\
=& \argmin[\gumbelmat\in \realnum^{\Nclass\times \Nt}] \underbrace{-\ln \prob(\yvec|\gumbelmat) + \ones^\trapo \gumbelmat \ones + \ones^\trapo e^{-\gumbelmat} \ones}_{\obf(\gumbelmat,\softmaxtemp)} \label{eq:concreteMAP} \eqpoint
\end{align}
\end{subequations}
However, due to the softmax and exponential terms in $\obf(\gumbelmat,\softmaxtemp)$, \eqref{eq:concreteMAP} has no analytical solution. Furthermore, $\obf(\gumbelmat,\softmaxtemp)$ may be non-convex: For real-valued model \eqref{eq:mimomodel}, $\Nt=1$, $\chmat=1$ and $\Nclass=2$, the first term is a vertically shifted, squared and scaled two-dimensional non-convex sigmoid function w.r.t. $\gumbel_1$ and $\gumbel_2$. The operations applied to the sigmoid do not change non-convexity. Also the sum of this non-convex term and convex functions, i.e., linear and exponential functions, remains non-convex.

\subsection{Gradient Descent Optimization}

One common strategy for solving the non-linear and non-analytical problem \eqref{eq:concreteMAP} is to use a variant of gradient descent based approaches. Since we aim to reduce complexity, we choose the most basic form steepest descent. The minimum is approached iteratively by taking gradient descent steps until the necessary condition
\begin{align}
\frac{\partial \obf(\gumbelmat,\softmaxtemp)}{\partial \gumbelmat} &= \zero
\end{align}
is fulfilled. We point out that convergence to the global solution depends heavily on the starting point initialization since the objective function may be non-convex. A reasonable choice of starting point value is $\xt^{(0)}=\evaltxt[\xvec]=\priorvec^T\cdot \dissetvec$, i.e., the expected value of the true discrete \rv\ $\xvec$. We achieve this by setting $\gumbelmat^{(0)}=\zero$ and $\tau = 1$. After some tensor/matrix calculus and by noting that every $\xtvar_{\indx}$ only depends on one $\gumbelvec_{\indx}$, the gradient descent step for \eqref{eq:concreteMAP} in iteration $\indgrad$ is:
\begin{subequations}
\begin{align}
\gumbelmat^{(\indgrad+1)} = & \gumbelmat^{(\indgrad)} - \delta^{(\indgrad)} \cdot \frac{\partial \obf(\gumbelmat,\softmaxtemp)}{\partial \gumbelmat} \bigg|_{\gumbelmat = \gumbelmat^{(\indgrad)}} \\
\frac{\partial \obf(\gumbelmat,\softmaxtemp)}{\partial \gumbelmat}= & -\begin{bmatrix}
	\frac{\partial \xtvar_1(\gumbelvec_1)}{\partial \gumbelvec_1} & \cdots & \frac{\partial \xtvar_{\Nt}(\gumbelvec_{\Nt})}{\partial \gumbelvec_{\Nt}}
\end{bmatrix} \nonumber \\ &\cdot \diag{\derloglike} + 1 - e^{-\gumbelmat} \label{eq:grad2} \\
\frac{\partial \xtvar_{\indx}(\gumbelvec_{\indx})}{\partial \gumbelvec_{\indx}} = & \frac{1}{\tau^{(\indgrad)}} \cdot \left[\diag{\softmax{\gumbelvec_{\indx}}} \cdot \dissetvec- \softmax{\gumbelvec_{\indx}} \cdot \xtvar_{\indx}(\gumbelvec_{\indx}) \right] \eqpoint
\end{align}
\label{eq:grad}
\end{subequations}
The operator $\diag{\plvec}$ creates a diagonal matrix with the vector $\plvec$ on its main diagonal. The step-size $\delta^{(\indgrad)}$ can be chosen adaptively in every iteration $\indgrad$ just as the parameter $\softmaxtemp^{(\indgrad)}$. For example, we can follow a heuristic schedule like in simulated annealing: We start with a large $\softmaxtemp^{(\indgrad)}$ and decrease until we approach the true prior pdf for $\softmaxtemp^{(\indgrad)}\rightarrow 0$. Finally, after the last iteration $\Ni$, we get as a result the continuous estimate $\gumbelmat^{(\Ni)}$. For approximate detection of $\xvec$ in \eqref{eq:map}, the estimate has to be transformed back to the discrete domain by quantizing $\xt$ onto the discrete set $\disset$:
\begin{align}
\xest=\argmin[\xvec\in \disset^{\Nt\times 1}]\norm{\xvec-\xt\left(\gumbelmat^{(\Ni)}\right)}_2 \eqpoint \label{eq:xquant}
\end{align}
In the following, we name this detection approach \cmdfull\ (\cmd). It is generic and applicable in any differentiable probabilistic non-linear model. For our guiding example of a linear Gaussian model \eqref{eq:mimomodel}, we are able to give the explicit expression of
\begin{equation}
	\derloglike=-\frac{2}{\noisestd^2} \cdot \lbb\chmat^\Hm \chmat \xt(\gumbelmat)- \chmat^\Hm \yvec \rbb
\end{equation}
in \eqref{eq:grad2}. This means that further only elementwise nonlinearities and matrix vector multiplications are present in this example. As a final remark, we note that our implementation of Section \ref{sec:num_res} relies on scaling of the objective function by the noise variance parameter, i.e., $\noisestd^2 \cdot \obf(\gumbelmat,\softmaxtemp)$. Although scaling does not change the optimization problem, we observed that this slightly modified version of \eqref{eq:grad} is numerically more stable.

\subsection{Special Case: Binary Random Variables}

Noting that the softmax function \eqref{eq:softmax} is normalized, we are able to eliminate one degree of freedom in matrix $\gumbelmat\in \realnum^{\Nclass\times \Nt}$ along dimension $\Nclass$. For the special case of binary \rv s or $\Nclass=2$ classes, this means that the matrix $\gumbelmat$ can be reduced to a vector $\logvec\in \realnum^{\Nt\times 1}$ of logistic \rv s to derive a different algorithm of low complexity. Here, we only briefly summarize the result of binary \cmd\ in a real-valued system model and refer the reader to \cite{beck_concrete_2020} for the complete derivation:
\begin{subequations}
	\begin{align}
	\logvec^{(\indgrad+1)} &= \logvec^{(\indgrad)} - \delta^{(\indgrad)} \cdot \frac{\partial \obf(\logvec,\softmaxtemp)}{\partial \logvec} \bigg|_{\logvec = \logvec^{(\indgrad)}} \\
	\frac{\partial \obf(\logvec,\softmaxtemp)}{\partial \logvec} &= -\frac{\partial \xt(\logvec)}{\partial \logvec} \cdot \derloglikes + \tanh\left(\frac{\logvec}{2}\right) \\
	\overset{\txt{\eqref{eq:mimomodel}}}&{=} \frac{1}{\noisestd^2} \cdot \frac{\partial \xt(\logvec)}{\partial \logvec} \cdot \lbb\chmat^T \chmat \xt(\logvec)- \chmat^T \yvec\rbb + \tanh\left(\frac{\logvec}{2}\right) \label{eq:reg_term}\\
	\frac{\partial \xt(\logvec)}{\partial \logvec} &= \frac{1}{2\tau^{(\indgrad)}}\cdot\diag{ 1-\xt^2(\logvec)} \\
	\xt(\logvec)&=\txt{tanh}\left(\frac{\ln\left(1/\prior-1\right)+\logvec}{2\tau^{(\indgrad)}}\right) \eqpoint
	\end{align}
	\label{eq:bin_grad}
\end{subequations}
The final step consists again of quantization - in this case it simplifies to the sign function: $\xest=\signum(\xt(\logvec^{(\Ni)}))$.

\section{Learning to Relax}
\label{sec:3}

Although being simple and computational efficient, using a gradient descent approach like \eqref{eq:grad} and \eqref{eq:bin_grad} leads to several inconveniences. Regarding theoretical properties, a major drawback becomes apparent: Convergence of the gradient descent steps to an optimum is slow since consecutive gradients are perpendicular. Also practical questions arise: How to choose the parameters $\tau^{(j)}$ and $\delta^{(j)}$ and the number of iterations $\Ni$ for a good complexity performance trade off? And how are we able to deliver soft information, e.g., probabilities, to a soft decoder which is standard in today's communication systems?

Our idea is to improve \cmd\ by learning and in particular the idea of deep unfolding to address these questions. This means we have to deal with
\begin{itemize}
	\item[A.] how learning is defined
	\item[B.] the application of deep unfolding to \cmd.
\end{itemize}

\subsection{Basic Problem of Learning}
To introduce our notation of learning, we revisit our basic task of MAP detection. Ideally, we would like to infer the most likely transmit signal $\xvec$ based on an a-posteriori pdf $\prob(\xvec|\yvec)$. But as pointed out earlier, evaluation of $\prob(\xvec|\yvec)$ has intractable complexity. For this reason, we propose to relax the MAP problem and \cmd, respectively.

Another idea to tackle this problem is to approximate this pdf $\tprob(\xvec|\yvec)$ by another computationally tractable pdf $\aprob(\xvec|\yvec)$, e.g., by calculation of $\aprob(\xvec|\yvec)$ using few samples/observations $\xvec$, and use this pdf for inference. Note that this approach includes cases where we do not know the pdf $\prob(\xvec|\yvec)$ completely. The quality of the approximation can be quantified by the information theoretic measure of Kullback-Leibler (KL) divergence:
\begin{align}
\label{eq:kl_divergence}
\dkl{\tprob}{\aprob}&= -\summ{\xvec\in\disset^{\Nt\times 1}} \tprob(\xvec|\yvec) \ln \frac{\tprob(\xvec|\yvec)}{\aprob(\xvec|\yvec)} \\
					&= \eval[\xvec\sim \tprob(\xvec|\yvec)]{\ln \frac{\tprob(\xvec|\yvec)}{\aprob(\xvec|\yvec)}} \eqpoint
\end{align}
Just as the Mean Square Error (MSE), the measure of KL divergence can be used to define an optimization problem targeting at a tight $\aprob(\xvec|\yvec)$ as a solution. This brings me to a crucial viewpoint of this article: \textbf{Learning is defined to be the optimization process aiming to derive a good approximation $\aprob(\xvec|\yvec)$ of $\tprob(\xvec|\yvec)$, i.e.,}
\begin{align}
\label{eq:learning_problem}
\aprob^{\ast}(\xvec|\yvec)= \argmin[\aprob] \dkl{\tprob}{\aprob} \eqpoint
\end{align}
This kind of problem is also referred to as Variational Inference (VI). We can rewrite the KL divergence into a sum of cross entropy $\HE[\tprob, \aprob]$ and entropy $\HE[\tprob]$:
\begin{align}
\dkl{\tprob}{\aprob}&=\eval[\xvec\sim \tprob(\xvec|\yvec)]{-\ln \aprob(\xvec|\yvec)} - \eval[\xvec\sim \tprob(\xvec|\yvec)]{-\ln \tprob(\xvec|\yvec)} \\
&=\HE[\tprob, \aprob] - \HE[\tprob] \eqpoint
\end{align}
Since we defined the basic learning problem \eqref{eq:learning_problem} w.r.t. approximation $\aprob$, we can neglect the entropy term $\HE[\tprob]$ independent of $\aprob$ and use cross entropy as the learning criterion. If we further restrict $\aprob$ to a model $\aprob(\xvec|\yvec,\params)$ with parameters $\params$, the optimization problem now reads:
\begin{align}
\params^{\ast}&= \argmin[\params] \HE[\tprob, \aprob] \eqpoint
\label{eq:cross_entropy}
\end{align}
We note that problem \eqref{eq:cross_entropy} is solved separately for each $\yvec$ and thus parameters $\params$ need to be continuously updated in an online learning procedure. Since this procedure is not computationally efficient, we follow an offline learning strategy known as Amortized Inference \cite{simeone2018brief} and define one inference distribution $\aprob(\xvec|\yvec,\params)$ for any value $\yvec$:
\begin{align}
\params^{\ast}&= \argmin[\params] \eval[\yvec\sim \prob(\yvec)]{\HE[\tprob(\xvec|\yvec), \aprob(\xvec|\yvec,\params)]} \label{eq:cross_entropy1} \\
&= \argmin[\params] \eval[\yvec\sim \prob(\yvec)]{\eval[\xvec\sim \prob(\xvec|\yvec)]{-\ln \aprob(\xvec|\yvec,\params)}} \\
&\approx \argmin[\params] -\frac{1}{\Nsamp} \summ[\Nsamp]{\idxi=1} \ln \aprob(\xvec_{\idxi}|\yvec_{\idxi},\params) \quad \text{, } \Nsamp \rightarrow \infty
\eqpoint
\label{eq:cross_entropy2}
\end{align}
Rewriting the optimization criterion of \eqref{eq:cross_entropy1} into
\begin{align}
	& \eval[\tobsvec\sim \prob(\tobsvec)]{\eval[\xvec\sim \prob(\xvec|\tobsvec)]{-\ln \aprob(\xvec|\tobsvec,\params)}} \\
	=& \eval[\noisestd^2\sim \prob(\noisestd^2)]{\eval[\chmat\sim \prob(\chmat)]{\eval[\yvec\sim \prob(\yvec|\chmat,\noisestd^2)]{\evalnb[\xvec\sim \prob(\xvec|\tobsvec)]{-\ln \aprob(\xvec|\tobsvec,\params)}}}} \nonumber
\end{align}
for our guiding example \eqref{eq:mimomodel}, we note that we are able to amortize across all observations $\tobsvec$ from \eqref{eq:tobs} and hence to obviate the need for online training also for each channel $\chmat$ and noise variance $\noisestd^2$ at the potential cost of accuracy.

The final result \eqref{eq:cross_entropy2} equals the maximum likelihood problem in supervised learning. We make use of it in the following since it allows for numerical optimization based on data points $\{\xvec_{\idxi},\yvec_{\idxi}\}$. Furthermore, it proves to be a Monte Carlo approximation of \eqref{eq:cross_entropy1} and is hence well motivated from information theory \cite{simeone2018brief}.

\subsection{Idea of Unfolding and Application to \cmd}
\label{subsec:unfolding}

Learning gives us the ability to obtain a tractable approximation $\aprob(\xvec|\yvec,\params)$. But it remains one question: How to choose a suitable functional form of $\aprob(\xvec|\yvec,\params)$ of low complexity and for good performance? We follow the idea of deep unfolding from \cite{monga_algorithm_2021, balatsoukas-stimming_deep_2019} and apply it to our model-based approach \cmd\ with parameters $\params=\lcb\softmaxtemp^{(0)}, \dots, \softmaxtemp^{(\Ni)}, \grstepsize^{(0)}, \dots, \grstepsize^{(\Ni-1)}\rcb\in \realnum^{(2\Ni+1)\times 1}$ able to relax tightly. Thereby, we combine strengths of DNNs and the latter: DNNs are known to be universal approximators \cite{hornik_multilayer_1989} and their fixed structure of parallel computations layer per layer allows to define a good performance complexity trade off at run time. But if the model is dynamic and changes, e.g., the channel or noise over time, reiterated optimization of \eqref{eq:cross_entropy}, i.e., possibly wasteful online training, is required and the benefit disappears. Fortunately, we know our model \eqref{eq:mimomodel}, a MIMO channel, well and are able to use generative model-based approaches which mostly rely on a suitable approximation of \eqref{eq:learning_problem} for computational tractability. For example, MFVI and AMP belong to this algorithm family. By model-based DNN design, we introduce varying model parameters like channel or noise explicitly and in a more sophisticated way into the DNN design and thus make efficient offline learning from \eqref{eq:cross_entropy2} at only a small cost of accuracy possible. Indeed, training of a DNN for our guiding example \eqref{eq:mimomodel} simply fed with inputs $\yvec$ and $\chmat$, reshaped as a vector, does not converge/lead to satisfactory results if trained offline \cite{samuel2019learning}.

\begin{figure}[!t]
	\centerline{\input{TikZ/CMD_layer.tikz}}
	\caption{One layer of the unfolded binary \cmd\ algorithm \cmdnet\ when applied to MIMO systems. In red: trainable parameters.}
	\label{fig3:cmd_unfolding}
\end{figure}
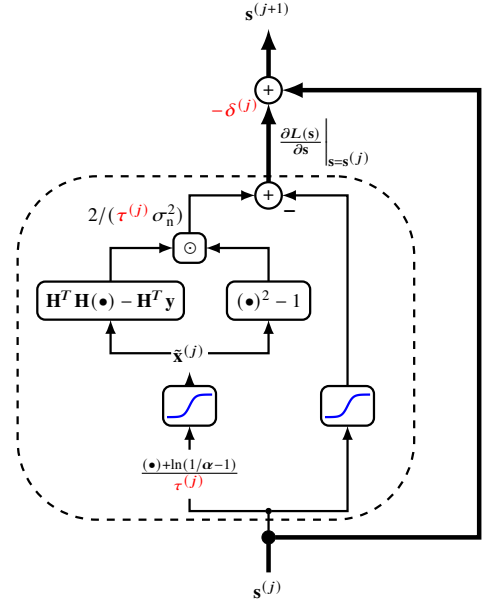

This means we unfold the iterations \eqref{eq:grad} of \cmd\ into a DNN by untying the parameters $\softmaxtemp^{(\grit)}$ and $\grstepsize^{(\grit)}$. Furthermore, we fix the complexity by setting the number of iterations $\Ni$. The resulting graph illustrated in Fig. \ref{fig3:cmd_unfolding} for binary \cmd\ and \eqref{eq:mimomodel} has a DNN-like structure which should be able to generalize and approximate well at the same time. Owing to the skip connection from $\logvec^{(\grit)}$ to  $\logvec^{(\grit +1)}$ on the right hand side, the structure resembles a Residual Network (ResNet) layer which is SotA in image processing \cite{he_deep_2016}. It is a result of the gradient descent approach which allows to interpret optimization of ResNets as learning gradient descent steps. The reason for the success of ResNet lies in the skip connection: The training error is able to backpropagate through it to early layers which allows for fast adaptation of early weights and hence fast training of DNNs. This makes \cmd\ especially suitable for online training proposed in \cite{khani_adaptive_2020} and allows for refinement in application.

As before, we have to define a final layer which is now also used for optimization. Usually, its output is chosen to be an continuous estimate of $\xvec$ and optimized w.r.t. the MSE criterion, see \cite{samuel2019learning, khani_adaptive_2020}. This viewpoint relaxes the estimate $\xest$ into $\realnum^{\Nt\times 1}$ and assumes a Gaussian distribution for errors at the output. In our case, the output would correspond to $\xt(\gumbelmat^{(\Ni)})$ from \eqref{eq:xquant}. But this is in contrast to our information theoretic viewpoint on learning which states that we want to approximate an output of the true pmf $\tprob(\vtvar|\vobs)$. Like in MFVI, we assume a factorization of the approximating posterior to make it computationally tractable and derive our learning criterion:
\begin{align}
\HE[\tprob, \aprob] &= -\summ{\vtvar\in\disset^{\Nt}} \tprob(\vtvar|\vobs) \cdot \ln \aprob(\vtvar|\vobs,\params) \\
					\overset{\txt{MFVI}}&{=} -\summ{\vtvar\in\disset^{\Nt}} \tprob(\vtvar|\vobs)  \cdot \ln \produ{\tvar_{\idxn}\in\disset} \aprob_{\idxn}(\tvar_{\idxn}|\vobs,\params) \\
 					&=	-\summ{\tvar_{\idxn}\in\disset} \ln \aprob_{\idxn}(\tvar_{\idxn}|\vobs,\params)  \cdot \summ{\tvar_{/\idxn}\in\disset^{\Nt-1}} \tprob(\vtvar|\vobs) \\
 					&=	-\summ{\tvar_{\idxn}\in\disset} \tprob(\tvar_{\idxn}|\vobs)  \cdot \ln \aprob_{\idxn}(\tvar_{\idxn}|\vobs,\params) \\
 					&=\summ{\tvar_{\idxn}\in\disset} \HE[\tprob(\tvar_{\idxn}|\vobs), \aprob_{\idxn}(\tvar_{\idxn}|\vobs,\params)] \eqpoint \label{eq:cross_entropy3}
\end{align}
This interesting result shows that assuming MFVI factorization leads to an optimization criterion w.r.t. the soft output $\tprob(\tvar_{\idxn}|\vobs)$ of the IO detector \eqref{eq:io_detector}. This soft output is required for subsequent decoding and thus exactly what we need.

The last step of our idea consists of inserting our unfolded \cmd\ structure into $\aprob_{\idxn}(\tvar_{\idxn}|\vobs,\params)$. Hence, we propose to use a softmax function for the last layer being a typical choice for classification in discriminative probabilistic models. Fortunately, \cmd\ already includes this softmax function as part of its structure so we rewrite
\begin{align}
\aprob_{\idxn}(\xvar_{\idxn}|\yvec,\params)=\produ[\Nclass]{\indc=1}\aprob_{\idxn,\indc}(\xvar_{\idxn}|\yvec,\params)^{(\xvar_{\idxn}=\dissetel_\indc)}=\produ[\Nclass]{\indc=1} \ohtvar_{\idxn,\indc}^{(\xvar_{\idxn}=\dissetel_\indc)}
\end{align}
with $\oht_{\indx}=\softmaxvar_{\softmaxtemp^{(\Ni)}}(\gumbelvec_{\idxn}^{(\Ni)})$ from the last iteration $\Ni$ of \eqref{eq:grad}. To summarize, we optimize the parameter set $\params$ of our approximating pdf $q(\xvec|\yvec,\params)$ based on \cmd:
\begin{align}
\params^{\ast}&= \argmin[\params] \eval[\yvec\sim \prob(\yvec)]{\HE[\tprob(\xvec|\yvec), \aprob(\xvec|\yvec,\params)]} \label{eq:cross_entropy4_1} \\
			&\approx \argmin[\params] -\frac{1}{\Nsamp} \summ[\Nsamp]{\idxi=1} \summ[\Nt]{\idxn=1} \begin{bmatrix}\xvar_{\indx}=\dissetel_1 \\ \vdots	\\ \xvar_{\indx}=\dissetel_{\Nclass}\end{bmatrix}^T \ln\lpa \softmaxvar_{\softmaxtemp^{(\Ni)}}(\gumbelvec_{\idxn}^{(\Ni)})\rpa \eqpoint
\label{eq:cross_entropy4_2}
\end{align}
As a side effect, we also learn to relax with \cmd\ by $\softmaxtemp^{(\indgrad)}$. We call this approach based on unfolding of \cmd\ \cmdnet. The optimization problem \eqref{eq:cross_entropy4_2} can be efficiently solved by variants of Stochastic Gradient Descent (SGD). Thanks to having a model, we are able to create infinite training and test data for reasonable approximation of \eqref{eq:cross_entropy4_1} by \eqref{eq:cross_entropy4_2} in every iteration of SGD. We notice that this is in contrast to classic data sets from the machine learning community.

\section{Numerical Results} \label{sec:num_res}

\subsection{Implementation Details / Settings}

\begin{table}[!t]
\renewcommand{\arraystretch}{1.3}
\caption{Simulation Scenarios}
\label{tab1:scenarios}
\centering
\begin{tabular}{l|lllll}
Scenario 					& Sys Dim 			& Mod 			& Corr.						& Coding	\\
\hline
Large MIMO 					& $32\times 32$ 	& QPSK			& no						& no		\\
MIMO 						& $8\times 8$ 		& QPSK			& no						& no		\\
Multi-class 				& $32\times 32$ 	& $16$-QAM			& no						& no		\\
massive MIMO One-Ring 	& $64\times 32$ 	& QPSK			& $20^{\circ}$				& no		\\
Soft Output					& $32\times 32$ 	& QPSK			& no			 			& LDPC		
\end{tabular}
\end{table}

In order to evaluate the performance of the proposed approaches \cmd\ and \cmdnet, we present numerical simulation results of application in our guiding example for different MIMO systems with $\Nt$ transmit and $\Nr$ receive antennas given in Tab. \ref{tab1:scenarios}. We assume an uplink scenario with multiple UEs, each transmitting one symbol $\xvar_{\indx}$ with equal a-priori probabilities $\prior_1=\ldots=\prior_{\Nclass}$ to one BS. As an example, we assume the number of iterations or layers to be $\Ni=\NL=2\Nt$. For numerical optimization of the parameters $\delta^{(\indgrad)}$ and $\tau^{(\indgrad)}$ of \cmdnet\ according to \eqref{eq:cross_entropy4_2}, we employ the Tensorflow framework in Python \cite{martin_abadi_tensorflow_2015}. Here, we use Adam (Adaptive Moment Estimation) as a popular variant of SGD with a default batch size of $\Nb=500$ and $\Ne=10^5$ training iterations. Although providing fast convergence and requiring little hyperparameter tuning, it is known to generalize poorly \cite{wilson_marginal_2017}. Since we are able to generate a sufficient amount of training data, i.e., $N=\Nb\cdot \Ne=5 \cdot 10^7$ to fulfill \eqref{eq:cross_entropy4_1} by \eqref{eq:cross_entropy4_2} approximately, we make sure that generalization to unseen data points is possible. As Tensorflow does not natively support computation with complex numbers, we transform the complex-valued system model \eqref{eq:sys_model} into its real-valued equivalent to allow for training and comparison to DNN-based approaches. This means, we restrict to QAM constellations with Gray encoding so that we have $\xvec\in \disset^{2\Nt \times 1}$. As a training default, we choose the noise variance statistics $\prob(\noisestd^2)$ such that $\ebnot=10\log_{10}(1/\noisestd^2)-10\log_{10}(\log_2(\Nclass))$ is uniformly distributed between $\lbb 4, 27\rbb$ dB. We set the default parameter starting point to $\params_{0}$ with constant $\grstepsize_{0}^{(\idxj)}=1$ and heuristically motivated and linear decreasing 
\begin{align}
\softmaxtemp_{0}^{(\idxj)}=\softmaxtemp_{\txt{max}} - (\softmaxtemp_{\txt{max}} - 0.1)/\Ni\cdot \idxj
\end{align}
with $\softmaxtemp_{\txt{max}}=1/(\Nclass-1)$, $j\in [0,\Ni]$. With this choice, $\prob(\xtvar)$ is always log-convex and hence reasonably approximating $\prob(\xvar)$ (see Fig. \ref{fig1:concrete_bern}). For training of DNN-based approaches DetNet and MMNet, we used the original implementations uploaded to GitHub (see \cite{samuel2019learning, khani_adaptive_2020}) with only minor modifications to parametrization if beneficial. Consequently, we trained MMNet with \cmdnet\ training SNR and layer number. Since we focus on offline derived or trained algorithms which are used for inference at run time, we used its \iid variant. We always used the soft output version of DetNet with output normalization to $1$ since we noted that performance is close to or better than the hard decision version. Furthermore, we compare \cmd\ and \cmdnet\ to several SotA approaches for MIMO detection (see Tab. \ref{tab3:algos}) choosing the number of Monte Carlo runs with data batches of size $10000$ so that always $1000$ errors are detected ($100$ for SD and SDR).

\begin{table}[!t]
	\renewcommand{\arraystretch}{1.3}
	\caption{Selected detection algorithms}
	\label{tab3:algos}
	\centering
	\begin{tabular}{l|lllll}
		Abbrev.			& Complexity 		 & Literature			\\
		\hline
		MAP / SD 		& $\bigo[\Nclass^{\gamma\Nt}]$, $\gamma\in(0,1]$ & \cite{jalden_complexity_2005}						\\
		SDR				& $\bigo[\max(\Nr,\Nt)^3 \Nt^{1/2}\log(1/\epsilon)]$	& \cite{luo_semidefinite_2010}						\\
		OAMPNet			& $\bigo[\NL\Nt^3]$					& \cite{he_model-driven_2018}						\\
		MMSE / MOSIC 	& $\bigo[\Nt^3]$					& \cite{khani_adaptive_2020, wubben2003mmse}		\\
		DetNet			& $\bigo[\NL(\Nt\Nr+\Nt^2\Nclass)]$					& \cite{samuel2017deep, samuel2019learning}			\\
		MMNet (iid)		& $\bigo[\NL\Nt(\Nt+\Nr+\Nclass)]$			& \cite{khani_adaptive_2020}						\\
		AMP				& $\bigo[\Ni\Nt(\Nr+\Nclass)]$							& \cite{jeon2015optimality}							\\
		\cmd / \cmdnet 			& $\bigo[\NL\Nt(\Nr+\Nclass)]$						& \cite{beck_concrete_2020}							\\
		MF				& $\bigo[\Nt\Nr]$	&				
	\end{tabular}
\end{table}

\subsection{Symmetric MIMO system}

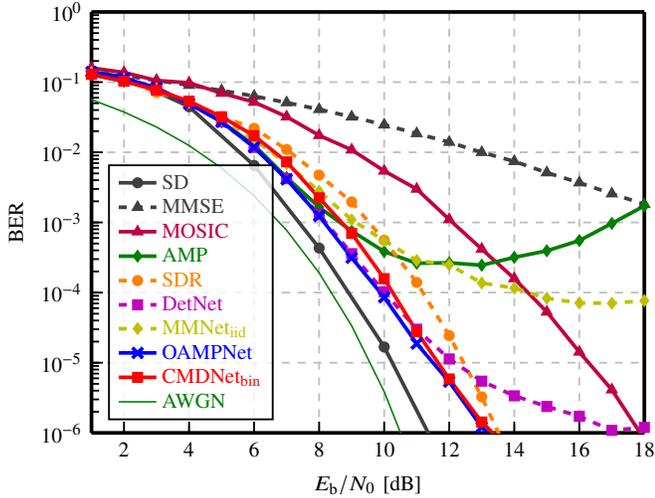
\begin{figure}[!t]
	\centerline{\input{TikZ/plot64x64_comparison_resized.tikz}}
	\caption{BER curves of several detection methods in a $32\times 32$ MIMO system with QPSK modulation. Effective system dimension is $64\times 64$ and for iterative algorithms $\Ni=\NL=64$.}
	\label{fig:results_64x64}
\end{figure}

First, we test application of \cmdnet\ in a large symmetric $32\times 32$ / $64\times 64$ MIMO system with i.i.d. Gaussian channel statistics $\prob(\chmat)$ and QPSK/BPSK modulation. Fig. \ref{fig:results_64x64} shows the results in terms of Bit Error Rate (BER) as a function of $\ebno$. Owing to near-optimal performance, the SD is always provided as a benchmark in the following. In addition, we give the AWGN curve as a reference since it shows the maximum accuracy if $\Nt=\Nr\rightarrow\infty$ \cite{jeon2015optimality}.

Linear detectors perform bad in this setup: Since the curve of the MF remains almost constant at BER $\approx20\%$ and the Zero Forcer performs even worse, both are not shown in the following. At least, MMSE equalization leads to an acceptable BER but the curve is still separated by a $7$ dB gap at $\ebno=13$ dB from SD's. In contrast, nonlinear SotA detectors like \mosic, AMP and SDR technique (see Sec. \ref{sec:1} for algorithm details) have a strikingly better accuracy. Whereas AMP runs into an error floor for high SNR since then the message statistics are not Gaussian anymore in finite small-dimensional MIMO systems \cite{jeon2015optimality}, SDR proves to be a close relaxation by only dropping the non-convex requirement of $\txt{rank}(\xvec\xvec^T)=1$ \cite{luo_semidefinite_2010}.

Notably, our approach \cmdnet\ in its binary version $\text{\cmdnet}_{\text{bin}}$ from \eqref{eq:bin_grad} performs even better than the latter, comparable to the best suboptimal approaches in this setup DetNet and OAMPNet. Further, $\text{\cmdnet}_{\text{bin}}$ does not run into an error floor in the simulated SNR range like AMP and DetNet. Setting the accuracy in context to complexity (see Tab. \ref{tab3:algos}), this is impressive: Note that our approach is similar in asymptotic complexity to the light-weight algorithm AMP with $\bigo[\NL\Nt(\Nr+\Nclass)]$ at inference run time after offline training whereas DetNet and OAMPNet are very complex DNN architectures. In particular, OAMPNet requires one costly matrix inversion per iteration resulting in high $\bigo[\NL\Nt^3]$. In Sec. \ref{subsec:comp} and Fig. \ref{fig:results_complexity}, we give a more detailed complexity analysis and comparison illustrating CMD's promising accuracy complexity trade-off more clearly. In contrast, the other DNN-based approach $\text{MMNet}_{\text{iid}}$ with comparable low complexity fails to beat $\text{\cmdnet}_{\text{bin}}$ and runs into an early error floor. Since we observed this behavior similar to AMP in all settings and MMNet is actually designed to perform well with fast online training, we omit further results. We conjecture that the denoising layers are insufficient expressive in the interference limited high SNR region with offline training.

\begin{figure}[!t]
	\centerline{\input{TikZ/plot16x16_comparison_resized.tikz}}
	\caption{BER curves of several detection methods in a $8\times 8$ MIMO system with QPSK modulation. Effective system dimension is $16\times 16$ and for iterative algorithms $\Ni=\NL=16$.}
	\label{fig:results_16x16}
\end{figure}
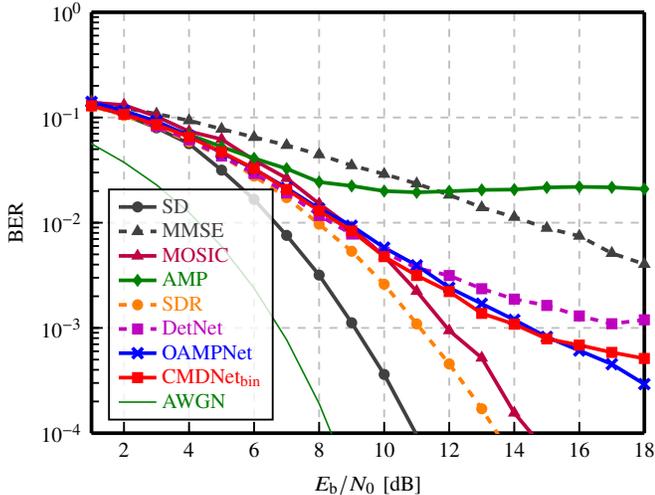

Results in a smaller $8\times 8$ MIMO system plotted in Fig. \ref{fig:results_16x16}, show that all soft non-linear approaches except for SDR and \mosic\ run into an error floor at lower SNR. Thus, we conjecture that they share the same suboptimality at finite system dimensions. They may rely on the statistics of the interference terms to be Gaussian like AMP which is only approximately true for large system dimensions. Apart from SDR and \mosic, $\text{\cmdnet}_{\text{bin}}$ manages to beat the more more expressive and complex DNN models, i.e., DetNet and OAMPNet, and is close in accuracy to SDR for $\ebno<10$ dB.

\subsection{Algorithm and Parametrization}

\begin{figure}[!t]
	\centerline{\input{TikZ/plot64x64_parameter_resized.tikz}}
	\caption{Parameters $\params$ of $\text{\cmdnet}_{\text{bin}}$ in a $32\times 32$ MIMO system with QPSK modulation. Effective system dimension is $64\times 64$.}
	\label{fig:results_64x64_parameter}
\end{figure}
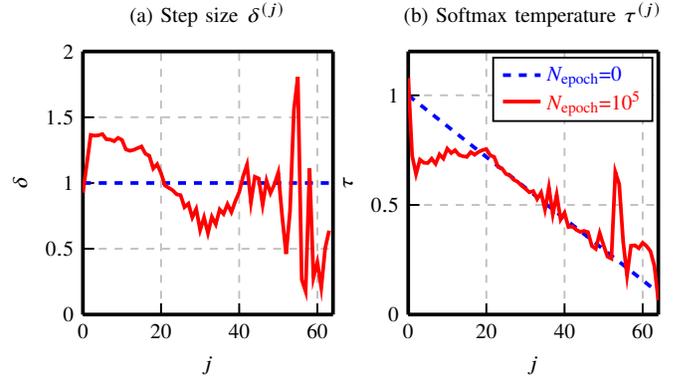

To investigate the influence of learning on $\text{\cmdnet}_{\text{bin}}$ and the values of its parameters $\params$, we visualize them per layer $\indgrad$ in Fig. \ref{fig:results_64x64_parameter} for the $32\times 32$ MIMO system considered before. Basically, we cannot observe any pattern after parameter optimization and interpretation seems very difficult.

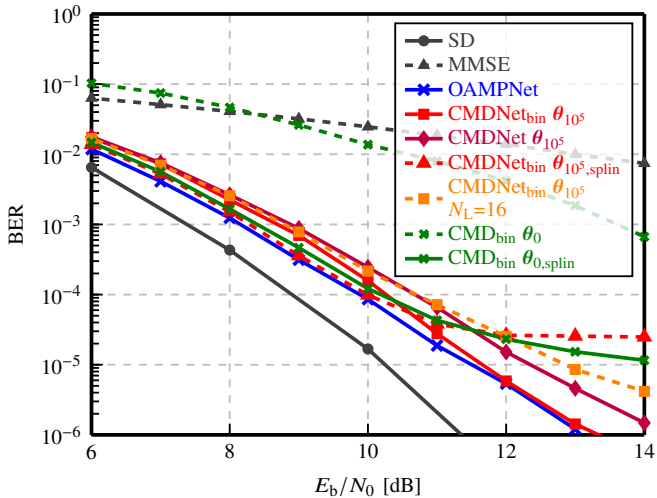
\begin{figure}[!t]
	\centerline{\input{TikZ/plot64x64_CMDcomparison_resized.tikz}}
	\caption{BER curves of \cmd\ and \cmdnet\ with different parametrization or algorithmic in a $32\times 32$ MIMO system with QPSK modulation. Effective system dimension is $64\times 64$. Default number of iterations or layers is $\Ni=\NL=64$.}
	\label{fig:results_64x64_CMD}
\end{figure}

Furthermore, we notice from Fig. \ref{fig:results_64x64_CMD} that starting point initialization $\params_{0}$ has a large impact on the optimum $\params_{10^5}$ found by SGD (after $\Ne=10^5$ iterations). If we use a starting point $\params_{0\text{,splin}}$ with linear decreasing
\begin{align}
	\softmaxtemp_{0\text{,splin}}^{(\idxj)}=\grstepsize_{0\text{,splin}}^{(\idxj)}= 1 - (1 - 0.01)/\Ni\cdot \idxj
\end{align}
for $j\in [0,\Ni]$, a solution $\params_{10^5\text{,splin}}$ is learned allowing \cmdnet\ to perform better in the low $\ebno$ region from $6$ to $10$ dB. Notably, \cmdnet\ even reaches the performance of the best suboptimal algorithm considered in this setup OAMPNet. To explain the error floor in the interference limited higher $\ebno$ region in contrast to \cmdnet\ with default training, we conjecture that a higher starting and correlating end step size (see Fig. \ref{fig:results_64x64_parameter}) allows \cmdnet\ to leave a local optimum with higher probability and to find a better one. On the contrary, a small step size enforces convergence to a local solution. In the noise limited $\ebno$ region, noise removal is crucial and hence convergence. This means \cmdnet\ can be optimized to different working points and is sensitive to starting point initialization. The result supports our view of a promising accuracy complexity trade-off: Since \cmdnet\ only has a small parameter set, we are able to load the $\params$ dynamically for each $\ebno$ to achieve the performance of the best suboptimal algorithm in all $\ebno$ regions.

In particular, we are able to further decrease the number of parameters with negligible performance loss: $\text{\cmdnet}_{\text{bin}}$ with only $\NL=16$ layers performs equally well compared to default \cmd\ with $\NL=64$ at low $\ebno$ and slightly worse at $\ebno=12$ dB by $1$ dB.

Without unfolding, heuristics for parameter selection are required similar to starting point initialization. The detection accuracy of \cmd\ with such heuristic parameters $\params_{0\text{,splin}}$ is quite impressive since the BER curve is close to that of learned \cmdnet\ with $\params_{10^5\text{,splin}}$. Therefore, we are able to use the plain algorithm \cmd\ for detection. We note that this is not true with default parameters $\params_{0}$ and that performance can be quite different after optimization ($\params_{10^5}$).

Finally, we compare the accuracy of algorithm $\text{\cmdnet}_{\text{bin}}$ for the special case of binary \rv\ from \eqref{eq:bin_grad} with that of the generic multi-class algorithm \cmdnet\ from \eqref{eq:grad} since both are different. From Fig. \ref{fig:results_64x64_CMD}, we observe that the performance is very similar and conjecture that \cmdnet\ is capable of achieving the same accuracy if training is parameterized correctly.

\subsection{Multi-class Detection}

So far, only BPSK modulation and hence two classes have been considered. To test multi-class detection with $\Nclass=4$ classes, we show numerical results in a $32\times 32$ MIMO system with $16$-QAM modulation being equivalent to a $64\times 64$ $4$-ASK MIMO system after transformation into the equivalent real-valued problem. Owing to now $3$ degrees of freedom in the soft-max function and denser symbol packing, we changed our batch size to $\Nb=1500$ and training SNR to higher $\ebno\in\lbb 10, 33\rbb$, respectively. Setting the default starting point with $\softmaxtemp_{\txt{max}}=2/(\Nclass-1)=2/3$ so that the MAP criterion $\ln \prob(\xt,\vobs)$ becomes convex for a couple of iterations proves to be crucial for successful training of \cmdnet\ with multiple classes. Without training parameter tuning, \cmdnet\ performs even worse than the MMSE detector.

\begin{figure}[!t]
	\centerline{\input{TikZ/plot64x64_qam16_resized.tikz}}
	\caption{BER curves of several detection methods in a $32\times 32$ MIMO system with $16$-QAM modulation. Effective system has dimension $64\times 64$ and 4-ASK modulation and for iterative algorithms $\Ni=\NL=64$.}
	\label{fig:results_64x64_qam16}
\end{figure}
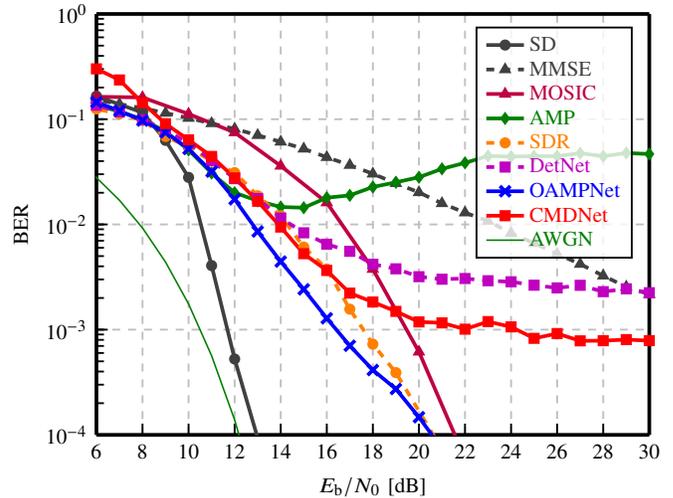

Fig. \ref{fig:results_64x64_qam16} shows BER curves in this system. Clearly, we can now observe a large gap between the BER curve of SD and that of all other suboptimal approaches. Comparing the latter, OAMPNet is superior over the whole SNR region. Observing a maximum $2$ dB curve shift, we note that \cmdnet\ is competitive to OAMPNet and SDR at $\ebno\in[10,17]$ and when BER$=[10^{-2},10^{-3}]$ which is a typical working point of decoders whereas being much less complex. At higher SNR, an error floor follows. Although using a more expressive DNN model, DetNet now trained for $\ebno\in\lbb 9, 16\rbb$ fails to beat \cmdnet\ especially in this region.

\subsection{Massive MIMO system}

Investigation in large symmetric MIMO systems reveals the potential and shortcomings of the algorithms. Rather in 5G, massive MIMO systems with $\Nr>\Nt$ are employed \cite{bjornson_massive_2017}. Assuming \iid Gaussian channels, we shortly report the results of a $64\times 32$ MIMO system with QPSK modulation: The BER curves of learning based approaches and SDR almost follow that of SD and thus suggest that they fit perfectly for application in massive MIMO.

However in practice, channels are spatially correlated at the receiver side due to good spatial resolution of BS' large arrays compared to the number of scattering clusters \cite{bjornson_massive_2017}. Hence, the results for \iid Gaussian channel statistics $\prob(\chmat)$ are less meaningful as noted in \cite{khani_adaptive_2020}. As a first and quick attempt towards a realistic channel model which captures its key characteristics, we test performance in the so-called One-ring model $\prob(\chmat)$ assuming a BS equipped with a uniform linear antenna array \cite{samuel2019learning, bjornson_massive_2017}. We parameterize the correlation matrices of every column in $\chmat$ with reasonable values: Assuming an urban cellular network, we set the angular spread to $20^{\circ}$ and sample the nominal angle uniformly from $[-60^{\circ}, 60^{\circ}]$, i.e., $120^{\circ}$ cell sector. Further, we place the antennas at half a wavelength distance.

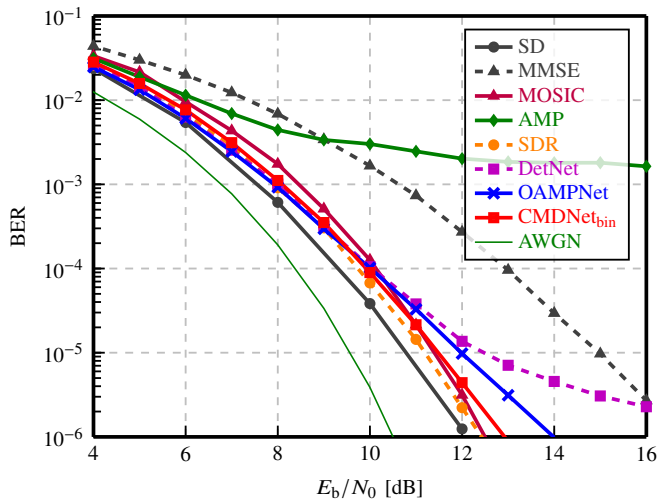
\begin{figure}[!t]
	\centerline{\input{TikZ/plot128x64_comparison_OneRing20_120_resized.tikz}}
	\caption{BER curves of several equalization methods in a correlated $64\times 32$ MIMO system with QPSK modulation. The correlation matrices were generated according to a One-Ring model with $20^{\circ}$ angular spread and $120^{\circ}$ cell sector. Effective system dimension is $128\times 64$ and for iterative algorithms $\Ni=\NL=64$.}
	\label{fig:results_128x64_OneRing20}
\end{figure}

From Fig. \ref{fig:results_128x64_OneRing20}, it becomes evident that the performance loss of learning based approaches compared to SD in such a One-Ring model of dimension $64\times 32$ is similar to the symmetric setting $32\times 32$ in Fig. \ref{fig:results_64x64}. Surprisingly, \mosic\ and SDR now prove to be comparable whereas the BER of AMP degrades since the \iid Gaussian channel assumption is not fulfilled anymore. Again, \cmdnet\ outperforms other learning-based approaches DetNet and OAMPNet and performs very close to the best suboptimal algorithm SDR whereas being much less complex (see Tab. \ref{tab3:algos} and Fig. \ref{fig:results_complexity}).

Considering the low complexity, we finally conclude that \cmdnet\ performs surprisingly well in all previous settings. Hence, it proves to be a generic and hence promising detection approach.

\subsection{Soft Output (Coded MIMO System)}

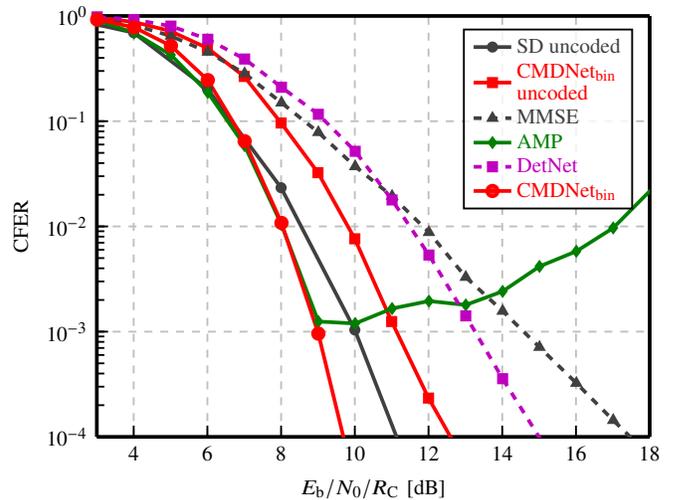
\begin{figure}[!t]
	\centerline{\input{TikZ/plot64x64_comparisonCodeLDPC_FER_resized.tikz}}
	\caption{CFER curves of a horizontally coded $32\times 32$ MIMO system with QPSK modulation. A $128\times 64$ LDPC code with belief propagation decoder was used. Effective system dimension is $64\times 64$ and for iterative algorithms $\Ni=\NL=64$.}
	\label{fig:results_64x64_LDPC}
\end{figure}

After investigation of detection performance in uncoded systems, we turn to an interleaved and horizontally coded $32\times32$ / $64\times64$ MIMO system with Rayleigh block fading reflecting our uplink model. We aim to verify whether not only hard decisions but also soft outputs generated by \cmdnet\ and the soft output version of DetNet have high quality. This is especially important in practice since coding is an essential component besides equalization in today's communication systems. Therefore, we use a $128\times 64$ LDPC code with rate $\rc=1/2$ from \cite{helmling_database_2019} and at receiver side a belief propagation decoder with $10$ iterations. The results in terms of Coded Frame Error Rate (CFER) as a function of $\ebno/\rc$ are shown in Fig. \ref{fig:results_64x64_LDPC}. Owing to overwhelming computational complexity, we refrained from using the MAP solution with coding as a benchmark and instead show uncoded \cmdnet\ and SD curves for reference. Strikingly, \cmdnet\ with coding beats the latter and allows for a coding gain. In contrast, AMP with coding runs into an error floor after $9$ dB: The output statistics become unreliable for high SNR in finite dimensional systems \cite{jeon2015optimality}. Surprisingly, although being one of the best detection methods in the uncoded setting, DetNet with coding performs close to MMSE equalization with soft outputs and thus worse than expected. Actually, the soft output version of DetNet should deliver accurate probabilities or Log Likelihood Ratios (LLRs) according to \cite{samuel2019learning} after optimization.

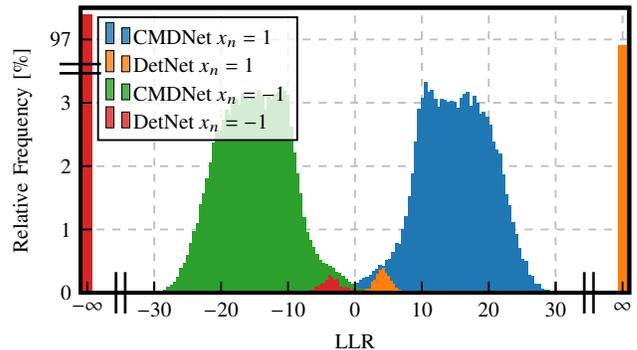
\begin{figure}[!t]
	\centerline{\input{TikZ/MIMOhistogram_QPSK_64x64_64_remastered2_resized.tikz}}
	\caption{Exemplary histogram showing the relative frequencies of LLRs of one symbol $\xvar_n$ in one random channel realization $\chmat$ at $\ebno=10$ dB.}
	\label{fig:results_64x64_LDPC_hist}
\end{figure}

Indeed, we visualize with an exemplary histogram of LLRs that this is not the case. In Fig. \ref{fig:results_64x64_LDPC_hist}, we show the relative frequencies of LLRs of one symbol $\xvar_n$ in one random channel realization $\chmat$ for $\ebno=10$ dB. First, we note the histograms for $\xvar_n=-1$ and $\xvar_n=1$ to be symmetric meaning that both algorithms fulfill a basic quality criterion. Furthermore, it can be clearly seen that DetNet mostly provides hard decisions with $\approx 97\%$ LLRs being $-\infty$ and $\infty$, respectively. Only a few values are close to $0$. In contrast, \cmdnet\ provides meaningful soft information resembling a mixture of Gaussians as expected from literature \cite{cirkic_approximating_2012} ranging from $-30$ to $30$. These results strongly indicate that the difference of soft output quality originates from different underlying optimization strategies: As pointed out in Section \ref{subsec:unfolding}, \cmdnet\ relies on minimization of KL divergence between IO a-posteriori and approximating softmax pdf whereas the one-hot representation in DetNet is optimized w.r.t. MSE. We conclude that our approach yields a better optimization strategy.

\subsection{Complexity Analysis}
\label{subsec:comp}

Since complexity is the main driver for development of suboptimal algorithms like \cmd\ instead of relying on MAP detection, we complete our numerical study by relating detection accuracy to results on the computational complexity given in Tab. \ref{tab3:algos}. With regard to \cmd\ and \cmdnet\ applied in our guiding example \eqref{eq:mimomodel}, the iterative asymptotic complexity of $\bigo[\Nt(2\Nr+4\Nclass)]$ or $\bigo[2\Nt\Nr]$ for binary \rv\ is dominated by the matrix vector multiplications in $\chmat^T \chmat \xt$, i.e., \cmd\ scales linearly with the input and output dimension as well as the number of classes. Clearly, \cmd\ and \cmdnet\ have very low complexity comparable to AMP and MMNet but with remarkable higher detection rate (see, e.g., Fig. \ref{fig:results_64x64}). In most analyzed scenarios, the accuracy is even higher than DetNet's as well as OAMPNet's and on par with SDR's.

\begin{figure}[!t]
	\centerline{\input{TikZ/complexity_64x64_resized.tikz}}
	\caption{Complexity of detection algorithms in terms of number of multiplicative operations in a $32\times32$ / $64\times64$ MIMO system: Light colored bars indicate a realistic low-complexity implementation with BPSK and dark colored bars the worst-case complexity with $16$-QAM modulation.}
	\label{fig:results_complexity}
\end{figure}
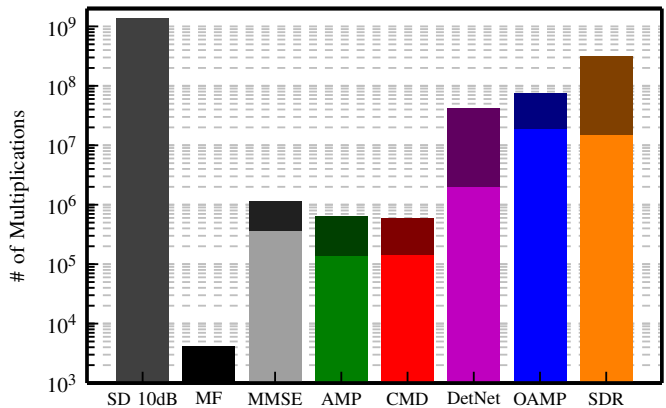

Besides qualitative $\bigo$ analysis, we capture complexity quantitatively by counting the number of Multiplicative OPerations (MOPs) for one iteration and channel realization being the most common and costly floating point operations. In Fig. \ref{fig:results_complexity}, we show the respective bar chart assuming a realistic low-complexity implementation in a $32\times32$ with QPSK ($\Nclass=2$) and $\NL=16$ and worst-case complexity implementation with $16$-QAM modulation ($\Nclass=4$) and $\NL=64$, respectively. For BPSK and the lower bar of MMSE equalization, we assumed Gaussian elimination to solve the linear equation system and, for higher order QAM and the higher bar, LU decomposition. We estimate the upper bound on SDR MOP count by unadapted $\bigo[\max(\Nr,\Nt)^4 \Nt^{1/2}\log(1/\epsilon)]$ and the lower bound on MOPs to account for half of the FLOPS from \cite{samuel2019learning} with inaccurate $\epsilon=0.1$. The expected number of visiting nodes $\bigo[\Nclass^{\gamma\Nt}]$ of the SD is SNR dependent with $\gamma\in(0,1]$ and was extracted from \cite{jalden_complexity_2005}.

Apparently, only the very basic MF beats \cmd\ and \cmdnet\ in complexity at considerably worse detection accuracy. Approaches with comparable accuracy like DetNet, OAMPNet and SDR are $10$-$100$ times more complex w.r.t. MOPs. We conclude that \cmdnet\ offers an excellent accuracy complexity trade-off and note that AMP, MMNet, DetNet and \cmdnet\ further come with the benefit of already delivering soft outputs.

As a final remark, note that complexity analysis depends on the assumptions made: If we, e.g., assume long channel coherence time intervals, MMSE and \mosic\ are able to reuse its computations with only one matrix vector multiplication remaining for any further detection inside the interval effectively decreasing complexity. For the same reason, online learning approaches do not require further training inside the interval and could be feasible. Comparing training cost of all unfolding algorithms in Tab. \ref{tab4:training_complexity}, we note that $\Nb$ and $\Ne$ lie in the same range. Hence, the forward pass of backpropagation in SGD and respectively run time complexity from Fig. \ref{fig:results_complexity} as well as the number of parameters $\abs[\params]$ to be optimized dominate training complexity. OAMPNet fails in the former and DetNet in the latter category with $\abs[\params]\in[10^5,10^7]$ assuming $\NL=\{16,64\}$ and \{QPSK, 16-QAM\}. In contrast, \cmdnet\ with low runtime complexity and only $\abs[\params]=\{33,129\}$ may be a promising online training approach similar to MMNet \cite{khani_adaptive_2020}.

\begin{table}[!t]
	\renewcommand{\arraystretch}{1.3}
	\caption{Training complexity}
	\label{tab4:training_complexity}
	\centering
	\begin{tabular}{l|lllll}
		Algo.		& $\sim\Nb$ 	& $\sim\Ne$			& $\abs[\params]$ \\
		\hline
		DetNet		& $2000$ 	& $10^5$	& $\NL[(\{2,4\}\Nclass+\{6,20\})\Nt^2$	\\
		 \{QPSK, 16-QAM\} &  -$5000$	& 	& $+(\Nclass+\{3,6\})\Nt + 2]$	\\
		OAMPNet		& $1000$	& $10^4$-$10^5$		& $2\NL$	\\
		MMNet \{iid, full\}		& $500$		& $10^4$-$10^5$		& $\{2\NL, \NL\Nt(\Nr+1)\}$	\\
		\cmdnet				& $500$		& $10^4$-$10^5$	& $2\NL+1$	\\
	\end{tabular}
\end{table}

\section{Conclusion}
\label{sec:conclusion}

In this article, we introduced the so called continuous relaxation of discrete \rv\ to the MAP detection problem. Allowing to replace exhaustive search by continuous optimization, we defined our classification approach Concrete MAP Detection (\cmd), e.g., based on gradient descent. By unfolding \cmd\ into a DNN \cmdnet, we further were able to optimize its low number of parameters and hence to improve detection accuracy while limiting it to low complexity. As a side effect, the resulting structure has the potential to allow for fast online training. Using the example of MIMO detection, simulations reveal \cmdnet\ to be a generic detection method competitive to SotA outperforming it in terms of complexity and other recently proposed ML-based approaches DetNet and MMNet in every considered scenario. Notably, we selected an optimization criterion grounded in information theory, i.e., cross entropy, and showed that it aims at learning an approximation of the individual optimal detector. By simulations in coded systems, we demonstrated its ability to provide reliable soft outputs as opposed to \cite{samuel2019learning}, being a requirement for soft decoding, a crucial component in today's communication systems.

All these findings prove \cmdnet\ to be a promising detection approach for application in future massive MIMO systems. Further research is required to evaluate its potential for fast online learning and to demonstrate its applicability to non-linear scenarios of other research domains.

\bibliographystyle{IEEEtran}
\bibliography{IEEEabrv, references}

\begin{IEEEbiography}[{\includegraphics[width=1in,height=1.25in,clip,keepaspectratio]{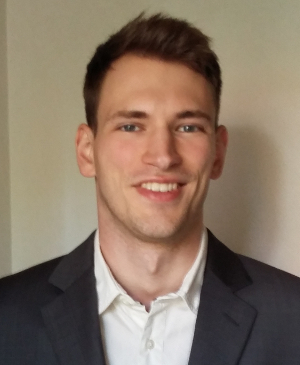}}]{Edgar Beck}
(GS'18) received both his B.Sc. and M.Sc. in electrical engineering (with honors) from the University of Bremen, Germany, in 2017, where he is currently pursuing his Ph.D. degree in electrical engineering at the Department of Communications Engineering (ANT). His research interests include several aspects of future 5G/6G systems: Cognitive radio, compressive sensing, massive MIMO systems and in particular the fertile application of machine learning in wireless communications.
\end{IEEEbiography}

\begin{IEEEbiography}[{\includegraphics[width=1in,height=1.25in,clip,keepaspectratio]{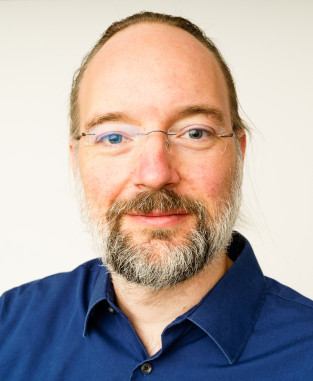}}]{Dr. Carsten Bockelmann}
(M'XX) received his Dipl.-Ing. and Ph.D. degrees in electrical engineering from the University of Bremen, Germany, in 2006 and 2012, respectively. Since 2012, he has been a Senior Research Group Leader with the University of Bremen coordinating research activities regarding the application of compressive sensing/sampling to communication problems. His research interests include communications in massive machine communication, ultra reliable low latency communications (5G) and industry 4.0, compressive sensing, channel coding, and transceiver design.
\end{IEEEbiography}

\begin{IEEEbiography}[{\includegraphics[width=1in,height=1.25in,clip,keepaspectratio]{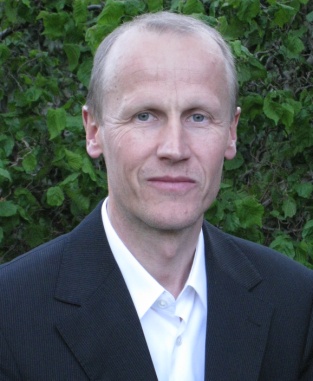}}]{Prof. Dr. Armin Dekorsy}
(SM'18) is the head of the Department of Communications Engineering, University of Bremen. He is distinguished by more than 10 years of industrial experience in leading research positions (DMTS at Bell Labs Europe, Head of Research Europe Qualcomm Nuremberg) and by conducting (inter)national research projects (25+BMBF/BMWI/EU projects) in affiliation with his scientific expertise shown by 200+ journal and conference publications and 19+ patents. He is a senior member of the IEEE Communications and Signal Processing Society, head of VDE/ITG Expert Committee “Information and System Theory”, and member of executive board of the Technologiezentrum Informatik und Informationstechnik (TZI) of the University of Bremen. Prof. Dekorsy investigates new lines of research in wireless communication and signal processing for the baseband of  transceivers of future communication systems, the results of which are transferred to the pre-development of industry through political and strategic activities. His current research focuses on distributed signal  processing, compressed  sampling, information bottleneck method, and machine learning leading to the further development of  communication technologies for 5G/6G, industrial wireless communications and NewSpace satellite communications.
\end{IEEEbiography}

\end{document}

%% file: commands.tex
\newcommand{\tvar}{\ensuremath{x}} 
\newcommand{\vtvar}{\ensuremath{\bo{x}}} 
\newcommand{\tprob}{\ensuremath{p}} 
\newcommand{\aprob}{\ensuremath{q}} 
\newcommand{\params}{\ensuremath{\bog{\theta}}} 
\newcommand{\vobs}{\ensuremath{\bo{y}}} 
\newcommand{\Nsamp}{\ensuremath{N}} 
\newcommand{\idxn}{\ensuremath n} 
\newcommand{\idxi}{\ensuremath i} 
\newcommand{\idxj}{\ensuremath j} 

\newcommand{\bo}[1]{\ensuremath{\mathbf{#1}}}
\newcommand{\bog}[1]{\ensuremath{\bm{#1}}}
\newcommand{\txt}[1]{\ensuremath{\textrm{#1}}}

\newcommand{\eqpoint}{\ensuremath{\,.}} 

\DeclareMathOperator*{\armax}{arg\,max}
\DeclareMathOperator*{\armin}{arg\,min}
\newcommand{\argmax}[1][\cdot]{\ensuremath{\underset{#1}{\armax}\ }}
\newcommand{\argmin}[1][\cdot]{\ensuremath{\underset{#1}{\armin}\ }}

\newcommand{\lpa}{\ensuremath\left(}
\newcommand{\rpa}{\ensuremath\right)}
\newcommand{\lbb}{\ensuremath\left[}
\newcommand{\rbb}{\ensuremath\right]}
\newcommand{\lcb}{\ensuremath\left\{}
\newcommand{\rcb}{\ensuremath\right\}}

\newcommand{\summ}[2][]{\ensuremath{\sum \limits_{#2}^{#1}}} 
\newcommand{\produ}[2][]{\ensuremath{\prod \limits_{#2}^{#1}}} 
\newcommand{\complexnum}{\ensuremath{\mathbb{C}}} 
\newcommand{\realnum}{\ensuremath{\mathbb{R}}} 

\DeclareMathOperator{\E}{E} 
\newcommand{\eval}[2][\cdot]{\ensuremath \E_{#1}\negthinspace\lbb#2\rbb} 
\newcommand{\evalnb}[2][\cdot]{\ensuremath \E_{#1}\negthinspace[#2]} 
\newcommand{\evaltxt}[1][\cdot]{\ensuremath \E[#1]} 
\newcommand{\dkl}[2]{\ensuremath D_{\text{KL}}\lpa#1 \parallel #2\rpa} 
\newcommand{\he}{\mathcal{H}}
\newcommand{\HE}[1][\cdot]{\ensuremath \he\lpa#1\rpa} 

\DeclareMathOperator{\dg}{diag}
\newcommand{\diag}[1]{\ensuremath{\dg\lcb #1\rcb}}

\newcommand{\rv}{RV}
\newcommand{\mosic}{MOSIC} 
\newcommand{\cmdfull}{Concrete MAP Detection}
\newcommand{\cmd}{CMD}
\newcommand{\cmdnet}{CMDNet}
\newcommand{\bigo}[1][\cdot]{\ensuremath{\mathcal{O}(#1)}} 
\newcommand{\signum}{\ensuremath{\txt{sign}}}
\newcommand{\prior}{\ensuremath{\alpha}}					
\newcommand{\priorvec}{\ensuremath{\bog{\alpha}}}			
\newcommand{\gumbel}{\ensuremath{g}}						
\newcommand{\gumbelvec}{\ensuremath{\bo{\gumbel}}}			
\newcommand{\gumbelmat}{\ensuremath{\bo{G}}}				
\newcommand{\gumbelest}{\ensuremath{\hat{\gumbelmat}}}		
\newcommand{\softmaxtemp}{\ensuremath{\tau}}				
\newcommand{\grstepsize}{\ensuremath{\delta}}
\newcommand{\grit}{\ensuremath{j}}
\newcommand{\softmax}[2][\softmaxtemp]{\ensuremath{\softmaxvar_{#1}\left(#2\right)}}
\newcommand{\softmaxvar}{\ensuremath{\sigma}}

\newcommand{\logvar}{\ensuremath{s}}
\newcommand{\logvec}{\ensuremath{\bo{\logvar}}}

\newcommand{\Nt}{\ensuremath{N_{\txt{T}}}} 
\newcommand{\Nr}{\ensuremath{N_{\txt{R}}}} 
\newcommand{\Ni}{\ensuremath{N_{\txt{it}}}} 
\newcommand{\Nb}{\ensuremath{N_{\txt{b}}}} 
\newcommand{\Ne}{\ensuremath{N_{\txt{epoch}}}} 
\newcommand{\NL}{\ensuremath{N_{\txt{L}}}} 
\newcommand{\noisestd}{\ensuremath{\sigma_{\txt{n}}}} 
\newcommand{\noisevar}{\ensuremath{n}} 
\newcommand{\noisevec}{\ensuremath{\bo{\noisevar}}} 
\newcommand{\Nclass}{\ensuremath{M}} 
\newcommand{\disset}{\ensuremath{\mathcal{\Nclass}}} 
\newcommand{\dissetel}{\ensuremath{m}} 
\newcommand{\dissetvec}{\ensuremath{\bo{\dissetel}}} 
\newcommand{\xvar}{\ensuremath{x}} 
\newcommand{\xvec}{\ensuremath{\bo{x}}} 
\newcommand{\xtvar}{\ensuremath{\tilde{x}}} 
\newcommand{\xt}{\ensuremath{\tilde{\bo{x}}}} 
\newcommand{\xest}{\ensuremath{\hat{\xvec}}} 
\newcommand{\xvarest}{\ensuremath{\hat{\xvar}}} 
\newcommand{\chtap}{\ensuremath{h}} 
\newcommand{\chmat}{\ensuremath{\bo{H}}} 
\newcommand{\tobsvec}{\ensuremath{\tilde{\bo{y}}}} 
\newcommand{\yvec}{\ensuremath{\bo{y}}} 
\newcommand{\prob}{\ensuremath{p}} 
\newcommand{\ohvar}{\ensuremath{z}} 
\newcommand{\oh}{\ensuremath{\bo{\ohvar}}} 
\newcommand{\oht}{\ensuremath{\tilde{\bo{\ohvar}}}} 
\newcommand{\ohtvar}{\ensuremath{\tilde{\ohvar}}} 
\newcommand{\ohfunc}{\ensuremath{\txt{one-hot}}} 
\newcommand{\indx}{\ensuremath{n}} 
\newcommand{\indy}{\ensuremath{m}} 
\newcommand{\indc}{\ensuremath{k}} 
\newcommand{\indgrad}{\ensuremath{j}} 
\newcommand{\indi}{\ensuremath{i}} 
\newcommand{\indl}{\ensuremath{l}} 
\newcommand{\indiopt}{\ensuremath{i^\ast}} 
\newcommand{\plvec}{\ensuremath{\bo{a}}} 
\newcommand{\obf}{\ensuremath{L}} 
\newcommand{\ebno}{\ensuremath{E_{\txt{b}}/N_0}} 
\newcommand{\ebnot}{\ensuremath{E_{\txt{b}}/N_0}} 
\newcommand{\iid}{i.i.d. } 
\newcommand{\rc}[1][]{\ensuremath \ifthenelse{\equal{#1}{}}{R_{\text{C}}}{R_{\text{C},#1}}} 
\newcommand{\derloglike}{\ensuremath{\frac{\partial \ln \prob(\yvec|\gumbelmat)}{\partial\xt}}} 
\newcommand{\derloglikes}{\ensuremath{\frac{\partial \ln \prob(\yvec|\logvec)}{\partial\xt}}} 

\newcommand{\abs}[1][\cdot]{\ensuremath{\left|#1\right|}}
\newcommand{\abstxt}[1][\cdot]{\ensuremath{|#1|}}
\newcommand{\norm}[1]{\ensuremath{\left\|#1\right\|}}

\newcommand{\trapo}{\ensuremath{T}} 
\newcommand{\Hm}{\ensuremath{H}} 
\newcommand{\eye}{\ensuremath{\bo{I}}} 
\newcommand{\ones}{\ensuremath{\bo{1}}} 
\newcommand{\zero}{\ensuremath{\bo{0}}} 
\newcommand{\normdis}{\ensuremath{\mathcal{N}}} 
\newcommand{\normdisc}{\ensuremath{\mathcal{C}\normdis}} 

%% file: TikZ/pres_concrete_bern_resized2.tikz
\begin{tikzpicture}[scale=1]
\pgfplotsset{
	every axis plot/.style={line width=1.4pt},
	every axis/.style={grid style={line width=0.7pt, dashed}, 
					 },
	axis background/.style={fill=white},
	every outer x axis line/.style={line width=1.4pt},
	every outer y axis line/.style={line width=1.4pt},
	every tick/.style={black,
					line width=0.7pt,
					},
	label style={font=\fontsize{8}{9}\selectfont},
	every y tick label/.style={font=\fontsize{8}{9}\color{black}},
	every x tick label/.style={font=\fontsize{8}{9}\color{black}},
	every extra x tick/.style={grid style={solid, violet, line width=2.8pt},
							x tick label style={/pgf/number format/.cd,precision=10}
							},
	legend style={
					line width=0.7pt,
					font=\fontsize{11.4}{12}\selectfont\color{black},
					legend cell align=left,
					align=left,
					fill=white,
					fill opacity=0.8, draw opacity=1, text opacity=1,
					nodes={scale=0.7, transform shape},
					},
	legend image code/.code={
		\draw[mark repeat=2,mark phase=2]
			plot coordinates {
				(0cm,0cm)
				(0.25cm,0cm)        
				(0.5cm,0cm)         
			};%
		},
	legend entries={ 
					{\color{red}$p(\tilde{x}|\boldsymbol{\alpha}=[0.5,0.5]^T,\tau=0.1)$},
					{\color{black} $p(\tilde{x}|\boldsymbol{\alpha}=[0.5,0.5]^T,\tau=2)$},
					{\color{orange} $p(\tilde{x}|\boldsymbol{\alpha}=[0.5,0.5]^T,\tau=1)$},
					{\color{darkgreen} $p(\tilde{x}|\boldsymbol{\alpha}=[0.8,0.2]^T,\tau=0.8)$},
					{\color{blue} $p(x|\boldsymbol{\alpha}=[0.5,0.5]^T)$},
					},
}

\begin{axis}[%
width=10cm, 
height=6cm, 
scale=0.75,
scale only axis,
separate axis lines,
tick pos=left,
xmin=-1.075,
xmax=1.075,
xminorticks=true,
xtick distance = {0.5},
ytick distance = {0.5},
xlabel={$\tilde{x}=-2\tilde{z}_1+1$}, 
ymin=0,
ymax=1.5,
ylabel={$p(\tilde{x}|\boldsymbol{\alpha},\tau)$},
xmajorgrids,
xminorgrids,
ymajorgrids,
]



\addplot [color=red, mark=*, only marks, mark size = 2*0.7]
table {%
-0.802010050251256 0.138447464799472
-0.404020100502512 0.0596441839211066
0 0.0500018000195638
0.404020100502513 0.0596441839211066
0.802010050251256 0.138447464799472
};


\addplot [color=darkgray, mark=triangle*, only marks, mark size = 3*0.7]
table {%
-1.2 -0.0739048642837947
-0.802010050251256 0.132132260854593
-0.404020100502512 0.6184036948055
0 0.99989091845656
0.404020100502513 0.618403694805499
0.802010050251256 0.132132260854593
1.2 -0.0739048642837947
};

\addplot [color=orange, mark=square*, dashed, mark options={solid}, only marks, mark size = 2*0.7]
table {%
-0.802010050251256 0.5
-0.404020100502512 0.5
0 0.5
0.404020100502513 0.5
0.802010050251256 0.5
};

\addplot [color=darkgreen, mark=*, only marks, mark size = 2*0.7]
table {%
-1.2 nan
-1.0070351758794 nan
-0.802010050251256 1.0815275507197
-0.404020100502512 0.423829679066559
0 0.254527765220321
0.404020100502513 0.189975284616529
0.802010050251256 0.176210402805603
1.0070351758794 nan
1.2 nan
};

\addplot [color=blue, mark=diamond*, only marks, mark size = 3*0.75] 
  table{%
-1 0.5
1 0.5
};


\addplot [color=darkgray]
table {%
-1.2 -0.0739048642837947
-1.18793969849246 -0.0707273102384636
-1.17587939698492 -0.0674083729640582
-1.16381909547739 -0.063943638918373
-1.15175879396985 -0.0603286073035779
-1.13969849246231 -0.0565586921955457
-1.12763819095477 -0.0526292250887914
-1.11557788944724 -0.0485354578893529
-1.1035175879397 -0.0442725663894772
-1.09145728643216 -0.0398356542595124
-1.07939698492462 -0.0352197575939114
-1.06733668341709 -0.0304198500497278
-1.05527638190955 -0.0254308486174111
-1.04321608040201 -0.0202476200650582
-1.03115577889447 -0.0148649880985452
-1.01909547738693 -0.00927774128111214
-1.0070351758794 -0.00348064175698919
-0.994974874371859 0.00253156517549934
-0.982914572864322 0.00876414059526605
-0.970854271356784 0.0152223405477133
-0.958793969849246 0.0219114036452372
-0.946733668341708 0.0288365376324575
-0.934673366834171 0.0360029048701372
-0.922613065326633 0.0434156066926381
-0.910552763819095 0.0510796665950776
-0.898492462311558 0.0590000122081559
-0.88643216080402 0.0671814560209769
-0.874371859296482 0.0756286748151215
-0.862311557788945 0.0843461877768379
-0.850251256281407 0.0933383332585147
-0.838190954773869 0.102609244165687
-0.826130653266332 0.112162821951731
-0.814070351758794 0.122002709209213
-0.802010050251256 0.132132260854593
-0.789949748743718 0.142554513911749
-0.777889447236181 0.153272155909595
-0.765829145728643 0.164287491919976
-0.753768844221105 0.175602410274074
-0.741708542713568 0.187218347008817
-0.72964824120603 0.199136249109196
-0.717587939698492 0.211356536628042
-0.705527638190955 0.223879063781702
-0.693467336683417 0.236703079138048
-0.681407035175879 0.249827185032501
-0.669346733668342 0.263249296368013
-0.657286432160804 0.276966598976264
-0.645226130653266 0.290975507739547
-0.633165829145728 0.305271624695774
-0.621105527638191 0.319849697372651
-0.609045226130653 0.334703577621012
-0.596984924623115 0.349826181241516
-0.584924623115578 0.365209448722927
-0.57286432160804 0.380844307433904
-0.560804020100502 0.396720635633149
-0.548743718592965 0.412827228684566
-0.536683417085427 0.429151767884409
-0.524623115577889 0.445680792325706
-0.512562814070352 0.462399674241155
-0.500502512562814 0.479292598278653
-0.488442211055276 0.496342545173128
-0.476381909547739 0.513531280283904
-0.464321608040201 0.530839347467859
-0.452261306532663 0.548246068754662
-0.440201005025125 0.565729550280837
-0.428140703517588 0.583266694923883
-0.41608040201005 0.600833222055669
-0.404020100502512 0.6184036948055
-0.391959798994975 0.63595155518727
-0.379899497487437 0.653449167401762
-0.367839195979899 0.670867869574269
-0.355778894472362 0.688178034129363
-0.343718592964824 0.705349136938815
-0.331658291457286 0.72234983530576
-0.319597989949749 0.739148054768499
-0.307537688442211 0.755711084621491
-0.295477386934673 0.772005681959781
-0.283417085427136 0.787998183957242
-0.271356783919598 0.803654627989685
-0.25929648241206 0.818940879112237
-0.247236180904522 0.833822764297869
-0.235175879396985 0.84826621274196
-0.223115577889447 0.86223740143802
-0.211055276381909 0.875702905133737
-0.198994974874372 0.888629849686159
-0.186934673366834 0.900986067751862
-0.174874371859296 0.912740255673997
-0.162814070351759 0.923862130365004
-0.150753768844221 0.934322584932966
-0.138693467336683 0.944093841762546
-0.126633165829146 0.953149601739515
-0.114572864321608 0.961465188301965
-0.10251256281407 0.969017685012276
-0.0904522613065326 0.97578606537226
-0.0783919597989948 0.981751313649762
-0.066331658291457 0.986896535548285
-0.0542713567839195 0.991207057631395
-0.0422110552763817 0.994670514509971
-0.0301507537688441 0.997276922911561
-0.0180904522613063 0.999018741875836
-0.00603015075376878 0.99989091845656
0.006030150753769 0.99989091845656
0.0180904522613068 0.999018741875836
0.0301507537688444 0.99727692291156
0.0422110552763821 0.994670514509971
0.0542713567839197 0.991207057631395
0.0663316582914575 0.986896535548285
0.078391959798995 0.981751313649762
0.0904522613065328 0.97578606537226
0.102512562814071 0.969017685012276
0.114572864321608 0.961465188301965
0.126633165829146 0.953149601739515
0.138693467336684 0.944093841762546
0.150753768844221 0.934322584932966
0.162814070351759 0.923862130365005
0.174874371859297 0.912740255673996
0.186934673366834 0.900986067751861
0.198994974874372 0.888629849686159
0.21105527638191 0.875702905133736
0.223115577889447 0.86223740143802
0.235175879396985 0.848266212741959
0.247236180904523 0.833822764297869
0.25929648241206 0.818940879112237
0.271356783919598 0.803654627989684
0.283417085427136 0.787998183957241
0.295477386934674 0.772005681959781
0.307537688442211 0.755711084621491
0.319597989949749 0.739148054768499
0.331658291457287 0.72234983530576
0.343718592964824 0.705349136938815
0.355778894472362 0.688178034129362
0.3678391959799 0.670867869574269
0.379899497487437 0.653449167401762
0.391959798994975 0.63595155518727
0.404020100502513 0.618403694805499
0.41608040201005 0.600833222055669
0.428140703517588 0.583266694923883
0.440201005025126 0.565729550280837
0.452261306532663 0.548246068754662
0.464321608040201 0.530839347467859
0.476381909547739 0.513531280283904
0.488442211055277 0.496342545173128
0.500502512562814 0.479292598278653
0.512562814070352 0.462399674241155
0.52462311557789 0.445680792325705
0.536683417085427 0.429151767884408
0.548743718592965 0.412827228684565
0.560804020100503 0.396720635633148
0.57286432160804 0.380844307433903
0.584924623115578 0.365209448722926
0.596984924623116 0.349826181241516
0.609045226130654 0.334703577621012
0.621105527638191 0.319849697372651
0.633165829145729 0.305271624695774
0.645226130653267 0.290975507739547
0.657286432160804 0.276966598976264
0.669346733668342 0.263249296368013
0.68140703517588 0.249827185032501
0.693467336683417 0.236703079138047
0.705527638190955 0.223879063781702
0.717587939698493 0.211356536628042
0.72964824120603 0.199136249109195
0.741708542713568 0.187218347008817
0.753768844221106 0.175602410274074
0.765829145728643 0.164287491919976
0.777889447236181 0.153272155909595
0.789949748743719 0.142554513911748
0.802010050251256 0.132132260854593
0.814070351758794 0.122002709209213
0.826130653266332 0.112162821951731
0.83819095477387 0.102609244165687
0.850251256281407 0.0933383332585147
0.862311557788945 0.0843461877768376
0.874371859296483 0.0756286748151213
0.88643216080402 0.0671814560209768
0.898492462311558 0.0590000122081557
0.910552763819096 0.0510796665950774
0.922613065326633 0.043415606692638
0.934673366834171 0.0360029048701371
0.946733668341709 0.0288365376324573
0.958793969849246 0.0219114036452371
0.970854271356784 0.0152223405477132
0.982914572864322 0.00876414059526582
0.99497487437186 0.00253156517549918
1.0070351758794 -0.00348064175698918
1.01909547738693 -0.00927774128111206
1.03115577889447 -0.0148649880985453
1.04321608040201 -0.0202476200650582
1.05527638190955 -0.025430848617411
1.06733668341709 -0.0304198500497279
1.07939698492462 -0.0352197575939114
1.09145728643216 -0.0398356542595124
1.1035175879397 -0.0442725663894771
1.11557788944724 -0.048535457889353
1.12763819095477 -0.0526292250887915
1.13969849246231 -0.0565586921955456
1.15175879396985 -0.060328607303578
1.16381909547739 -0.0639436389183731
1.17587939698492 -0.0674083729640582
1.18793969849246 -0.0707273102384635
1.2 -0.0739048642837947
};

\addplot [color=red]
table {%
-1.2 nan
-1.18793969849246 nan
-1.17587939698492 nan
-1.16381909547739 nan
-1.15175879396985 nan
-1.13969849246231 nan
-1.12763819095477 nan
-1.11557788944724 nan
-1.1035175879397 nan
-1.09145728643216 nan
-1.07939698492462 nan
-1.06733668341709 nan
-1.05527638190955 nan
-1.04321608040201 nan
-1.03115577889447 nan
-1.01909547738693 nan
-1.0070351758794 nan
-0.994974874371859 4.56640809625937
-0.982914572864322 1.39549609820692
-0.970854271356784 0.832917635823674
-0.958793969849246 0.596939695635688
-0.946733668341708 0.466901925356314
-0.934673366834171 0.384473503767726
-0.922613065326633 0.327531073703357
-0.910552763819095 0.285829582879547
-0.898492462311558 0.253970556908183
-0.88643216080402 0.228838014169995
-0.874371859296482 0.208506917536678
-0.862311557788945 0.191723542880118
-0.850251256281407 0.177636168868083
-0.838190954773869 0.165645827597872
-0.826130653266332 0.155318975395175
-0.814070351758794 0.146334044024211
-0.802010050251256 0.138447464799472
-0.789949748743718 0.131471360359655
-0.777889447236181 0.125258481369274
-0.765829145728643 0.119691783335354
-0.753768844221105 0.11467705696158
-0.741708542713568 0.110137616703102
-0.72964824120603 0.106010406552403
-0.717587939698492 0.102243100544353
-0.705527638190955 0.0987919135732199
-0.693467336683417 0.0956199274202756
-0.681407035175879 0.0926957958348834
-0.669346733668342 0.089992732147813
-0.657286432160804 0.0874877100045734
-0.645226130653266 0.0851608266391714
-0.633165829145728 0.082994791380652
-0.621105527638191 0.0809745115626929
-0.609045226130653 0.0790867548585516
-0.596984924623115 0.077319872074127
-0.584924623115578 0.0756635681348509
-0.57286432160804 0.074108711766053
-0.560804020100502 0.0726471764487125
-0.548743718592965 0.0712717068149643
-0.536683417085427 0.0699758058602121
-0.524623115577889 0.068753639284924
-0.512562814070352 0.0675999540073439
-0.500502512562814 0.0665100084586179
-0.488442211055276 0.065479512721341
-0.476381909547739 0.0645045769290509
-0.464321608040201 0.0635816666286254
-0.452261306532663 0.0627075640357189
-0.440201005025125 0.061879334297407
-0.428140703517588 0.061094296025377
-0.41608040201005 0.0603499954845087
-0.404020100502512 0.0596441839211066
-0.391959798994975 0.0589747975967623
-0.379899497487437 0.0583399401612643
-0.367839195979899 0.0577378670538625
-0.355778894472362 0.0571669716686922
-0.343718592964824 0.0566257730589861
-0.331658291457286 0.0561129049872442
-0.319597989949749 0.0556271061558911
-0.307537688442211 0.0551672114760432
-0.295477386934673 0.0547321442515512
-0.283417085427136 0.0543209091720798
-0.271356783919598 0.0539325860231296
-0.25929648241206 0.053566324032978
-0.247236180904522 0.0532213367868698
-0.235175879396985 0.05289689764768
-0.223115577889447 0.0525923356299353
-0.211055276381909 0.0523070316807027
-0.198994974874372 0.0520404153265947
-0.186934673366834 0.0517919616511281
-0.174874371859296 0.0515611885710222
-0.162814070351759 0.0513476543838229
-0.150753768844221 0.0511509555625746
-0.138693467336683 0.0509707247761943
-0.126633165829146 0.050806629116794
-0.114572864321608 0.0506583685174933
-0.10251256281407 0.0505256743463081
-0.0904522613065326 0.0504083081635301
-0.0783919597989948 0.050306060631653
-0.066331658291457 0.0502187505683898
-0.0542713567839195 0.0501462241346715
-0.0422110552763817 0.0500883541507616
-0.0301507537688441 0.0500450395347604
-0.0180904522613063 0.0500162048588414
-0.00603015075376878 0.0500018000195638
0.006030150753769 0.0500018000195638
0.0180904522613068 0.0500162048588414
0.0301507537688444 0.0500450395347604
0.0422110552763821 0.0500883541507616
0.0542713567839197 0.0501462241346715
0.0663316582914575 0.0502187505683898
0.078391959798995 0.050306060631653
0.0904522613065328 0.0504083081635301
0.102512562814071 0.0505256743463081
0.114572864321608 0.0506583685174933
0.126633165829146 0.050806629116794
0.138693467336684 0.0509707247761943
0.150753768844221 0.0511509555625746
0.162814070351759 0.0513476543838229
0.174874371859297 0.0515611885710222
0.186934673366834 0.0517919616511281
0.198994974874372 0.0520404153265947
0.21105527638191 0.0523070316807027
0.223115577889447 0.0525923356299353
0.235175879396985 0.05289689764768
0.247236180904523 0.0532213367868698
0.25929648241206 0.053566324032978
0.271356783919598 0.0539325860231296
0.283417085427136 0.0543209091720797
0.295477386934674 0.0547321442515512
0.307537688442211 0.0551672114760432
0.319597989949749 0.0556271061558911
0.331658291457287 0.0561129049872442
0.343718592964824 0.0566257730589861
0.355778894472362 0.0571669716686922
0.3678391959799 0.0577378670538626
0.379899497487437 0.0583399401612643
0.391959798994975 0.0589747975967623
0.404020100502513 0.0596441839211066
0.41608040201005 0.0603499954845088
0.428140703517588 0.0610942960253771
0.440201005025126 0.061879334297407
0.452261306532663 0.0627075640357189
0.464321608040201 0.0635816666286254
0.476381909547739 0.064504576929051
0.488442211055277 0.065479512721341
0.500502512562814 0.0665100084586179
0.512562814070352 0.0675999540073439
0.52462311557789 0.068753639284924
0.536683417085427 0.0699758058602121
0.548743718592965 0.0712717068149643
0.560804020100503 0.0726471764487125
0.57286432160804 0.074108711766053
0.584924623115578 0.0756635681348509
0.596984924623116 0.077319872074127
0.609045226130654 0.0790867548585517
0.621105527638191 0.080974511562693
0.633165829145729 0.0829947913806521
0.645226130653267 0.0851608266391714
0.657286432160804 0.0874877100045735
0.669346733668342 0.089992732147813
0.68140703517588 0.0926957958348835
0.693467336683417 0.0956199274202757
0.705527638190955 0.0987919135732199
0.717587939698493 0.102243100544353
0.72964824120603 0.106010406552403
0.741708542713568 0.110137616703102
0.753768844221106 0.114677056961581
0.765829145728643 0.119691783335354
0.777889447236181 0.125258481369274
0.789949748743719 0.131471360359655
0.802010050251256 0.138447464799472
0.814070351758794 0.146334044024212
0.826130653266332 0.155318975395176
0.83819095477387 0.165645827597872
0.850251256281407 0.177636168868083
0.862311557788945 0.191723542880118
0.874371859296483 0.208506917536678
0.88643216080402 0.228838014169995
0.898492462311558 0.253970556908184
0.910552763819096 0.285829582879548
0.922613065326633 0.327531073703357
0.934673366834171 0.384473503767726
0.946733668341709 0.466901925356317
0.958793969849246 0.596939695635691
0.970854271356784 0.832917635823678
0.982914572864322 1.39549609820696
0.99497487437186 4.56640809625966
1.0070351758794 nan
1.01909547738693 nan
1.03115577889447 nan
1.04321608040201 nan
1.05527638190955 nan
1.06733668341709 nan
1.07939698492462 nan
1.09145728643216 nan
1.1035175879397 nan
1.11557788944724 nan
1.12763819095477 nan
1.13969849246231 nan
1.15175879396985 nan
1.16381909547739 nan
1.17587939698492 nan
1.18793969849246 nan
1.2 nan
};

\addplot [color=orange, mark=square*, dashed, mark options={solid}, no marks]
table {%
-1.0 0.5
-0.802010050251256 0.5
-0.404020100502512 0.5
0 0.5
0.404020100502513 0.5
0.802010050251256 0.5
1.0 0.5
};  

\addplot [darkgreen, dashed]
table {%
-1.2 nan
-1.18793969849246 nan
-1.17587939698492 nan
-1.16381909547739 nan
-1.15175879396985 nan
-1.13969849246231 nan
-1.12763819095477 nan
-1.11557788944724 nan
-1.1035175879397 nan
-1.09145728643216 nan
-1.07939698492462 nan
-1.06733668341709 nan
-1.05527638190955 nan
-1.04321608040201 nan
-1.03115577889447 nan
-1.01909547738693 nan
-1.0070351758794 nan
-0.994974874371859 4.98393916031338
-0.982914572864322 3.55058647739312
-0.970854271356784 2.95840824644346
-0.958793969849246 2.58380826303512
-0.946733668341708 2.31214134367065
-0.934673366834171 2.10094976852139
-0.922613065326633 1.92965688094224
-0.910552763819095 1.78667117046493
-0.898492462311558 1.66479586595569
-0.88643216080402 1.55924723968481
-0.874371859296482 1.4666806496218
-0.862311557788945 1.38466471572039
-0.850251256281407 1.31137640736023
-0.838190954773869 1.24541399324281
-0.826130653266332 1.18567694948182
-0.814070351758794 1.13128590286645
-0.802010050251256 1.0815275507197
-0.789949748743718 1.03581573206344
-0.777889447236181 0.993663268858605
-0.765829145728643 0.95466118245112
-0.753768844221105 0.918463079180142
-0.741708542713568 0.884773233936559
-0.72964824120603 0.853337367744284
-0.717587939698492 0.823935420158299
-0.705527638190955 0.796375820500731
-0.693467336683417 0.770490900257579
-0.681407035175879 0.746133184808218
-0.669346733668342 0.72317237019657
-0.657286432160804 0.701492838963893
-0.645226130653266 0.680991604102154
-0.633165829145728 0.661576595924868
-0.621105527638191 0.643165225780676
-0.609045226130653 0.625683174905907
-0.596984924623115 0.609063367618229
-0.584924623115578 0.593245096407314
-0.57286432160804 0.578173272933565
-0.560804020100502 0.563797783974578
-0.548743718592965 0.5500729353065
-0.536683417085427 0.536956969628368
-0.524623115577889 0.524411647121777
-0.512562814070352 0.512401879228158
-0.500502512562814 0.500895407829585
-0.488442211055276 0.489862523318601
-0.476381909547739 0.479275816101528
-0.464321608040201 0.469109956946835
-0.452261306532663 0.459341502303788
-0.440201005025125 0.449948721306505
-0.428140703517588 0.440911441668355
-0.41608040201005 0.432210912080041
-0.404020100502512 0.423829679066559
-0.391959798994975 0.415751476545478
-0.379899497487437 0.407961126571243
-0.367839195979899 0.400444449955209
-0.355778894472362 0.393188185625166
-0.343718592964824 0.386179917736419
-0.331658291457286 0.379408009673127
-0.319597989949749 0.372861544187219
-0.307537688442211 0.366530269015444
-0.295477386934673 0.360404547395547
-0.283417085427136 0.354475312972003
-0.271356783919598 0.348734028641895
-0.25929648241206 0.343172648943778
-0.247236180904522 0.337783585637822
-0.235175879396985 0.33255967616519
-0.223115577889447 0.327494154709281
-0.211055276381909 0.322580625611842
-0.198994974874372 0.317813038923631
-0.186934673366834 0.313185667892762
-0.174874371859296 0.308693088214542
-0.162814070351759 0.304330158884889
-0.150753768844221 0.300092004515553
-0.138693467336683 0.295973998983688
-0.126633165829146 0.291971750301048
-0.114572864321608 0.288081086599373
-0.10251256281407 0.284298043138608
-0.0904522613065326 0.280618850253597
-0.0783919597989948 0.277039922162903
-0.066331658291457 0.273557846570647
-0.0542713567839195 0.270169374998648
-0.0422110552763817 0.266871413791996
-0.0301507537688441 0.26366101574637
-0.0180904522613063 0.260535372310123
-0.00603015075376878 0.257491806318389
0.006030150753769 0.254527765220321
0.0180904522613068 0.251640814764011
0.0301507537688444 0.24882863310683
0.0422110552763821 0.246089005321763
0.0542713567839197 0.243419818272959
0.0663316582914575 0.240819055836074
0.078391959798995 0.238284794441193
0.0904522613065328 0.235815198918126
0.102512562814071 0.233408518625707
0.114572864321608 0.231063083848463
0.126633165829146 0.228777302445604
0.138693467336684 0.226549656738753
0.150753768844221 0.224378700626261
0.162814070351759 0.222263056913241
0.174874371859297 0.220201414847711
0.186934673366834 0.218192527854454
0.198994974874372 0.216235211459318
0.21105527638191 0.214328341397852
0.223115577889447 0.212470851903247
0.235175879396985 0.210661734169676
0.247236180904523 0.208900034988234
0.25929648241206 0.207184855553812
0.271356783919598 0.205515350442404
0.283417085427136 0.203890726759577
0.295477386934674 0.202310243462123
0.307537688442211 0.200773210856281
0.319597989949749 0.199278990277422
0.331658291457287 0.197826993957716
0.343718592964824 0.1964166850901
0.355778894472362 0.195047578098868
0.3678391959799 0.193719239129446
0.379899497487437 0.192431286772478
0.391959798994975 0.191183393040212
0.404020100502513 0.189975284616529
0.41608040201005 0.188806744405752
0.428140703517588 0.187677613409839
0.440201005025126 0.186587792968713
0.452261306532663 0.185537247404532
0.464321608040201 0.184526007117864
0.476381909547739 0.183554172192103
0.488442211055277 0.182621916572521
0.500502512562814 0.181729492898245
0.512562814070352 0.180877238079806
0.52462311557789 0.180065579732093
0.536683417085427 0.179295043593456
0.548743718592965 0.178566262087016
0.560804020100503 0.1778799842113
0.57286432160804 0.177237086985339
0.584924623115578 0.176638588720513
0.596984924623116 0.176085664449825
0.609045226130654 0.175579663918452
0.621105527638191 0.175122132631351
0.633165829145729 0.174714836570181
0.645226130653267 0.17435979134038
0.657286432160804 0.174059296700122
0.669346733668342 0.173815977670197
0.68140703517588 0.173632833746924
0.693467336683417 0.173513298166098
0.705527638190955 0.173461309732915
0.717587939698493 0.173481400495318
0.72964824120603 0.173578803575425
0.741708542713568 0.173759586901558
0.753768844221106 0.174030820575053
0.765829145728643 0.174400788423728
0.777889447236181 0.174879258342495
0.789949748743719 0.17547783193827
0.802010050251256 0.176210402805603
0.814070351758794 0.177093766148686
0.826130653266332 0.178148443286642
0.83819095477387 0.179399817797384
0.850251256281407 0.180879734601615
0.862311557788945 0.182628805835295
0.874371859296483 0.184699830376017
0.88643216080402 0.187163033773239
0.898492462311558 0.190114415806453
0.910552763819096 0.19368968673835
0.922613065326633 0.198088916284011
0.934673366834171 0.203623431751231
0.946733668341709 0.210814004575595
0.958793969849246 0.220625344832652
0.970854271356784 0.235147062955713
0.982914572864322 0.260350973120134
0.99497487437186 0.331234019674563
1.0070351758794 nan
1.01909547738693 nan
1.03115577889447 nan
1.04321608040201 nan
1.05527638190955 nan
1.06733668341709 nan
1.07939698492462 nan
1.09145728643216 nan
1.1035175879397 nan
1.11557788944724 nan
1.12763819095477 nan
1.13969849246231 nan
1.15175879396985 nan
1.16381909547739 nan
1.17587939698492 nan
1.18793969849246 nan
1.2 nan
};

\addplot [color=blue] 
  table{%
-1 0.5
-1 0
};
\addplot [color=blue]
  table{%
1 0.5
1 0
};

\end{axis}
\end{tikzpicture}%

%% file: TikZ/pres_concrete_MAP_resized.tikz
\begin{tikzpicture}[scale=1]
\pgfplotsset{
	every axis plot/.style={line width=1.4pt},
	every axis/.style={grid style={line width=0.7pt, dashed}, 
					 },
	axis background/.style={fill=white},
	every outer x axis line/.style={line width=1.4pt},
	every outer y axis line/.style={line width=1.4pt},
	every tick/.style={black,
					line width=0.7pt,
					},
	label style={font=\fontsize{8}{9}\selectfont},
	every y tick label/.style={font=\fontsize{8}{9}\color{black}},
	every x tick label/.style={font=\fontsize{8}{9}\color{black}},
	every extra x tick/.style={grid style={solid, violet, line width=2.8pt},
							x tick label style={/pgf/number format/.cd,precision=10}
							},
	legend style={
					line width=0.7pt,
					font=\fontsize{11.4}{12}\selectfont\color{black},
					legend cell align=left,
					align=left,
					fill=white,
					nodes={scale=0.7, transform shape},
					},
	legend image code/.code={
		\draw[mark repeat=2,mark phase=2]
			plot coordinates {
				(0cm,0cm)
				(0.25cm,0cm)        
				(0.5cm,0cm)         
			};%
		},
	legend entries={	{\color{red} $-\ln p(\tilde{x})$},
					{\color{darkgray}$-\ln p(y|\tilde{x})$},
					{\color{darkgreen}$-\ln p(\tilde{x},y)$},
					{\color{blue}$-\ln p(x,y)$},
					},
}

\begin{axis}[%
width=10cm,
height=5cm, 
scale=0.75,
scale only axis,
separate axis lines,
tick pos=left,
xmin=-1.2,
xmax=1.2,
xminorticks=true,
xtick distance = {0.5},
ytick distance = {2},
xlabel={$\tilde{x}$}, 
ymin=-1,
ymax=7,
ylabel={$-\ln p$},
xmajorgrids,
xminorgrids,
ymajorgrids,
]



\addplot [color=red, mark=*,only marks, mark size=2*0.75]
table {%
-1.2 nan
-0.802010050251256 1.97726434084035
-0.404020100502512 2.81935863861014
0.006030150753769 2.99569627381071
0.404020100502513 2.81935863861014
0.802010050251256 1.97726434084035
1.2 nan
};

\addplot [color=darkgray, mark=triangle*,only marks, mark size=2*0.75] 
table {%
-0.802010050251256 3.80859485501473
-0.404020100502512 2.21183517722881
0.006030150753769 1.22936301743487
0.404020100502513 0.918970855620773
0.802010050251256 1.24216269421071
};

\addplot [color=darkgreen, mark=square*,only marks, mark size=2*0.75]
table {%
-1.2 nan
-0.802010050251256 5.78585919585508
-0.404020100502512 5.03119381583895
0.006030150753769 4.22505929124558
0.404020100502513 3.73832949423092
0.802010050251256 3.21942703505106
1.2 nan
};

\addplot [color=blue, mark=diamond*,only marks, mark size = 3*0.75] 
table {%
-1 5.5039955140122405
1 2.32007591602229
};





\addplot [color=red]
table {%
-1.2 nan
-1.18793969849246 nan
-1.17587939698492 nan
-1.16381909547739 nan
-1.15175879396985 nan
-1.13969849246231 nan
-1.12763819095477 nan
-1.11557788944724 nan
-1.1035175879397 nan
-1.09145728643216 nan
-1.07939698492462 nan
-1.06733668341709 nan
-1.05527638190955 nan
-1.04321608040201 nan
-1.03115577889447 nan
-1.01909547738693 nan
-1.0070351758794 nan
-0.994974874371859 -1.51872692130585
-0.982914572864322 -0.333249978008191
-0.970854271356784 0.18282051826612
-0.958793969849246 0.515939183027156
-0.946733668341708 0.76163605331207
-0.934673366834171 0.955880403288078
-0.922613065326633 1.11617234687083
-0.910552763819095 1.25235950981735
-0.898492462311558 1.37053693636689
-0.88643216080402 1.47474088742931
-0.874371859296482 1.5677830606597
-0.862311557788945 1.65170082541164
-0.850251256281407 1.72801781562213
-0.838190954773869 1.79790333861339
-0.826130653266332 1.86227437087507
-0.814070351758794 1.92186329792939
-0.802010050251256 1.97726434084035
-0.789949748743718 2.02896624302621
-0.777889447236181 2.07737582578661
-0.765829145728643 2.12283531258923
-0.753768844221105 2.16563530134492
-0.741708542713568 2.20602463417708
-0.72964824120603 2.24421801466625
-0.717587939698492 2.28040196267471
-0.705527638190955 2.31473952400282
-0.693467336683417 2.34737403482781
-0.681407035175879 2.37843215981328
-0.669346733668342 2.40802636582583
-0.657286432160804 2.43625695257418
-0.645226130653266 2.46321373210905
-0.633165829145728 2.48897742760466
-0.621105527638191 2.51362084589486
-0.609045226130653 2.53720986628829
-0.596984924623115 2.55980427914251
-0.584924623115578 2.58145850076699
-0.57286432160804 2.60222218589761
-0.560804020100502 2.62214075484551
-0.548743718592965 2.64125584917978
-0.536683417085427 2.65960572724519
-0.524623115577889 2.67722560878303
-0.512562814070352 2.69414797629819
-0.500502512562814 2.71040283950768
-0.488442211055276 2.72601796814799
-0.476381909547739 2.74101909755764
-0.464321608040201 2.75543011074733
-0.452261306532663 2.76927320009129
-0.440201005025125 2.7825690112956
-0.428140703517588 2.79533677190256
-0.41608040201005 2.80759440626033
-0.404020100502512 2.81935863861014
-0.391959798994975 2.83064508571188
-0.379899497487437 2.84146834023276
-0.367839195979899 2.85184204595854
-0.355778894472362 2.86177896574609
-0.343718592964824 2.87129104301631
-0.331658291457286 2.88038945748433
-0.319597989949749 2.88908467573584
-0.307537688442211 2.89738649718322
-0.295477386934673 2.90530409587002
-0.283417085427136 2.91284605853598
-0.271356783919598 2.9200204193061
-0.25929648241206 2.92683469132466
-0.247236180904522 2.93329589561821
-0.235175879396985 2.93941058743912
-0.223115577889447 2.94518488031278
-0.211055276381909 2.9506244679869
-0.198994974874372 2.95573464445894
-0.186934673366834 2.96052032223859
-0.174874371859296 2.9649860489848
-0.162814070351759 2.96913602264168
-0.150753768844221 2.97297410518374
-0.138693467336683 2.97650383506885
-0.126633165829146 2.97972843848628
-0.114572864321608 2.98265083947726
-0.10251256281407 2.9852736689966
-0.0904522613065326 2.98759927297569
-0.0783919597989948 2.98962971943985
-0.066331658291457 2.99136680472607
-0.0542713567839195 2.99281205884101
-0.0422110552763817 2.9939667499931
-0.0301507537688441 2.99483188832724
-0.0180904522613063 2.99540822888531
-0.00603015075376878 2.99569627381071
0.006030150753769 2.99569627381071
0.0180904522613068 2.99540822888531
0.0301507537688444 2.99483188832724
0.0422110552763821 2.9939667499931
0.0542713567839197 2.99281205884101
0.0663316582914575 2.99136680472607
0.078391959798995 2.98962971943985
0.0904522613065328 2.98759927297569
0.102512562814071 2.9852736689966
0.114572864321608 2.98265083947726
0.126633165829146 2.97972843848628
0.138693467336684 2.97650383506885
0.150753768844221 2.97297410518374
0.162814070351759 2.96913602264168
0.174874371859297 2.9649860489848
0.186934673366834 2.96052032223859
0.198994974874372 2.95573464445894
0.21105527638191 2.9506244679869
0.223115577889447 2.94518488031278
0.235175879396985 2.93941058743912
0.247236180904523 2.93329589561821
0.25929648241206 2.92683469132466
0.271356783919598 2.9200204193061
0.283417085427136 2.91284605853598
0.295477386934674 2.90530409587002
0.307537688442211 2.89738649718322
0.319597989949749 2.88908467573584
0.331658291457287 2.88038945748433
0.343718592964824 2.8712910430163
0.355778894472362 2.86177896574609
0.3678391959799 2.85184204595854
0.379899497487437 2.84146834023276
0.391959798994975 2.83064508571188
0.404020100502513 2.81935863861014
0.41608040201005 2.80759440626033
0.428140703517588 2.79533677190256
0.440201005025126 2.78256901129559
0.452261306532663 2.76927320009129
0.464321608040201 2.75543011074732
0.476381909547739 2.74101909755764
0.488442211055277 2.72601796814799
0.500502512562814 2.71040283950768
0.512562814070352 2.69414797629819
0.52462311557789 2.67722560878303
0.536683417085427 2.65960572724519
0.548743718592965 2.64125584917978
0.560804020100503 2.62214075484551
0.57286432160804 2.60222218589761
0.584924623115578 2.58145850076699
0.596984924623116 2.55980427914251
0.609045226130654 2.53720986628829
0.621105527638191 2.51362084589486
0.633165829145729 2.48897742760466
0.645226130653267 2.46321373210905
0.657286432160804 2.43625695257418
0.669346733668342 2.40802636582583
0.68140703517588 2.37843215981328
0.693467336683417 2.34737403482781
0.705527638190955 2.31473952400282
0.717587939698493 2.28040196267471
0.72964824120603 2.24421801466625
0.741708542713568 2.20602463417708
0.753768844221106 2.16563530134492
0.765829145728643 2.12283531258923
0.777889447236181 2.07737582578661
0.789949748743719 2.02896624302621
0.802010050251256 1.97726434084035
0.814070351758794 1.92186329792939
0.826130653266332 1.86227437087507
0.83819095477387 1.79790333861339
0.850251256281407 1.72801781562213
0.862311557788945 1.65170082541164
0.874371859296483 1.5677830606597
0.88643216080402 1.47474088742931
0.898492462311558 1.37053693636689
0.910552763819096 1.25235950981735
0.922613065326633 1.11617234687083
0.934673366834171 0.955880403288078
0.946733668341709 0.761636053312063
0.958793969849246 0.51593918302715
0.970854271356784 0.182820518266115
0.982914572864322 -0.333249978008216
0.99497487437186 -1.51872692130592
1.0070351758794 nan
1.01909547738693 nan
1.03115577889447 nan
1.04321608040201 nan
1.05527638190955 nan
1.06733668341709 nan
1.07939698492462 nan
1.09145728643216 nan
1.1035175879397 nan
1.11557788944724 nan
1.12763819095477 nan
1.13969849246231 nan
1.15175879396985 nan
1.16381909547739 nan
1.17587939698492 nan
1.18793969849246 nan
1.2 nan
};

\addplot [color=darkgray] 
table {%
-1.2 6.03893853320467
-1.18793969849246 5.96204350530134
-1.17587939698492 5.88573028088781
-1.16381909547739 5.8099988599641
-1.15175879396985 5.7348492425302
-1.13969849246231 5.6602814285861
-1.12763819095477 5.58629541813182
-1.11557788944724 5.51289121116735
-1.1035175879397 5.44006880769269
-1.09145728643216 5.36782820770784
-1.07939698492462 5.2961694112128
-1.06733668341709 5.22509241820758
-1.05527638190955 5.15459722869216
-1.04321608040201 5.08468384266655
-1.03115577889447 5.01535226013076
-1.01909547738693 4.94660248108478
-1.0070351758794 4.8784345055286
-0.994974874371859 4.81084833346224
-0.982914572864322 4.74384396488569
-0.970854271356784 4.67742139979895
-0.958793969849246 4.61158063820202
-0.946733668341708 4.5463216800949
-0.934673366834171 4.48164452547759
-0.922613065326633 4.4175491743501
-0.910552763819095 4.35403562671241
-0.898492462311558 4.29110388256454
-0.88643216080402 4.22875394190647
-0.874371859296482 4.16698580473822
-0.862311557788945 4.10579947105978
-0.850251256281407 4.04519494087114
-0.838190954773869 3.98517221417232
-0.826130653266332 3.92573129096332
-0.814070351758794 3.86687217124412
-0.802010050251256 3.80859485501473
-0.789949748743718 3.75089934227515
-0.777889447236181 3.69378563302538
-0.765829145728643 3.63725372726543
-0.753768844221105 3.58130362499528
-0.741708542713568 3.52593532621495
-0.72964824120603 3.47114883092443
-0.717587939698492 3.41694413912371
-0.705527638190955 3.36332125081281
-0.693467336683417 3.31028016599172
-0.681407035175879 3.25782088466044
-0.669346733668342 3.20594340681897
-0.657286432160804 3.15464773246732
-0.645226130653266 3.10393386160547
-0.633165829145728 3.05380179423343
-0.621105527638191 3.00425153035121
-0.609045226130653 2.9552830699588
-0.596984924623115 2.90689641305619
-0.584924623115578 2.8590915596434
-0.57286432160804 2.81186850972042
-0.560804020100502 2.76522726328725
-0.548743718592965 2.71916782034389
-0.536683417085427 2.67369018089034
-0.524623115577889 2.6287943449266
-0.512562814070352 2.58448031245267
-0.500502512562814 2.54074808346855
-0.488442211055276 2.49759765797425
-0.476381909547739 2.45502903596975
-0.464321608040201 2.41304221745507
-0.452261306532663 2.3716372024302
-0.440201005025125 2.33081399089513
-0.428140703517588 2.29057258284988
-0.41608040201005 2.25091297829444
-0.404020100502512 2.21183517722881
-0.391959798994975 2.17333917965299
-0.379899497487437 2.13542498556699
-0.367839195979899 2.09809259497079
-0.355778894472362 2.0613420078644
-0.343718592964824 2.02517322424783
-0.331658291457286 1.98958624412106
-0.319597989949749 1.95458106748411
-0.307537688442211 1.92015769433697
-0.295477386934673 1.88631612467963
-0.283417085427136 1.85305635851211
-0.271356783919598 1.8203783958344
-0.25929648241206 1.7882822366465
-0.247236180904522 1.75676788094842
-0.235175879396985 1.72583532874014
-0.223115577889447 1.69548458002167
-0.211055276381909 1.66571563479302
-0.198994974874372 1.63652849305417
-0.186934673366834 1.60792315480514
-0.174874371859296 1.57989962004591
-0.162814070351759 1.5524578887765
-0.150753768844221 1.5255979609969
-0.138693467336683 1.49931983670711
-0.126633165829146 1.47362351590713
-0.114572864321608 1.44850899859696
-0.10251256281407 1.4239762847766
-0.0904522613065326 1.40002537444606
-0.0783919597989948 1.37665626760532
-0.066331658291457 1.35386896425439
-0.0542713567839195 1.33166346439328
-0.0422110552763817 1.31003976802197
-0.0301507537688441 1.28899787514048
-0.0180904522613063 1.2685377857488
-0.00603015075376878 1.24865949984693
0.006030150753769 1.22936301743487
0.0180904522613068 1.21064833851262
0.0301507537688444 1.19251546308018
0.0422110552763821 1.17496439113755
0.0542713567839197 1.15799512268474
0.0663316582914575 1.14160765772173
0.078391959798995 1.12580199624854
0.0904522613065328 1.11057813826515
0.102512562814071 1.09593608377158
0.114572864321608 1.08187583276781
0.126633165829146 1.06839738525386
0.138693467336684 1.05550074122972
0.150753768844221 1.04318590069539
0.162814070351759 1.03145286365087
0.174874371859297 1.02030163009617
0.186934673366834 1.00973220003127
0.198994974874372 0.999744573456181
0.21105527638191 0.990338750370906
0.223115577889447 0.981514730775441
0.235175879396985 0.973272514669787
0.247236180904523 0.965612102053944
0.25929648241206 0.958533492927912
0.271356783919598 0.952036687291691
0.283417085427136 0.94612168514528
0.295477386934674 0.940788486488681
0.307537688442211 0.936037091321892
0.319597989949749 0.931867499644914
0.331658291457287 0.928279711457747
0.343718592964824 0.925273726760391
0.355778894472362 0.922849545552846
0.3678391959799 0.921007167835111
0.379899497487437 0.919746593607188
0.391959798994975 0.919067822869075
0.404020100502513 0.918970855620773
0.41608040201005 0.919455691862282
0.428140703517588 0.920522331593602
0.440201005025126 0.922170774814733
0.452261306532663 0.924401021525675
0.464321608040201 0.927213071726427
0.476381909547739 0.93060692541699
0.488442211055277 0.934582582597365
0.500502512562814 0.93914004326755
0.512562814070352 0.944279307427546
0.52462311557789 0.950000375077353
0.536683417085427 0.95630324621697
0.548743718592965 0.963187920846399
0.560804020100503 0.970654398965638
0.57286432160804 0.978702680574689
0.584924623115578 0.98733276567355
0.596984924623116 0.996544654262222
0.609045226130654 1.0063383463407
0.621105527638191 1.016713841909
0.633165829145729 1.0276711409671
0.645226130653267 1.03921024351502
0.657286432160804 1.05133114955274
0.669346733668342 1.06403385908028
0.68140703517588 1.07731837209763
0.693467336683417 1.09118468860479
0.705527638190955 1.10563280860176
0.717587939698493 1.12066273208854
0.72964824120603 1.13627445906513
0.741708542713568 1.15246798953153
0.753768844221106 1.16924332348775
0.765829145728643 1.18660046093377
0.777889447236181 1.20453940186961
0.789949748743719 1.22306014629525
0.802010050251256 1.24216269421071
0.814070351758794 1.26184704561598
0.826130653266332 1.28211320051105
0.83819095477387 1.30296115889594
0.850251256281407 1.32439092077064
0.862311557788945 1.34640248613515
0.874371859296483 1.36899585498948
0.88643216080402 1.39217102733361
0.898492462311558 1.41592800316755
0.910552763819096 1.44026678249131
0.922613065326633 1.46518736530487
0.934673366834171 1.49068975160825
0.946733668341709 1.51677394140144
0.958793969849246 1.54343993468443
0.970854271356784 1.57068773145724
0.982914572864322 1.59851733171986
0.99497487437186 1.62692873547229
1.0070351758794 1.65592194271453
1.01909547738693 1.68549695344659
1.03115577889447 1.71565376766845
1.04321608040201 1.74639238538012
1.05527638190955 1.77771280658161
1.06733668341709 1.8096150312729
1.07939698492462 1.84209905945401
1.09145728643216 1.87516489112493
1.1035175879397 1.90881252628566
1.11557788944724 1.9430419649362
1.12763819095477 1.97785320707655
1.13969849246231 2.01324625270671
1.15175879396985 2.04922110182668
1.16381909547739 2.08577775443646
1.17587939698492 2.12291621053605
1.18793969849246 2.16063647012546
1.2 2.19893853320467
};

\addplot [color=darkgreen]
table {%
-1.2 nan
-1.18793969849246 nan
-1.17587939698492 nan
-1.16381909547739 nan
-1.15175879396985 nan
-1.13969849246231 nan
-1.12763819095477 nan
-1.11557788944724 nan
-1.1035175879397 nan
-1.09145728643216 nan
-1.07939698492462 nan
-1.06733668341709 nan
-1.05527638190955 nan
-1.04321608040201 nan
-1.03115577889447 nan
-1.01909547738693 nan
-1.0070351758794 nan
-0.994974874371859 3.29212141215639
-0.982914572864322 4.4105939868775
-0.970854271356784 4.86024191806507
-0.958793969849246 5.12751982122918
-0.946733668341708 5.30795773340697
-0.934673366834171 5.43752492876567
-0.922613065326633 5.53372152122093
-0.910552763819095 5.60639513652976
-0.898492462311558 5.66164081893143
-0.88643216080402 5.70349482933578
-0.874371859296482 5.73476886539792
-0.862311557788945 5.75750029647142
-0.850251256281407 5.77321275649328
-0.838190954773869 5.78307555278571
-0.826130653266332 5.78800566183839
-0.814070351758794 5.78873546917351
-0.802010050251256 5.78585919585508
-0.789949748743718 5.77986558530136
-0.777889447236181 5.77116145881199
-0.765829145728643 5.76008903985466
-0.753768844221105 5.74693892634021
-0.741708542713568 5.73195996039203
-0.72964824120603 5.71536684559068
-0.717587939698492 5.69734610179842
-0.705527638190955 5.67806077481563
-0.693467336683417 5.65765420081953
-0.681407035175879 5.63625304447373
-0.669346733668342 5.61396977264481
-0.657286432160804 5.5909046850415
-0.645226130653266 5.56714759371452
-0.633165829145728 5.54277922183809
-0.621105527638191 5.51787237624607
-0.609045226130653 5.49249293624708
-0.596984924623115 5.4667006921987
-0.584924623115578 5.44055006041039
-0.57286432160804 5.41409069561803
-0.560804020100502 5.38736801813276
-0.548743718592965 5.36042366952367
-0.536683417085427 5.33329590813553
-0.524623115577889 5.30601995370963
-0.512562814070352 5.27862828875086
-0.500502512562814 5.25115092297623
-0.488442211055276 5.22361562612224
-0.476381909547739 5.19604813352739
-0.464321608040201 5.1684723282024
-0.452261306532663 5.14091040252149
-0.440201005025125 5.11338300219073
-0.428140703517588 5.08590935475244
-0.41608040201005 5.05850738455477
-0.404020100502512 5.03119381583895
-0.391959798994975 5.00398426536488
-0.379899497487437 4.97689332579975
-0.367839195979899 4.94993464092933
-0.355778894472362 4.92312097361049
-0.343718592964824 4.89646426726413
-0.331658291457286 4.86997570160539
-0.319597989949749 4.84366574321995
-0.307537688442211 4.81754419152019
-0.295477386934673 4.79162022054965
-0.283417085427136 4.76590241704809
-0.271356783919598 4.7403988151405
-0.25929648241206 4.71511692797116
-0.247236180904522 4.69006377656663
-0.235175879396985 4.66524591617925
-0.223115577889447 4.64066946033445
-0.211055276381909 4.61634010277992
-0.198994974874372 4.59226313751312
-0.186934673366834 4.56844347704373
-0.174874371859296 4.54488566903071
-0.162814070351759 4.52159391141818
-0.150753768844221 4.49857206618064
-0.138693467336683 4.47582367177596
-0.126633165829146 4.45335195439341
-0.114572864321608 4.43115983807422
-0.10251256281407 4.4092499537732
-0.0904522613065326 4.38762464742175
-0.0783919597989948 4.36628598704517
-0.066331658291457 4.34523576898047
-0.0542713567839195 4.32447552323429
-0.0422110552763817 4.30400651801507
-0.0301507537688441 4.28382976346773
-0.0180904522613063 4.26394601463411
-0.00603015075376878 4.24435577365764
0.006030150753769 4.22505929124558
0.0180904522613068 4.20605656739793
0.0301507537688444 4.18734735140742
0.0422110552763821 4.16893114113065
0.0542713567839197 4.15080718152574
0.0663316582914575 4.1329744624478
0.078391959798995 4.11543171568839
0.0904522613065328 4.09817741124084
0.102512562814071 4.08120975276818
0.114572864321608 4.06452667224507
0.126633165829146 4.04812582374014
0.138693467336684 4.03200457629857
0.150753768844221 4.01616000587913
0.162814070351759 4.00058888629255
0.174874371859297 3.98528767908096
0.186934673366834 3.97025252226986
0.198994974874372 3.95547921791513
0.21105527638191 3.94096321835781
0.223115577889447 3.92669961108822
0.235175879396985 3.9126831021089
0.247236180904523 3.89890799767216
0.25929648241206 3.88536818425257
0.271356783919598 3.87205710659779
0.283417085427136 3.85896774368126
0.295477386934674 3.8460925823587
0.307537688442211 3.83342358850512
0.319597989949749 3.82095217538076
0.331658291457287 3.80866916894208
0.343718592964824 3.7965647697767
0.355778894472362 3.78462851129893
0.3678391959799 3.77284921379366
0.379899497487437 3.76121493383995
0.391959798994975 3.74971290858096
0.404020100502513 3.73832949423092
0.41608040201005 3.72705009812261
0.428140703517588 3.71585910349616
0.440201005025126 3.70473978611033
0.452261306532663 3.69367422161697
0.464321608040201 3.68264318247375
0.476381909547739 3.67162602297463
0.488442211055277 3.66060055074535
0.500502512562814 3.64954288277523
0.512562814070352 3.63842728372573
0.52462311557789 3.62722598386038
0.536683417085427 3.61590897346216
0.548743718592965 3.60444377002618
0.560804020100503 3.59279515381115
0.57286432160804 3.5809248664723
0.584924623115578 3.56879126644054
0.596984924623116 3.55634893340473
0.609045226130654 3.54354821262899
0.621105527638191 3.53033468780386
0.633165829145729 3.51664856857176
0.645226130653267 3.50242397562407
0.657286432160804 3.48758810212692
0.669346733668342 3.47206022490612
0.68140703517588 3.45575053191091
0.693467336683417 3.4385587234326
0.705527638190955 3.42037233260457
0.717587939698493 3.40106469476325
0.72964824120603 3.38049247373138
0.741708542713568 3.35849262370861
0.753768844221106 3.33487862483267
0.765829145728643 3.309435773523
0.777889447236181 3.28191522765621
0.789949748743719 3.25202638932146
0.802010050251256 3.21942703505106
0.814070351758794 3.18371034354536
0.826130653266332 3.14438757138612
0.83819095477387 3.10086449750933
0.850251256281407 3.05240873639278
0.862311557788945 2.99810331154679
0.874371859296483 2.93677891564918
0.88643216080402 2.86691191476292
0.898492462311558 2.78646493953444
0.910552763819096 2.69262629230866
0.922613065326633 2.5813597121757
0.934673366834171 2.44657015489633
0.946733668341709 2.2784099947135
0.958793969849246 2.05937911771158
0.970854271356784 1.75350824972336
0.982914572864322 1.26526735371165
0.99497487437186 0.108201814166376
1.0070351758794 nan
1.01909547738693 nan
1.03115577889447 nan
1.04321608040201 nan
1.05527638190955 nan
1.06733668341709 nan
1.07939698492462 nan
1.09145728643216 nan
1.1035175879397 nan
1.11557788944724 nan
1.12763819095477 nan
1.13969849246231 nan
1.15175879396985 nan
1.16381909547739 nan
1.17587939698492 nan
1.18793969849246 nan
1.2 nan
};

\end{axis}
\end{tikzpicture}%

%% file: TikZ/CMD_layer.tikz

\begin{tikzpicture}[scale=0.7,
					outer xsep=0pt, outer ysep=0pt,
					>=latex,
					triangle/.style={regular polygon,shape border rotate=180,regular polygon sides=3,line width=0.30mm,draw=black},
					normalrec/.style={rectangle,line width=0.30mm,draw=black,fill=none},
					nonlin/.style={rectangle,line width=0.30mm,draw=black,fill=none, rounded corners=3},%
					circle1/.style={circle,line width=0.30mm,draw=black,fill=none, inner sep=0pt, minimum height=10},	
					arrowthick/.style={->,line width=0.60mm},
					arrowthin/.style={->,line width=0.30mm},
					linethin/.style={-,line width=0.30mm},
					linethick/.style={-,line width=0.60mm},
					linesplit/.style={circle, draw=black, fill=black, inner sep=0pt, minimum height=2},
					dashedbox/.style={line width=0.30mm, draw=black, dash pattern=on 2.00mm off 1.00mm},
					font={\fontsize{7}{12}\selectfont},
					tanh/.pic={
						\draw[scale=1, domain=-2:2, smooth,variable=\x, draw=blue, fill=white, line width=0.30mm]
										plot ({\x}, {2 * (1 / (1+exp(-4*\x))) - 1});
					},
					]

\node[normalrec, minimum width=150, minimum height=130, dashed, rounded corners=30] (grad_block) at (-1,2.1) {};

\node[normalrec, draw=none] (g_i) at (0,-2.5) {$\mathbf{s}^{(j)}$};
\node[linesplit, minimum height=5] (hnode2) at (0,-1.5) {};
\node[linesplit] (hnode1) at (0,-1) {};
\node[normalrec, minimum height=5, draw=none] (llr_scale) at (-1.5,-0.3) {$\frac{(\bullet)+\ln(1/\boldsymbol{\alpha}-1)}{\color{red}\tau^{\color{red}(j)}}$};
\node[nonlin, minimum height=15, minimum width=20] (tanh1) at (-1.5,1) {};
\node[nonlin, minimum height=15, minimum width=20] (tanh2) at (1.5,1) {};
\draw (tanh1) pic[shift={(0,0)}, scale=0.15] {tanh};
\draw (tanh2) pic[shift={(0,0)}, scale=0.15] {tanh};
\node[circle1, draw=none] (x_i) at (-1.5,2) {$\tilde{\mathbf{x}}^{(j)}$};
\node[nonlin, minimum height=5] (quadr) at (0,3) {$(\bullet)^2-1$};
\node[nonlin, minimum height=5] (lin) at (-3,3) {$\mathbf{H}^T\mathbf{H}(\bullet)-\mathbf{H}^T\mathbf{y}$};
\node[nonlin, minimum height=5] (el_prod) at (-1.5,4) {$\odot$}; 
\node[circle1] (reg_con) at (0,5) {$+$};
\node[circle1] (res_con) at (0,7) {$+$};
\node[normalrec, draw=none] (g_i+1) at (0,8.5) {$\mathbf{s}^{(j+1)}$};


\draw[linethick] (g_i) -- (hnode2);
\draw[linethin] (hnode2) -- (hnode1);
\draw[arrowthin] (hnode1) -| (tanh2);
\draw[linethin] (hnode1) -| (llr_scale);
\draw[arrowthin] (llr_scale) -- (tanh1);
\draw[arrowthin] (tanh1) -- (x_i);
\draw[arrowthin] (x_i) -| (quadr);
\draw[arrowthin] (quadr) |- (el_prod);
\draw[arrowthin] (x_i) -| (lin);
\draw[arrowthin] (lin) |- (el_prod);
\draw[arrowthin] (el_prod) |- (reg_con) node[below left, xshift=-1cm]{$2/({\color{red}\tau^{(j)}}\sigma_{\text{n}}^2)$};
\draw[arrowthin] (tanh2) |- (reg_con) node[below right, xshift=0.075cm, yshift=-0.075cm]{$\bm{-}$};
\draw[arrowthick] (reg_con) -- (res_con) node[below left]{${\color{red}-\delta^{(j)}}$};
\draw[arrowthick, draw=none] (reg_con) -- (res_con) node[midway, right]{$\frac{\partial L(\mathbf{s})}{\partial \mathbf{s}} \bigg|_{\mathbf{s} = \mathbf{s}^{(j)}}$};
\draw[arrowthick] (res_con) -- (g_i+1);
\draw[arrowthick] (hnode2) -- node {} ++(4,0)   |- (res_con.east);


\end{tikzpicture}%


%% file: TikZ/plot64x64_comparison_resized.tikz
\begin{tikzpicture}[scale=1]
\pgfplotsset{
	every axis plot/.style={line width=1.4pt},
	every axis/.style={grid style={line width=0.7pt, dashed}, 
					 },
	axis background/.style={fill=white},
	every outer x axis line/.style={line width=1.4pt},
	every outer y axis line/.style={line width=1.4pt},
	every tick/.style={black,
					line width=0.7pt,
					},
	label style={font=\fontsize{8}{9}\selectfont},
	every y tick label/.style={font=\fontsize{8}{9}\color{black}},
	every x tick label/.style={font=\fontsize{8}{9}\color{black}},
	every extra x tick/.style={grid style={solid, violet, line width=2.8pt},
							x tick label style={/pgf/number format/.cd,precision=10}
							},
	legend style={
					line width=0.7pt,
					font=\fontsize{11.4}{12}\selectfont\color{black},
					legend cell align=left,
					align=left,
					fill=white,
					fill opacity=0.8, draw opacity=1, text opacity=1,
					nodes={scale=0.7, transform shape},
					},
	legend image code/.code={
		\draw[mark repeat=2,mark phase=2]
			plot coordinates {
				(0cm,0cm)
				(0.25cm,0cm)        
				(0.5cm,0cm)         
			};%
		},
	legend pos={south west},
	legend entries={{\color{\clsd} SD},
					{\color{\clmmse} MMSE},
					{\color{\clmosic} MOSIC},
					{\color{\clamp} AMP}, 
					{\color{\clsdr} SDR},
					{\color{\cldetnet} DetNet}, 
					{\color{\clmmnet} $\text{MMNet}_{\text{iid}}$}, 
					{\color{\cloamp} OAMPNet},
					{\color{\clcmd} $\text{CMDNet}_{\text{bin}}$}, 
					{\color{\clawgn} AWGN},
					},
}
\begin{axis}[%
width=10.5cm,
height=8cm, 
scale=0.7,
scale only axis,
separate axis lines,
tick pos=left,
xmin=1,
xmax=18,
ymode=log,
xminorticks=true,
xtick distance = {2},
ytick distance = {10},
xlabel={$E_{\text{b}}/N_0$ [dB]}, 
ymin=0.000001,
ymax=1,
ylabel={BER},
xmajorgrids,
xminorgrids,
ymajorgrids,
]



\addplot [ptsd]
table {%
-4 0.262451171875
-2 0.208413461538462
0 0.164713541666667
2 0.119514627659574
4 0.0446013931888545
6 0.00649971461187215
8 0.000431358421357456
10 1.68004425778524e-05
12 2.6041449654586e-07
};
\addplot [ptmmse]
table {%
-6 0.2730390625
-5 0.255496875
-4 0.23509375
-3 0.21550625
-2 0.1973953125
-1 0.1774640625
0 0.158225
1 0.139621875
2 0.123746875
3 0.1057859375
4 0.09038125
5 0.075959375
6 0.06318125
7 0.051059375
8 0.0408828125
9 0.0320109375
10 0.024646875
11 0.018515625
12 0.013928125
13 0.0099703125
14 0.007421875
15 0.0051234375
16 0.0036828125
17 0.0025671875
18 0.0018171875
19 0.001225
20 0.00082421875
21 0.00058125
22 0.0004
23 0.000285677083333333
24 0.0002076171875
25 0.000142897727272727
26 0.000112083333333333
27 8.33059210526316e-05
28 5.81597222222222e-05
29 4.90234375e-05
30 3.29752604166667e-05
31 2.65364583333333e-05
32 2.1e-05
33 1.70006793478261e-05
34 1.2376968503937e-05
35 1.00806451612903e-05
36 8.05412371134021e-06
};
\addplot [ptmosic]
table {%
-6 0.288068181818182
-5 0.273706896551724
-4 0.253024193548387
-3 0.214255136986301
-2 0.208470394736842
-1 0.190323795180723
0 0.163015463917526
1 0.159598214285714
2 0.137364130434783
3 0.105574324324324
4 0.09775390625
5 0.06982421875
6 0.0518936258278146
7 0.032116273100616
8 0.017405282079646
9 0.0108272821576764
10 0.00542346407497396
11 0.00299271545178436
12 0.00110304396449291
13 0.000417209244205931
14 0.000158483025834
15 5.29657426051345e-05
16 1.42406936142511e-05
17 4.12121864616234e-06
18 8.1254407485075e-07
};
\addplot [ptamp]
table {%
-6 0.2720453125
-5 0.252784375
-4 0.233028125
-3 0.2145140625
-2 0.193340625
-1 0.1715015625
0 0.14971875
1 0.1275609375
2 0.1028546875
3 0.077434375
4 0.05076875
5 0.0271890625
6 0.0126734375
7 0.004375
8 0.00165
9 0.000754166666666667
10 0.000384375
11 0.000261979166666667
12 0.000265625
13 0.000245758928571429
14 0.0003175
15 0.00039453125
16 0.000551041666666667
17 0.00096875
18 0.001721875
19 0.00168828125
20 0.00180703125
21 0.001875
22 0.00196875
23 0.0018015625
24 0.0017171875
25 0.002253125
26 0.0020671875
27 0.00180625
28 0.0019234375
29 0.00200625
30 0.00255
31 0.002121875
32 0.0020890625
33 0.001746875
34 0.00179375
35 0.00255625
36 0.0019640625
};
\addplot [ptsdr]
table {%
-6 0.28796875
-5 0.25703125
-4 0.24234375
-3 0.21109375
-2 0.20140625
-1 0.16359375
0 0.1596875
1 0.1290625
2 0.107421875
3 0.07125
4 0.0526953125
5 0.032125
6 0.021953125
7 0.0109479166666667
8 0.00472426470588235
9 0.0019609375
10 0.000556605871886121
11 0.000139508928571429
12 2.43380062305296e-05
13 3.25860271115746e-06
14 3.27293674067868e-07
};
\addplot [ptdetnet]
table {%
-6 0.28168125
-5 0.264075
-4 0.24350625
-3 0.223678125
-2 0.202959375
-1 0.1814828125
0 0.15836875
1 0.1337796875
2 0.1066453125
3 0.07731875
4 0.04951875
5 0.0270546875
6 0.0117828125
7 0.00420625
8 0.0012953125
9 0.0003590625
10 0.00010263671875
11 3.00480769230769e-05
12 1.13224637681159e-05
13 5.46328671328671e-06
14 3.37810475161987e-06
15 2.38061263318113e-06
16 1.72271223814774e-06
17 1.0867214532872e-06
18 1.19716298552932e-06
19 8.82269904009034e-07
20 8.40638383297644e-07
21 5.17775603392042e-07
22 6.14763969171484e-07
};
\addplot [ptmmnet]
table {%
-6 0.274875
-5 0.258209375
-4 0.2406265625
-3 0.221940625
-2 0.202671875
-1 0.183628125
0 0.1603578125
1 0.1372671875
2 0.1096421875
3 0.08030625
4 0.0519703125
5 0.0298875
6 0.0152859375
7 0.006865625
8 0.002778125
9 0.00108984375
10 0.0005515625
11 0.000287239583333333
12 0.000249330357142857
13 0.000136848958333333
14 0.000115848214285714
15 8.29769736842105e-05
16 7.11277173913043e-05
17 7.03804347826087e-05
18 7.67857142857143e-05
19 5.61383928571429e-05
20 4.96212121212121e-05
21 5.09072580645161e-05
22 4.40878378378378e-05
23 4.65992647058824e-05
24 4.05048076923077e-05
25 4.63235294117647e-05
26 4.75378787878788e-05
27 4.34121621621622e-05
28 5.04536290322581e-05
29 5.05040322580645e-05
30 3.75744047619048e-05
31 4.47767857142857e-05
32 4.40538194444444e-05
33 4.98046875e-05
34 4.17763157894737e-05
35 4.1858552631579e-05
36 4.34461805555556e-05
};
\addplot [ptoamp]
table {%
-6 0.2738015625
-5 0.2549609375
-4 0.2372171875
-3 0.2192015625
-2 0.19913125
-1 0.1794421875
0 0.1592015625
1 0.1389328125
2 0.11398125
3 0.082875
4 0.049428125
5 0.0272046875
6 0.0118296875
7 0.0041109375
8 0.0012375
9 0.0003134375
10 8.58552631578947e-05
11 1.87686011904762e-05
12 5.36941580756014e-06
13 1.20843000773395e-06
};
\addplot [ptcmd]
table {%
-6 0.2789734375
-5 0.2601125
-4 0.239846875
-3 0.220659375
-2 0.199109375
-1 0.1776140625
0 0.1526890625
1 0.1287890625
2 0.1030671875
3 0.0772
4 0.0535609375
5 0.0320625
6 0.017253125
7 0.0073109375
8 0.00224375
9 0.000697916666666667
10 0.00015796875
11 2.77686403508772e-05
12 5.86961610486891e-06
13 1.42861212397448e-06
14 4.97136493795737e-07
15 2.38029693195626e-07
16 1.45065453532634e-07
};
\addplot [ptawgn]
table {%
-6 0.239228710767672
-5 0.21322801835762
-4 0.186113817483389
-3 0.158368318809598
-2 0.130644488522829
-1 0.103759095953406
0 0.0786496035251426
1 0.0562819519765415
2 0.037506128358926
3 0.0228784075610853
4 0.0125008180407376
5 0.00595386714777866
6 0.00238829078093281
7 0.000772674815378445
8 0.000190907774075993
9 3.36272284196175e-05
10 3.87210821552204e-06
11 2.6130679535752e-07
12 9.00601035062875e-09
13 1.33293101753005e-10
14 6.81018912878077e-13
15 9.1239573626281e-16
};

\end{axis}
\end{tikzpicture}%

%% file: TikZ/plot16x16_comparison_resized.tikz
\begin{tikzpicture}[scale=1]
\pgfplotsset{
	every axis plot/.style={line width=1.4pt},
	every axis/.style={grid style={line width=0.7pt, dashed}, 
					 },
	axis background/.style={fill=white},
	every outer x axis line/.style={line width=1.4pt},
	every outer y axis line/.style={line width=1.4pt},
	every tick/.style={black,
					line width=0.7pt,
					},
	label style={font=\fontsize{8}{9}\selectfont},
	every y tick label/.style={font=\fontsize{8}{9}\color{black}},
	every x tick label/.style={font=\fontsize{8}{9}\color{black}},
	every extra x tick/.style={grid style={solid, violet, line width=2.8pt},
							x tick label style={/pgf/number format/.cd,precision=10}
							},
	legend style={
					line width=0.7pt,
					font=\fontsize{11.4}{12}\selectfont\color{black},
					legend cell align=left,
					align=left,
					fill=white,
					fill opacity=0.8, draw opacity=1, text opacity=1,
					nodes={scale=0.7, transform shape},
					},
	legend image code/.code={
		\draw[mark repeat=2,mark phase=2]
			plot coordinates {
				(0cm,0cm)
				(0.25cm,0cm)        
				(0.5cm,0cm)         
			};%
		},
	legend pos={south west},
	legend entries={{\color{\clsd} SD},
					{\color{\clmmse} MMSE},
					{\color{\clmosic} MOSIC},
					{\color{\clamp} AMP}, 
					{\color{\clsdr} SDR},
					{\color{\cldetnet} DetNet}, 
					{\color{\cloamp} OAMPNet},
					{\color{\clcmd} $\text{CMDNet}_{\text{bin}}$}, 
					{\color{\clawgn} AWGN},
					},
}
\begin{axis}[%
width=10.5cm,
height=8cm, 
scale=0.7,
scale only axis,
separate axis lines,
tick pos=left,
xmin=1,
xmax=18,
ymode=log,
xminorticks=true,
xtick distance = {2},
ytick distance = {10},
xlabel={$E_{\text{b}}/N_0$ [dB]}, 
ymin=0.0001,
ymax=1,
ylabel={BER},
xmajorgrids,
xminorgrids,
ymajorgrids,
]



\addplot [ptsd]
table {%
-6 0.274945175438597
-5 0.269935344827586
-4 0.246062992125984
-3 0.242277992277992
-2 0.205592105263158
-1 0.180655619596542
0 0.171232876712329
1 0.129639175257732
2 0.10603813559322
3 0.079675572519084
4 0.0563570784490532
5 0.0316135558927668
6 0.0166489078316463
7 0.00757965834746789
8 0.00318017605454638
9 0.00111997132873399
10 0.000361386567984036
11 9.57370218893117e-05
12 2.58476905191904e-05
13 6.83832279246917e-06
14 1.690831352017e-06
};
\addplot [ptmmse]
table {%
-6 0.2726625
-5 0.2547625
-4 0.23651875
-3 0.21499375
-2 0.19801875
-1 0.1788625
0 0.15893125
1 0.141825
2 0.1232375
3 0.10826875
4 0.09355
5 0.07768125
6 0.06491875
7 0.05449375
8 0.04443125
9 0.034975
10 0.028975
11 0.0235
12 0.018375
13 0.01395625
14 0.0113
15 0.0088875
16 0.007475
17 0.00513125
18 0.0040375
19 0.003446875
20 0.00269166666666667
21 0.002028125
22 0.0016265625
23 0.00130375
24 0.001046875
25 0.00081875
26 0.000729861111111111
27 0.000488942307692308
28 0.000423046875
29 0.000336842105263158
30 0.00026328125
31 0.000206048387096774
32 0.000168918918918919
33 0.000124754901960784
34 0.000112165178571429
35 8.04588607594937e-05
36 6.38392857142857e-05
};
\addplot [ptmosic]
table {%
-6 0.279854910714286
-5 0.269396551724138
-4 0.242732558139535
-3 0.238788167938931
-2 0.204918032786885
-1 0.184742647058824
0 0.173130193905817
1 0.138719512195122
2 0.131302521008403
3 0.102124183006536
4 0.0737898465171194
5 0.0621279761904762
6 0.039332913782253
7 0.0264494286923402
8 0.0151341227646206
9 0.00896182495344505
10 0.0047995699585317
11 0.00223437723437724
12 0.000945825530115378
13 0.000519008154656133
14 0.000155818383078872
15 6.67877324719148e-05
16 3.11624233781888e-05
17 1.39416067512171e-05
18 4.99450574609083e-06
19 2.09385990792967e-06
20 1.14110069877146e-06
};
\addplot [ptamp]
table {%
-6 0.273325
-5 0.254375
-4 0.23553125
-3 0.2171375
-2 0.19445625
-1 0.17295
0 0.15216875
1 0.1316125
2 0.1104875
3 0.08755625
4 0.06876875
5 0.0526875
6 0.040925
7 0.03261875
8 0.02434375
9 0.02236875
10 0.01999375
11 0.01944375
12 0.01989375
13 0.0205
14 0.02065625
15 0.021625
16 0.0219125
17 0.0216875
18 0.02090625
19 0.02233125
20 0.0209
21 0.02145625
22 0.02065
23 0.0224125
24 0.02008125
25 0.02275
26 0.02233125
27 0.0215
28 0.01928125
29 0.01996875
30 0.0202375
31 0.02253125
32 0.02230625
33 0.02120625
34 0.02000625
35 0.0217875
36 0.02166875
};
\addplot [ptsdr]
table {%
-6 0.277975
-5 0.2587
-4 0.2391125
-3 0.2189375
-2 0.19563125
-1 0.1739125
0 0.1520125
1 0.12805625
2 0.10443125
3 0.08205
4 0.05885625
5 0.04311875
6 0.027725
7 0.017325
8 0.00968125
9 0.005353125
10 0.00260208333333333
11 0.00109375
12 0.000455357142857143
13 0.000170439189189189
14 5.53650442477876e-05
};
\addplot [ptdetnet]
table {%
-6 0.28209375
-5 0.261925
-4 0.24168125
-3 0.22044375
-2 0.2017
-1 0.17893125
0 0.1539625
1 0.1306875
2 0.10805
3 0.08264375
4 0.06135625
5 0.04305625
6 0.02940625
7 0.01918125
8 0.01171875
9 0.0078
10 0.0052375
11 0.003709375
12 0.003146875
13 0.00235833333333333
14 0.001878125
15 0.0016359375
16 0.0013
17 0.00109583333333333
18 0.00119479166666667
19 0.00102946428571429
20 0.000921428571428571
21 0.000938392857142857
22 0.00081015625
23 0.0008734375
24 0.000741666666666667
25 0.000728472222222222
26 0.00068625
27 0.000757638888888889
28 0.000704166666666667
29 0.00075
30 0.000736111111111111
31 0.000765277777777778
32 0.000669375
33 0.000724305555555556
34 0.000703472222222222
35 0.00067625
36 0.000727777777777778
};
\addplot [ptoamp] 
table {%
-6 0.27683125
-5 0.2568625
-4 0.23725625
-3 0.2206125
-2 0.20229375
-1 0.18210625
0 0.161325
1 0.14050625
2 0.11558125
3 0.0919375
4 0.06683125
5 0.04626875
6 0.03238125
7 0.02135
8 0.01359375
9 0.00935
10 0.00579375
11 0.00393125
12 0.0024125
13 0.001696875
14 0.00119270833333333
15 0.00082109375
16 0.000610227272727273
17 0.000452678571428571
18 0.000292613636363636
19 0.000237962962962963
20 0.000180535714285714
21 0.000124509803921569
22 0.0001046875
23 8.42927631578947e-05
24 6.20668316831683e-05
25 4.73958333333333e-05
26 3.75374251497006e-05
27 3.1095297029703e-05
28 2.34550561797753e-05
29 1.92307692307692e-05
30 1.41751126126126e-05
31 1.16822429906542e-05
32 9.87776025236593e-06
33 7.96178343949045e-06
34 6.13959764474975e-06
35 4.79294478527607e-06
36 3.81310827250608e-06
};
\addplot [ptcmd]
table {%
-6 0.29076875
-5 0.2655375
-4 0.24115
-3 0.22045625
-2 0.1968
-1 0.17360625
0 0.15236875
1 0.129025
2 0.10738125
3 0.08570625
4 0.0654
5 0.0470375
6 0.03263125
7 0.0207375
8 0.01298125
9 0.00835625
10 0.004759375
11 0.00315416666666667
12 0.00221666666666667
13 0.001375
14 0.00108541666666667
15 0.00078828125
16 0.000686875
17 0.000585227272727273
18 0.0005125
19 0.000424583333333333
20 0.00041640625
21 0.000372426470588235
22 0.0003984375
23 0.000334210526315789
24 0.0003265625
25 0.000290909090909091
26 0.000300892857142857
27 0.00030625
28 0.00031625
29 0.000290909090909091
30 0.000304761904761905
31 0.000279891304347826
32 0.000308630952380952
33 0.00026796875
34 0.000292045454545455
35 0.000292613636363636
36 0.000263020833333333
};
\addplot [ptawgn]
table {%
-6 0.239228710767672
-5 0.21322801835762
-4 0.186113817483389
-3 0.158368318809598
-2 0.130644488522829
-1 0.103759095953406
0 0.0786496035251426
1 0.0562819519765415
2 0.037506128358926
3 0.0228784075610853
4 0.0125008180407376
5 0.00595386714777866
6 0.00238829078093281
7 0.000772674815378445
8 0.000190907774075993
9 3.36272284196175e-05
10 3.87210821552204e-06
11 2.6130679535752e-07
12 9.00601035062875e-09
13 1.33293101753005e-10
14 6.81018912878077e-13
15 9.1239573626281e-16
16 2.26739584445442e-19
17 6.75896977065469e-24
18 1.39601431090675e-29
19 1.00107397357086e-36
20 1.04424379188127e-45
21 5.29969722887035e-57
22 3.29608811928447e-71
23 4.42763849843478e-89
24 1.44441906577261e-111
25 7.3069691846479e-140
26 1.79516322712009e-175
27 2.73597996566885e-220
28 1.06846073755636e-276
29 0
30 0
31 0
32 0
33 0
34 0
35 0
36 0
};

\end{axis}
\end{tikzpicture}%

%% file: TikZ/plot64x64_parameter_resized.tikz
\begin{tikzpicture}[scale=1]
\pgfplotsset{
	every axis plot/.style={line width=1.4pt},
	every axis/.style={grid style={line width=0.7pt, dashed}, 
					 },
	axis background/.style={fill=white},
	every outer x axis line/.style={line width=1.4pt},
	every outer y axis line/.style={line width=1.4pt},
	every tick/.style={black,
					line width=0.7pt,
					},
	title style={font=\fontsize{8}{9}\selectfont},
	label style={font=\fontsize{8}{9}\selectfont},
	every y tick label/.style={font=\fontsize{8}{9}\color{black}},
	every x tick label/.style={font=\fontsize{8}{9}\color{black}},
	every extra x tick/.style={grid style={solid, violet, line width=2.8pt},
							x tick label style={/pgf/number format/.cd,precision=10}
							},
	legend style={
					line width=0.7pt,
					font=\fontsize{11.4}{12}\selectfont\color{black},
					legend cell align=left,
					align=left,
					fill=white,
					nodes={scale=0.7, transform shape},
					},
	legend image code/.code={
		\draw[mark repeat=2,mark phase=2]
			plot coordinates {
				(0cm,0cm)
				(0.25cm,0cm)        
				(0.5cm,0cm)         
			};%
		},
	group style={columns= 2,
      				horizontal sep=1.65cm,
  				},
}
\begin{groupplot}[group style={group size=2 by 2}]


\nextgroupplot[%
width=4.75cm, 
height=5cm, 
scale=0.7,
scale only axis,
separate axis lines,
xmajorgrids,
xminorgrids,
ymajorgrids,
yminorgrids,
label style={font=\fontsize{8}{9}\selectfont},
title style={font=\fontsize{8}{9}\selectfont},
title=(a) Step size $\delta^{(j)}$,
tick pos=left,
xlabel={$j$},
xmin=0, xmax=64,
ylabel={$\delta$},
ymin=0, ymax=2,
]
\addplot [blue, dashed]
table {%
0 1
1 1
2 1
3 1
4 1
5 1
6 1
7 1
8 1
9 1
10 1
11 1
12 1
13 1
14 1
15 1
16 1
17 1
18 1
19 1
20 1
21 1
22 1
23 1
24 1
25 1
26 1
27 1
28 1
29 1
30 1
31 1
32 1
33 1
34 1
35 1
36 1
37 1
38 1
39 1
40 1
41 1
42 1
43 1
44 1
45 1
46 1
47 1
48 1
49 1
50 1
51 1
52 1
53 1
54 1
55 1
56 1
57 1
58 1
59 1
60 1
61 1
62 1
63 1
};
\addplot [red]
table {%
0 0.926926661937276
1 1.14300635716731
2 1.36557457448963
3 1.35999866744671
4 1.36298968064167
5 1.37194787195499
6 1.33256528133945
7 1.33021536354176
8 1.31768125205468
9 1.34290279598178
10 1.32776809620997
11 1.25582534134201
12 1.24487343776914
13 1.25189492660024
14 1.26533419378712
15 1.27754953854564
16 1.22258553409777
17 1.20712098289812
18 1.11002364612034
19 1.14299731389665
20 1.07425344379286
21 0.981487569252474
22 0.962806170679641
23 0.944493933155169
24 0.913694702221414
25 0.907480617456451
26 0.81638560190082
27 0.843125360802956
28 0.738424225112491
29 0.802514225846077
30 0.642490932667019
31 0.738420507122938
32 0.618996650638142
33 0.743442971704476
34 0.685752516074649
35 0.798839614720423
36 0.764894969081882
37 0.83862861829322
38 0.778747446058338
39 0.861423675990279
40 0.9305223799628
41 1.05050184389656
42 1.13936899964366
43 0.852097074201487
44 1.04749531668162
45 1.03403144385366
46 0.834635068613189
47 1.01705544703823
48 0.923606273920179
49 0.989508367491394
50 1.07926748914684
51 0.744772275343414
52 0.460911620368776
53 0.784153745494417
54 1.55718180428467
55 1.80652391560901
56 0.265291383506766
57 0.175565978262876
58 1.11108402838621
59 0.280357612507997
60 0.398287509853236
61 0.204362915160497
62 0.504377093985211
63 0.637708066162835
};


\nextgroupplot[%
width=4.75cm, 
height=5cm, 
scale=0.7,
scale only axis,
separate axis lines,
xmajorgrids,
xminorgrids,
ymajorgrids,
yminorgrids,
label style={font=\fontsize{8}{9}\selectfont},
title style={font=\fontsize{8}{9}\selectfont},
title=(b) Softmax temperature $\tau^{(j)}$,
legend entries={
				{\color{blue} $N_{\text{epoch}}$=$0$}, 
				{\color{red} $N_{\text{epoch}}$=$10^5$}, 
				},
tick pos=left,
xlabel={$j$},
xmin=0, xmax=64,
ylabel={$\tau$},
ymin=0, ymax=1.2,
]
\addplot [blue, dashed]
table {%
0 1
1 0.9859375
2 0.971875
3 0.9578125
4 0.94375
5 0.9296875
6 0.915625
7 0.9015625
8 0.8875
9 0.8734375
10 0.859375
11 0.8453125
12 0.83125
13 0.8171875
14 0.803125
15 0.7890625
16 0.775
17 0.7609375
18 0.746875
19 0.7328125
20 0.71875
21 0.7046875
22 0.690625
23 0.6765625
24 0.6625
25 0.6484375
26 0.634375
27 0.6203125
28 0.60625
29 0.5921875
30 0.578125
31 0.5640625
32 0.55
33 0.5359375
34 0.521875
35 0.5078125
36 0.49375
37 0.4796875
38 0.465625
39 0.4515625
40 0.4375
41 0.4234375
42 0.409375
43 0.3953125
44 0.38125
45 0.3671875
46 0.353125
47 0.3390625
48 0.325
49 0.3109375
50 0.296875
51 0.2828125
52 0.26875
53 0.2546875
54 0.240625
55 0.2265625
56 0.2125
57 0.1984375
58 0.184375
59 0.1703125
60 0.15625
61 0.1421875
62 0.128125
63 0.1140625
64 0.1
};
\addplot [red]
table {%
0 1.07883400172135
1 0.724866367034514
2 0.640252770916701
3 0.706458322498883
4 0.69232122471398
5 0.69106142790923
6 0.711159828725259
7 0.687341955450878
8 0.730475791276257
9 0.711466991640422
10 0.756730438056114
11 0.736686651413441
12 0.749602987153104
13 0.728205237916739
14 0.722082266074449
15 0.719008251810797
16 0.74125589265408
17 0.728989041139578
18 0.741853245194698
19 0.749825052108369
20 0.754762686730571
21 0.726322582834746
22 0.719089117363251
23 0.673413844247968
24 0.666747536681074
25 0.64157794138721
26 0.634389328589137
27 0.614861989499299
28 0.605842004881037
29 0.588156449883625
30 0.574016420847029
31 0.557696714120026
32 0.563127631324778
33 0.526782608899503
34 0.55650380820058
35 0.479414027521416
36 0.579654452490004
37 0.457786076925434
38 0.546397851386705
39 0.433874751774896
40 0.464447293515029
41 0.399121082638323
42 0.402356890632018
43 0.386679109692632
44 0.377126492683442
45 0.381597572718742
46 0.376002265963322
47 0.312589730684557
48 0.299545594477144
49 0.371458302449728
50 0.3158099628035
51 0.264261048006097
52 0.255009841297158
53 0.650000121312888
54 0.594208579774995
55 0.333596674217768
56 0.177361361212279
57 0.313837526919387
58 0.315922177525531
59 0.302649910957671
60 0.327608246613228
61 0.312206447039398
62 0.288412984376891
63 0.221827349489266
64 0.0658408341229779
};

\end{groupplot}
\end{tikzpicture}%

%% file: TikZ/plot64x64_CMDcomparison_resized.tikz
\begin{tikzpicture}[scale=1]
\pgfplotsset{
	every axis plot/.style={line width=1.4pt},
	every axis/.style={grid style={line width=0.7pt, dashed}, 
					 },
	axis background/.style={fill=white},
	every outer x axis line/.style={line width=1.4pt},
	every outer y axis line/.style={line width=1.4pt},
	every tick/.style={black,
					line width=0.7pt,
					},
	label style={font=\fontsize{8}{9}\selectfont},
	every y tick label/.style={font=\fontsize{8}{9}\color{black}},
	every x tick label/.style={font=\fontsize{8}{9}\color{black}},
	every extra x tick/.style={grid style={solid, violet, line width=2.8pt},
							x tick label style={/pgf/number format/.cd,precision=10}
							},
	legend style={
					line width=0.7pt,
					font=\fontsize{11.4}{12}\selectfont\color{black},
					legend cell align=left,
					align=left,
					fill=white,
					fill opacity=0.8, draw opacity=1, text opacity=1,
					nodes={scale=0.7, transform shape},
					},
	legend image code/.code={
		\draw[mark repeat=2,mark phase=2]
			plot coordinates {
				(0cm,0cm)
				(0.25cm,0cm)        
				(0.5cm,0cm)         
			};%
		},
	legend entries={{\color{\clsd} SD},
					{\color{\clmmse} MMSE},
					{\color{\cloamp} OAMPNet},
					{\color{\clcmd} $\text{CMDNet}_{\text{bin}}$ $\bm{\theta}_{10^5}$},
					{\color{purple} CMDNet $\bm{\theta}_{10^5}$},
					{\color{\clcmd} $\text{CMDNet}_{\text{bin}}$ $\bm{\theta}_{10^5\text{,splin}}$}, 
					{\color{orange} $\text{CMDNet}_{\text{bin}}$ $\bm{\theta}_{10^5}$\\ \color{orange} $N_{\text{L}}$=$16$}, 
					{\color{darkgreen} $\text{CMD}_{\text{bin}}$ $\bm{\theta}_{0}$},
					{\color{darkgreen} $\text{CMD}_{\text{bin}}$ $\bm{\theta}_{0\text{,splin}}$},
					},
}
\begin{axis}[%
width=10.5cm,
height=8cm, 
scale=0.7,
scale only axis,
separate axis lines,
tick pos=left,
xmin=6,
xmax=14,
ymode=log,
xminorticks=true,
xtick distance = {2},
ytick distance = {10},
xlabel={$E_{\text{b}}/N_0$ [dB]}, 
ymin=0.000001,
ymax=1,
ylabel={BER},
xmajorgrids,
xminorgrids,
ymajorgrids,
]



\addplot [ptsd]
table {%
-4 0.262451171875
-2 0.208413461538462
0 0.164713541666667
2 0.119514627659574
4 0.0446013931888545
6 0.00649971461187215
8 0.000431358421357456
10 1.68004425778524e-05
12 2.6041449654586e-07
};
\addplot [ptmmse]
table {%
-6 0.2730390625
-5 0.255496875
-4 0.23509375
-3 0.21550625
-2 0.1973953125
-1 0.1774640625
0 0.158225
1 0.139621875
2 0.123746875
3 0.1057859375
4 0.09038125
5 0.075959375
6 0.06318125
7 0.051059375
8 0.0408828125
9 0.0320109375
10 0.024646875
11 0.018515625
12 0.013928125
13 0.0099703125
14 0.007421875
15 0.0051234375
16 0.0036828125
17 0.0025671875
18 0.0018171875
19 0.001225
20 0.00082421875
21 0.00058125
22 0.0004
23 0.000285677083333333
24 0.0002076171875
25 0.000142897727272727
26 0.000112083333333333
27 8.33059210526316e-05
28 5.81597222222222e-05
29 4.90234375e-05
30 3.29752604166667e-05
31 2.65364583333333e-05
32 2.1e-05
33 1.70006793478261e-05
34 1.2376968503937e-05
35 1.00806451612903e-05
36 8.05412371134021e-06
};
\addplot [ptoamp]
table {%
-6 0.2738015625
-5 0.2549609375
-4 0.2372171875
-3 0.2192015625
-2 0.19913125
-1 0.1794421875
0 0.1592015625
1 0.1389328125
2 0.11398125
3 0.082875
4 0.049428125
5 0.0272046875
6 0.0118296875
7 0.0041109375
8 0.0012375
9 0.0003134375
10 8.58552631578947e-05
11 1.87686011904762e-05
12 5.36941580756014e-06
13 1.20843000773395e-06
};
\addplot [ptcmd]
table {%
-6 0.2789734375
-5 0.2601125
-4 0.239846875
-3 0.220659375
-2 0.199109375
-1 0.1776140625
0 0.1526890625
1 0.1287890625
2 0.1030671875
3 0.0772
4 0.0535609375
5 0.0320625
6 0.017253125
7 0.0073109375
8 0.00224375
9 0.000697916666666667
10 0.00015796875
11 2.77686403508772e-05
12 5.86961610486891e-06
13 1.42861212397448e-06
14 4.97136493795737e-07
15 2.38029693195626e-07
16 1.45065453532634e-07
};
\addplot [purple, mark=diamond*, mark size=3*0.7, mark options={solid}]
table {%
-6 0.6245546875
-5 0.618240625
-4 0.572690625
-3 0.5044828125
-2 0.371109375
-1 0.2418015625
0 0.1833578125
1 0.1418546875
2 0.1100578125
3 0.0827171875
4 0.05551875
5 0.03406875
6 0.01771875
7 0.0075375
8 0.0026390625
9 0.0008765625
10 0.000244642857142857
11 6.56901041666667e-05
12 1.52002427184466e-05
13 4.60914454277286e-06
14 1.48244781783681e-06
15 5.03707285622179e-07
};
\addplot [red, mark=triangle*, mark size=3*0.7, mark options={solid}, dashed]
table {%
-6 0.281903125
-5 0.263821875
-4 0.244475
-3 0.2240296875
-2 0.20246875
-1 0.1808640625
0 0.156540625
1 0.1309546875
2 0.1032015625
3 0.0772046875
4 0.05004375
5 0.02885
6 0.0138671875
7 0.00525625
8 0.0015109375
9 0.00035875
10 9.86328125e-05
11 3.80208333333333e-05
12 2.63020833333333e-05
13 2.55796370967742e-05
14 2.48511904761905e-05
15 2.32536764705882e-05
16 2.50248015873016e-05
17 2.33442164179105e-05
18 2.84090909090909e-05
19 2.48263888888889e-05
20 2.66154661016949e-05
21 2.69396551724138e-05
22 2.85511363636364e-05
23 2.61197916666667e-05
24 2.89488636363636e-05
25 2.96580188679245e-05
26 2.77138157894737e-05
27 2.82366071428571e-05
28 2.72090517241379e-05
29 2.9369212962963e-05
30 2.56915983606557e-05
31 2.91956018518519e-05
32 3.01682692307692e-05
33 3.05588942307692e-05
34 2.5922131147541e-05
35 2.54284274193548e-05
36 2.82087053571429e-05
};
\addplot [orange, dashed, mark=square*, mark options={solid}, mark size=2*0.7]
table {%
-6 0.3135828125
-5 0.2676578125
-4 0.2418953125
-3 0.2198265625
-2 0.1992
-1 0.1760859375
0 0.1529890625
1 0.1295765625
2 0.1034796875
3 0.07855
4 0.05289375
5 0.03211875
6 0.0167671875
7 0.006990625
8 0.0025640625
9 0.000783333333333333
10 0.0002162109375
11 7.16619318181818e-05
12 2.54284274193548e-05
13 8.56386612021858e-06
14 4.175e-06
15 2.27769679300292e-06
16 1.43876611418048e-06
17 9.89912974683544e-07
18 7.78525161933234e-07
19 5.89845224613061e-07
};
\addplot [darkgreen, dashed, mark=x, mark size=3*0.7, mark options={solid}]
table {%
-6 0.2994328125
-5 0.2896703125
-4 0.27251875
-3 0.2509625
-2 0.2303859375
-1 0.2124953125
0 0.197765625
1 0.1866640625
2 0.1739828125
3 0.160753125
4 0.14526875
5 0.1265328125
6 0.1028546875
7 0.074625
8 0.0465765625
9 0.0262140625
10 0.0137296875
11 0.00799375
12 0.0041578125
13 0.0018546875
14 0.000668229166666667
15 0.00015984375
16 3.25846354166667e-05
17 7.10227272727273e-06
18 2.54413555194805e-06
19 2.04248366013072e-06
20 2.14950478142077e-06
21 1.61399330587024e-06
22 1.5517453294002e-06
23 1.52309650824442e-06
24 1.67787447257384e-06
25 1.3695652173913e-06
26 1.28600823045267e-06
27 1.52154126213592e-06
28 1.52764306498545e-06
29 1.35828675838349e-06
30 1.56875e-06
31 1.46028037383178e-06
32 1.15638766519824e-06
33 1.34675171526587e-06
34 1.14659090909091e-06
};
\addplot [darkgreen, mark=x, mark size=3*0.7, mark options={solid}]
table {%
-6 0.2820859375
-5 0.2631578125
-4 0.241653125
-3 0.2218125
-2 0.201734375
-1 0.178728125
0 0.15485
1 0.1299109375
2 0.1023
3 0.0759921875
4 0.050721875
5 0.0293390625
6 0.01475
7 0.00563125
8 0.0016609375
9 0.000465234375
10 0.000121514423076923
11 4.29476351351351e-05
12 2.31617647058824e-05
13 1.5306432038835e-05
14 1.15740740740741e-05
15 7.89141414141414e-06
16 7.68995098039216e-06
17 6.1761811023622e-06
18 5.53224381625442e-06
19 4.82211538461538e-06
20 4.76848323170732e-06
21 4.34461805555556e-06
22 4.24712059620596e-06
23 4.30440771349862e-06
24 4.09440445026178e-06
25 3.9140625e-06
26 3.91796875e-06
27 3.70260663507109e-06
28 3.59088302752294e-06
29 3.71140142517815e-06
30 3.53860294117647e-06
31 3.45685840707965e-06
32 3.75600961538462e-06
33 3.58371559633028e-06
34 3.49469866071429e-06
35 3.60854503464203e-06
36 3.64219114219114e-06
};
\end{axis}
\end{tikzpicture}%

%% file: TikZ/plot64x64_QAM16_resized.tikz
\begin{tikzpicture}[scale=1]
\pgfplotsset{
	every axis plot/.style={line width=1.4pt},
	every axis/.style={grid style={line width=0.7pt, dashed}, 
					 },
	axis background/.style={fill=white},
	every outer x axis line/.style={line width=1.4pt},
	every outer y axis line/.style={line width=1.4pt},
	every tick/.style={black,
					line width=0.7pt,
					},
	label style={font=\fontsize{8}{9}\selectfont},
	every y tick label/.style={font=\fontsize{8}{9}\color{black}},
	every x tick label/.style={font=\fontsize{8}{9}\color{black}},
	every extra x tick/.style={grid style={solid, violet, line width=2.8pt},
							x tick label style={/pgf/number format/.cd,precision=10}
							},
	legend style={
					line width=0.7pt,
					font=\fontsize{11.4}{12}\selectfont\color{black},
					legend cell align=left,
					align=left,
					fill=white,
					fill opacity=0.8, draw opacity=1, text opacity=1,
					nodes={scale=0.7, transform shape},
					},
	legend image code/.code={
		\draw[mark repeat=2,mark phase=2]
			plot coordinates {
				(0cm,0cm)
				(0.25cm,0cm)        
				(0.5cm,0cm)         
			};%
		},
	legend pos={north east},
	legend entries={{\color{\clsd} SD},
					{\color{\clmmse} MMSE},
					{\color{\clmosic} MOSIC},
					{\color{\clamp} AMP}, 
					{\color{\clsdr} SDR},
					{\color{\cldetnet} DetNet}, 
					{\color{\cloamp} OAMPNet},
					{\color{\clcmd} CMDNet}, 
					{\color{\clawgn} AWGN},
					},
}
\begin{axis}[%
width=10.5cm,
height=8cm, 
scale = 0.7,
scale only axis,
separate axis lines,
tick pos=left,
xmin=6,
xmax=30,
ymode=log,
xminorticks=true,
xtick distance = {2},
ytick distance = {10},
xlabel={$E_{\text{b}}/N_0$ [dB]}, 
ymin=0.0001,
ymax=1,
ylabel={BER},
xmajorgrids,
xminorgrids,
ymajorgrids,
]



\addplot [ptsd]
table {%
4 0.1671875
6 0.165540540540541
8 0.115648674242424
9 0.062962962962963
10 0.028003589527027
11 0.00406364468864469
12 0.000526017271662763
13 9.32835820895522e-05
};
\addplot [ptmmse]
table {%
-6 0.33450390625
-5 0.3183859375
-4 0.3028265625
-3 0.286959375
-2 0.27115859375
-1 0.2537359375
0 0.23851015625
1 0.22325546875
2 0.207525
3 0.19287734375
4 0.179778125
5 0.164734375
6 0.1527921875
7 0.13922578125
8 0.1273546875
9 0.1144859375
10 0.102453125
11 0.0911984375
12 0.08089609375
13 0.07038046875
14 0.06092890625
15 0.05231484375
16 0.04348203125
17 0.03659921875
18 0.03027890625
19 0.0245921875
20 0.02015390625
21 0.01584921875
22 0.01293515625
23 0.01077265625
24 0.00826171875
25 0.00663125
26 0.0052515625
27 0.00419921875
28 0.0032515625
29 0.0025515625
30 0.0021671875
31 0.00166171875
32 0.0013375
33 0.00099296875
34 0.0008015625
35 0.0006890625
36 0.00052109375
};
\addplot [ptmosic]
table {%
-6 0.329427083333333
-4 0.303485576923077
-2 0.279017857142857
0 0.250732421875
2 0.215371621621622
4 0.186941964285714
6 0.163411458333333
8 0.161033163265306
10 0.112165178571429
12 0.0747767857142857
14 0.0359869909502263
16 0.0161352040816326
18 0.00376199040767386
20 0.000615321883351673
22 5.97717616462797e-05
24 6.9550504399973e-06
};
\addplot [ptamp]
table {%
-6 0.33392734375
-5 0.31721484375
-4 0.3014
-3 0.28422734375
-2 0.26797109375
-1 0.2503640625
0 0.23247578125
1 0.216484375
2 0.1990984375
3 0.18383984375
4 0.16697109375
5 0.1504703125
6 0.13578203125
7 0.11834140625
8 0.09955390625
9 0.07646484375
10 0.05098125
11 0.03074765625
12 0.0200296875
13 0.01675859375
14 0.014646875
15 0.01441015625
16 0.01793515625
17 0.01889609375
18 0.022684375
19 0.02513125
20 0.0280875
21 0.03374453125
22 0.038471875
23 0.0447203125
24 0.044125
25 0.044525
26 0.0445265625
27 0.0475375
28 0.044425
29 0.04771796875
30 0.0466484375
31 0.04517421875
32 0.04285390625
33 0.04614453125
34 0.04690703125
35 0.04563359375
36 0.045009375
};
\addplot [ptsdr]
table {%
-6 0.348359375
-5 0.334921875
-4 0.31421875
-3 0.3015625
-2 0.28171875
-1 0.255625
0 0.24234375
1 0.23234375
2 0.19640625
3 0.188046875
4 0.1703125
5 0.15734375
6 0.125078125
7 0.1121875
8 0.097421875
9 0.0684765625
10 0.0544921875
11 0.0398697916666667
12 0.0310677083333333
13 0.018734375
14 0.0120870535714286
15 0.00609933035714286
16 0.00376860119047619
17 0.001565625
18 0.000727719907407407
19 0.000390625 
20 0.000113224637681159
21 7.24943693693694e-05
22 4.64154411764706e-05
};
\addplot [ptdetnet]
table {%
-6 0.34416640625
-5 0.32867890625
-4 0.31322421875
-3 0.29748671875
-2 0.27856171875
-1 0.26142109375
0 0.24378984375
1 0.224753125
2 0.206296875
3 0.18825859375
4 0.1704296875
5 0.1532390625
6 0.13565
7 0.1177796875
8 0.0988140625
9 0.0795734375
10 0.05889765625
11 0.04046640625
12 0.0272046875
13 0.01763046875
14 0.011503125
15 0.008325
16 0.00648984375
17 0.00555234375
18 0.00418359375
19 0.00378671875
20 0.00318671875
21 0.00301484375
22 0.00306171875
23 0.0029171875
24 0.0028515625
25 0.0026453125
26 0.00249296875
27 0.0026390625
28 0.00229609375
29 0.00244375
30 0.00224375
31 0.00254609375
32 0.002409375
33 0.0021890625
34 0.00231171875
35 0.002046875
36 0.0024125
};
\addplot [ptoamp]
table {%
-6 0.3775078125
-5 0.3706515625
-4 0.36229921875
-3 0.3548890625
-2 0.34847265625
-1 0.34093203125
0 0.32909765625
1 0.30695859375
2 0.2725015625
3 0.23592109375
4 0.20199609375
5 0.17152421875
6 0.14522421875
7 0.11997109375
8 0.097125
9 0.0747828125
10 0.0521984375
11 0.031796875
12 0.0173328125
13 0.00856796875
14 0.00443984375
15 0.00241796875
16 0.0012828125
17 0.000703125
18 0.000413541666666667
19 0.000272916666666667
20 0.000147265625
21 7.39346590909091e-05
22 4.65073529411765e-05
23 3.63991477272727e-05
24 2.54150390625e-05
25 2.47472426470588e-05
26 1.578125e-05
27 1.015625e-05
28 6.66666666666667e-06
29 3.60659246575342e-06
30 3.87917698019802e-06
31 2.45150862068966e-06
32 2.65990802675585e-06
33 2.15916895604396e-06
34 1.25222507911392e-06
};
\addplot [ptcmd]
table {%
-6 0.55440078125
-5 0.55376875
-4 0.542409375
-3 0.5146390625
-2 0.49888359375
-1 0.48144296875
0 0.46631484375
1 0.452053125
2 0.43054296875
3 0.4032859375
4 0.36733046875
5 0.33243828125
6 0.301046875
7 0.2361140625
8 0.14333984375
9 0.09058125
10 0.06371953125
11 0.04427890625
12 0.027690625
13 0.01659296875
14 0.0094390625
15 0.00527109375
16 0.00365078125
17 0.00222421875
18 0.00183203125
19 0.00149140625
20 0.001184375
21 0.001159375
22 0.00101015625
23 0.00119296875
24 0.0010625
25 0.0008265625
26 0.00091796875
27 0.0007828125
28 0.00078515625
29 0.000804296875
30 0.0007859375
31 0.000790625
32 0.000780078125
33 0.000810546875
34 0.0007984375
35 0.000740625
36 0.00087578125
};
\addplot [ptawgn]
table {%
-6 0.245232895351461
-5 0.230618622666212
-4 0.214694657268862
-3 0.197474682988887
-2 0.179030058591435
-1 0.159509494456034
0 0.139160013571012
1 0.118345838849974
2 0.0975593523767089
3 0.0774154342655597
4 0.0586184574192509
5 0.0418923031812634
6 0.0278713063196607
7 0.0169667338966386
8 0.00924721373751743
9 0.00439033608734222
10 0.00175415061789272
11 0.000564706106481743
12 0.000138658688812619
13 2.42337854663159e-05
14 2.76320800168778e-06
15 1.84185551109448e-07
16 6.25020082774196e-09
17 9.07162538956511e-11
18 4.52230900481845e-13
19 5.87418096013832e-16
20 1.40403651907608e-19
21 3.98512492728761e-24
22 7.7384040456981e-30
23 5.13440790628314e-37
24 4.85685047723102e-46
25 2.17939904671361e-57
26 1.16084518268593e-71
27 1.28295037394233e-89
28 3.27380613621827e-112
29 1.21562020252509e-140
30 2.02344960435296e-176
31 1.88908890552234e-221
32 3.98049722149799e-278
33 0
34 0
35 0
36 0
};

\end{axis}
\end{tikzpicture}%

%% file: TikZ/plot128x64_comparison_OneRing20_120_resized.tikz
\begin{tikzpicture}[scale=1]
\pgfplotsset{
	every axis plot/.style={line width=1.4pt},
	every axis/.style={grid style={line width=0.7pt, dashed}, 
					 },
	axis background/.style={fill=white},
	every outer x axis line/.style={line width=1.4pt},
	every outer y axis line/.style={line width=1.4pt},
	every tick/.style={black,
					line width=0.7pt,
					},
	label style={font=\fontsize{8}{9}\selectfont},
	every y tick label/.style={font=\fontsize{8}{9}\color{black}},
	every x tick label/.style={font=\fontsize{8}{9}\color{black}},
	every extra x tick/.style={grid style={solid, violet, line width=2.8pt},
							x tick label style={/pgf/number format/.cd,precision=10}
							},
	legend style={
					line width=0.7pt,
					font=\fontsize{11.4}{12}\selectfont\color{black},
					legend cell align=left,
					align=left,
					fill=white,
					fill opacity=0.8, draw opacity=1, text opacity=1,
					nodes={scale=0.7, transform shape},
					},
	legend image code/.code={
		\draw[mark repeat=2,mark phase=2]
			plot coordinates {
				(0cm,0cm)
				(0.25cm,0cm)        
				(0.5cm,0cm)         
			};%
		},
	legend pos={north east},
	legend entries={{\color{\clsd} SD},
					{\color{\clmmse} MMSE},
					{\color{\clmosic} MOSIC},
					{\color{\clamp} AMP}, 
					{\color{\clsdr} SDR},
					{\color{\cldetnet} DetNet}, 
					{\color{\cloamp} OAMPNet},
					{\color{\clcmd} $\text{CMDNet}_{\text{bin}}$}, 
					{\color{\clawgn} AWGN},
					},
}
\begin{axis}[%
width=10.5cm,
height=8cm, 
scale=0.7,
scale only axis,
separate axis lines,
tick pos=left,
xmin=4,
xmax=16,
ymode=log,
xminorticks=true,
xtick distance = {2},
ytick distance = {10},
xlabel={$E_{\text{b}}/N_0$ [dB]}, 
ymin=0.000001,
ymax=0.1,
ylabel={BER},
xmajorgrids,
xminorgrids,
ymajorgrids,
]



\addplot [ptsd]
table {%
-6 0.251736111111111
-4 0.224553571428571
-2 0.170346467391304
0 0.129132231404959
2 0.0714183789954338
4 0.0237101669195751
6 0.00542911744266852
8 0.000610828772478496
10 3.83515579936927e-05
12 1.24311453774911e-06
};
\addplot [ptmmse]
table {%
-6 0.2591609375
-5 0.2378078125
-4 0.21380625
-3 0.19058125
-2 0.166984375
-1 0.144125
0 0.12110625
1 0.0989703125
2 0.0783515625
3 0.0597125
4 0.043840625
5 0.0300140625
6 0.0199015625
7 0.0123
8 0.0068734375
9 0.0034140625
10 0.0016609375
11 0.000736458333333333
12 0.000271875
13 9.57720588235294e-05
14 2.94560185185185e-05
15 9.71467391304348e-06
16 2.6986183074266e-06
17 7.05622467357046e-07
};
\addplot [ptmosic]
table {%
-6 0.277960526315789
-5 0.238399621212121
-4 0.217230902777778
-3 0.197784810126582
-2 0.181686046511628
-1 0.148584905660377
0 0.114356884057971
1 0.0996218152866242
2 0.0784375
3 0.0542534722222222
4 0.0335300429184549
5 0.021551724137931
6 0.00958001226241569
7 0.00434269038354643
8 0.0017440562562786
9 0.000512068655054998
10 0.000126299367896924
11 2.14582549391745e-05
12 3.14433830947061e-06
13 3.14833969157944e-07
};
\addplot [ptamp]
table {%
-6 0.2586421875
-5 0.2356453125
-4 0.213396875
-3 0.188840625
-2 0.1645390625
-1 0.1407390625
0 0.116203125
1 0.0917203125
2 0.069796875
3 0.048846875
4 0.0313921875
5 0.018884375
6 0.01145625
7 0.006928125
8 0.0044296875
9 0.003371875
10 0.0030140625
11 0.002465625
12 0.0020203125
13 0.0018484375
14 0.001803125
15 0.00180625
16 0.0016375
17 0.001846875
18 0.001821875
19 0.0017640625
20 0.001996875
21 0.001978125
22 0.0019796875
23 0.0021125
24 0.002315625
25 0.002315625
26 0.002253125
27 0.0025328125
28 0.00235625
29 0.0020125
30 0.002259375
31 0.0023546875
32 0.002215625
33 0.00213125
34 0.002475
35 0.002234375
36 0.0022515625
};
\addplot [ptsdr]
table {%
-6 0.28171875
-5 0.24234375
-4 0.225
-3 0.1934375
-2 0.17296875
-1 0.140546875
0 0.1153125
1 0.08171875
2 0.06375
3 0.0430859375
4 0.0259821428571429
5 0.0147585227272727
6 0.00694972826086957
7 0.00290219907407407
8 0.000910404624277457
9 0.000303988326848249
10 6.74412393162393e-05
11 1.43876611418048e-05
12 2.22356624448556e-06
13 3.41604722343682e-07
};
\addplot [ptdetnet]
table {%
-6 0.263509375
-5 0.2428078125
-4 0.2198453125
-3 0.1941875
-2 0.1704046875
-1 0.1445453125
0 0.1175765625
1 0.0919890625
2 0.06610625
3 0.04458125
4 0.026340625
5 0.014290625
6 0.0063078125
7 0.002546875
8 0.00100703125
9 0.00031875
10 0.000105208333333333
11 3.7983630952381e-05
12 1.36548913043478e-05
13 7.09134615384615e-06
14 4.55577761627907e-06
15 3.06372549019608e-06
16 2.28102189781022e-06
17 1.99134371029225e-06
18 1.45504182156134e-06
19 1.41019855595668e-06
20 1.11926934097421e-06
21 1.02863726135616e-06
22 9.91624525916561e-07
23 9.3265503875969e-07
24 8.31559340074508e-07
};
\addplot [ptoamp]
table {%
-6 0.2592140625
-5 0.238478125
-4 0.21521875
-3 0.1933171875
-2 0.1678140625
-1 0.1425984375
0 0.116878125
1 0.0909953125
2 0.06568125
3 0.0428421875
4 0.0248265625
5 0.0133234375
6 0.00608125
7 0.002459375
8 0.00092421875
9 0.000296614583333333
10 0.00010166015625
11 3.28776041666667e-05
12 9.775390625e-06
13 3.125e-06
14 9.85804416403785e-07
};
\addplot [ptcmd]
table {%
-6 0.4371375
-5 0.3149015625
-4 0.2603078125
-3 0.21948125
-2 0.1815453125
-1 0.1469609375
0 0.1164609375
1 0.09014375
2 0.0663015625
3 0.045865625
4 0.0283328125
5 0.0157828125
6 0.00766875
7 0.003128125
8 0.00111171875
9 0.000351875
10 8.94965277777778e-05
11 2.14683219178082e-05
12 4.40580985915493e-06
13 8.85770975056689e-07
};
%
\addplot [ptawgn]
table {%
-6 0.239228710767672
-5 0.21322801835762
-4 0.186113817483389
-3 0.158368318809598
-2 0.130644488522829
-1 0.103759095953406
0 0.0786496035251426
1 0.0562819519765415
2 0.037506128358926
3 0.0228784075610853
4 0.0125008180407376
5 0.00595386714777866
6 0.00238829078093281
7 0.000772674815378445
8 0.000190907774075993
9 3.36272284196175e-05
10 3.87210821552204e-06
11 2.6130679535752e-07
12 9.00601035062875e-09
13 1.33293101753005e-10
14 6.81018912878077e-13
};
\end{axis}
\end{tikzpicture}%

%% file: TikZ/plot64x64_comparisonCodeLDPC_FER_resized.tikz
\begin{tikzpicture}[scale=1]
\pgfplotsset{
	every axis plot/.style={line width=1.4pt},
	every axis/.style={grid style={line width=0.7pt, dashed}, 
					 },
	axis background/.style={fill=white},
	every outer x axis line/.style={line width=1.4pt},
	every outer y axis line/.style={line width=1.4pt},
	every tick/.style={black,
					line width=0.7pt,
					},
	label style={font=\fontsize{8}{9}\selectfont},
	every y tick label/.style={font=\fontsize{8}{9}\color{black}},
	every x tick label/.style={font=\fontsize{8}{9}\color{black}},
	every extra x tick/.style={grid style={solid, violet, line width=2.8pt},
							x tick label style={/pgf/number format/.cd,precision=10}
							},
	legend style={
					line width=0.7pt,
					font=\fontsize{11.4}{12}\selectfont\color{black},
					legend cell align=left,
					align=left,
					fill=white,
					fill opacity=0.8, draw opacity=1, text opacity=1,
					nodes={scale=0.7, transform shape},
					},
	legend pos={north east},
	legend image code/.code={
		\draw[mark repeat=2,mark phase=2]
			plot coordinates {
				(0cm,0cm)
				(0.25cm,0cm)        
				(0.5cm,0cm)         
			};%
		},
		legend entries={{\color{\clsd} SD uncoded},
						{\color{\clcmd}$\text{CMDNet}_{\text{bin}}$\\\color{\clcmd}uncoded}, 
						{\color{\clmmse} MMSE},
						{\color{\clamp} AMP}, 
						{\color{\cldetnet} DetNet}, 
						{\color{red}  $\text{CMDNet}_{\text{bin}}$}, 
						},
}
\begin{axis}[%
width=10.5cm,
height=8cm, 
scale = 0.7,
scale only axis,
separate axis lines,
tick pos=left,
xmin=3,
xmax=18,
ymode=log,
xminorticks=true,
xtick distance = {2},
ytick distance = {10},
xlabel={$E_{\text{b}}/N_0/R_{\text{C}}$ [dB]}, 
ymin=0.0001,
ymax=1,
ylabel={CFER},
xmajorgrids,
xminorgrids,
ymajorgrids,
]



\addplot [ptsd]
table {%
-4 1
-2 1
0 1
2 0.98936170212766
4 0.693498452012384
6 0.203196347031963
8 0.0233321382277504
10 0.00103856155260988
12 1.66665277789351e-05
};
\addplot [ptcmd]
table {%
-6 1
-5 1
-4 1
-3 1
-2 1
-1 0.9999
0 0.9997
1 0.9975
2 0.9904
3 0.9594
4 0.8785
5 0.7179
6 0.4951
7 0.2663
8 0.0962
9 0.0324333333333333
10 0.00763
11 0.00124736842105263
12 0.000233707865168539
13 5.62443026435734e-05
14 2.07445116131085e-05
15 1.18772782503038e-05
16 7.81728715996658e-06
};
\addplot [ptmmse]
table {%
-2.98970004336019 1
-1.98970004336019 1
-0.989700043360188 1
0.0102999566398121 0.9996337890625
1.01029995663981 0.9964599609375
2.01029995663981 0.9783935546875
3.01029995663981 0.9287109375
4.01029995663981 0.817626953125
5.01029995663981 0.64501953125
6.01029995663981 0.45458984375
7.01029995663981 0.2838134765625
8.01029995663981 0.149658203125
9.01029995663981 0.078369140625
10.0102999566398 0.037109375
11.0102999566398 0.0194091796875
12.0102999566398 0.0087890625
13.0102999566398 0.0032958984375
14.0102999566398 0.00156947544642857
15.0102999566398 0.000712076822916667
16.0102999566398 0.000324249267578125
17.0102999566398 0.00014380886130137
18.0102999566398 6.55749612603306e-05
19.0102999566398 3.47042871900826e-05
20.0102999566398 1.74243103978671e-05
21.0102999566398 9.54126080648982e-06
};
\addplot [ptamp]
table {%
-2.98970004336019 1
-1.98970004336019 1
-0.989700043360188 1
0.0102999566398121 0.9998779296875
1.01029995663981 0.9986572265625
2.01029995663981 0.984375
3.01029995663981 0.9078369140625
4.01029995663981 0.691162109375
5.01029995663981 0.4266357421875
6.01029995663981 0.18798828125
7.01029995663981 0.05859375
8.01029995663981 0.0104166666666667
9.01029995663981 0.00124782986111111
10.0102999566398 0.00119527180989583
11.0102999566398 0.00165812174479167
12.0102999566398 0.001953125
13.0102999566398 0.00179406368371212
14.0102999566398 0.00241400271045918
15.0102999566398 0.00416782924107143
16.0102999566398 0.005804443359375
17.0102999566398 0.009674072265625
18.0102999566398 0.0218017578125
19.0102999566398 0.0271240234375
20.0102999566398 0.0514729817708333
21.0102999566398 0.05596923828125
22.0102999566398 0.07037353515625
23.0102999566398 0.0809326171875
24.0102999566398 0.0919189453125
25.0102999566398 0.1236572265625
26.0102999566398 0.1021728515625
27.0102999566398 0.08758544921875
28.0102999566398 0.10430908203125
29.0102999566398 0.096435546875
30.0102999566398 0.1199951171875
31.0102999566398 0.0938720703125
32.0102999566398 0.106689453125
33.0102999566398 0.1270751953125
34.0102999566398 0.08245849609375
35.0102999566398 0.127197265625
36.0102999566398 0.1065673828125
37.0102999566398 0.1053466796875
38.0102999566398 0.09991455078125
39.0102999566398 0.1273193359375
};
\addplot [ptdetnet]
table {%
-2.98970004336019 1
-1.98970004336019 1
-0.989700043360188 1
0.0102999566398121 0.999755859375
1.01029995663981 0.99951171875
2.01029995663981 0.994873046875
3.01029995663981 0.9805908203125
4.01029995663981 0.9237060546875
5.01029995663981 0.801513671875
6.01029995663981 0.6011962890625
7.01029995663981 0.389892578125
8.01029995663981 0.210205078125
9.01029995663981 0.1163330078125
10.0102999566398 0.0517578125
11.0102999566398 0.0178571428571429
12.0102999566398 0.00534986413043478
13.0102999566398 0.00141232512718023
14.0102999566398 0.000357313684402332
15.0102999566398 9.83865008589329e-05
16.0102999566398 4.33893058265342e-05
};
\addplot [color=red, mark=*]
table {%
-2.98970004336019 1
-1.98970004336019 1
-0.989700043360188 1
0.0102999566398121 0.9998779296875
1.01029995663981 0.9967041015625
2.01029995663981 0.980712890625
3.01029995663981 0.9237060546875
4.01029995663981 0.779296875
5.01029995663981 0.52099609375
6.01029995663981 0.245849609375
7.01029995663981 0.0645751953125
8.01029995663981 0.01080322265625
9.01029995663981 0.000958251953125
10.0102999566398 3.45899402800659e-05
11.0102999566398 6.94332763407367e-07
};

\end{axis}
\end{tikzpicture}%

%% file: TikZ/MIMOhistogram_QPSK_64x64_64_remastered2_resized.tikz
\begin{tikzpicture}[scale=1]
\definecolor{color0}{rgb}{0.12156862745098,0.466666666666667,0.705882352941177}
\definecolor{color1}{rgb}{1,0.498039215686275,0.0549019607843137}
\definecolor{color2}{rgb}{0.172549019607843,0.627450980392157,0.172549019607843}
\definecolor{color3}{rgb}{0.83921568627451,0.152941176470588,0.156862745098039}
\pgfplotsset{
	every axis plot/.style={line width=1.4pt},
	every axis/.style={grid style={line width=0.7pt, dashed}, 
					 },
	every outer x axis line/.style={line width=1.4pt},
	every outer y axis line/.style={line width=1.4pt},
	every tick/.style={black,
					line width=0.7pt,
					},
	label style={font=\fontsize{8}{9}\selectfont},
	every y tick label/.style={font=\fontsize{8}{9}\color{black}},
	every x tick label/.style={font=\fontsize{8}{9}\color{black}},
	legend style={
					line width=0.7pt,
					font=\fontsize{11.4}{12}\selectfont\color{black},
					legend cell align=left,
					align=left,
					fill=white,
					fill opacity=0.8, draw opacity=1, text opacity=1,
					nodes={scale=0.7, transform shape},
					},
	legend image code/.code={
		\draw[mark repeat=2,mark phase=2]
			plot coordinates {
				(0cm,0cm)
				(0.25cm,0cm)        
				(0.5cm,0cm)         
			};%
		},
	legend pos={north west},
}
\begin{axis}[
width=12.0cm,
height=7cm, 
scale=0.7,
tick pos=left,
xmin=-41, xmax=41,
ymin=0, ymax=4.5,
xtick = {0, -20, 20, -10, 10, -30, 30},
ymin=0, ymax=4.5,
ytick = {0, 1, 2, 3},
extra x ticks={-40, 40}, extra x tick labels = {$-\infty$, $\infty$},
every extra y tick/.style={y tick label style={font=\fontsize{8}{9}\selectfont\color{black}}},
extra y ticks={4},  extra y tick labels = {97},
xmajorgrids,
xminorgrids,
ymajorgrids,
ylabel = {Relative Frequency [$\%$]},
xlabel={LLR},
]

\addlegendimage{ybar, ybar legend, draw=none, fill=color0, line width=0pt};
\addlegendentry{CMDNet $x_n=1$}
\draw[draw=none,fill=color0] (axis cs:-16.7194765921092,0) rectangle (axis cs:-16.2363887652594,0.00401445202729827);
\draw[draw=none,fill=color0] (axis cs:-16.2363887652594,0) rectangle (axis cs:-15.7533009384097,0.00200722601364914);
\draw[draw=none,fill=color0] (axis cs:-15.7533009384097,0) rectangle (axis cs:-15.27021311156,0);
\draw[draw=none,fill=color0] (axis cs:-15.27021311156,0) rectangle (axis cs:-14.7871252847103,0);
\draw[draw=none,fill=color0] (axis cs:-14.7871252847102,0) rectangle (axis cs:-14.3040374578605,0);
\draw[draw=none,fill=color0] (axis cs:-14.3040374578605,0) rectangle (axis cs:-13.8209496310108,0);
\draw[draw=none,fill=color0] (axis cs:-13.8209496310108,0) rectangle (axis cs:-13.3378618041611,0);
\draw[draw=none,fill=color0] (axis cs:-13.3378618041611,0) rectangle (axis cs:-12.8547739773113,0);
\draw[draw=none,fill=color0] (axis cs:-12.8547739773113,0) rectangle (axis cs:-12.3716861504616,0);
\draw[draw=none,fill=color0] (axis cs:-12.3716861504616,0) rectangle (axis cs:-11.8885983236119,0.00200722601364914);
\draw[draw=none,fill=color0] (axis cs:-11.8885983236119,0) rectangle (axis cs:-11.4055104967622,0);
\draw[draw=none,fill=color0] (axis cs:-11.4055104967622,0) rectangle (axis cs:-10.9224226699124,0);
\draw[draw=none,fill=color0] (axis cs:-10.9224226699124,0) rectangle (axis cs:-10.4393348430627,0.00200722601364914);
\draw[draw=none,fill=color0] (axis cs:-10.4393348430627,0) rectangle (axis cs:-9.95624701621297,0);
\draw[draw=none,fill=color0] (axis cs:-9.95624701621297,0) rectangle (axis cs:-9.47315918936324,0.00200722601364914);
\draw[draw=none,fill=color0] (axis cs:-9.47315918936324,0) rectangle (axis cs:-8.99007136251352,0.00401445202729827);
\draw[draw=none,fill=color0] (axis cs:-8.99007136251351,0) rectangle (axis cs:-8.50698353566379,0.00401445202729827);
\draw[draw=none,fill=color0] (axis cs:-8.50698353566379,0) rectangle (axis cs:-8.02389570881406,0);
\draw[draw=none,fill=color0] (axis cs:-8.02389570881406,0) rectangle (axis cs:-7.54080788196433,0.00200722601364914);
\draw[draw=none,fill=color0] (axis cs:-7.54080788196433,0) rectangle (axis cs:-7.0577200551146,0.00200722601364914);
\draw[draw=none,fill=color0] (axis cs:-7.0577200551146,0) rectangle (axis cs:-6.57463222826488,0.00602167804094741);
\draw[draw=none,fill=color0] (axis cs:-6.57463222826488,0) rectangle (axis cs:-6.09154440141515,0.00200722601364914);
\draw[draw=none,fill=color0] (axis cs:-6.09154440141515,0) rectangle (axis cs:-5.60845657456542,0);
\draw[draw=none,fill=color0] (axis cs:-5.60845657456542,0) rectangle (axis cs:-5.12536874771569,0.00401445202729827);
\draw[draw=none,fill=color0] (axis cs:-5.12536874771569,0) rectangle (axis cs:-4.64228092086596,0.0100361300682457);
\draw[draw=none,fill=color0] (axis cs:-4.64228092086596,0) rectangle (axis cs:-4.15919309401624,0.00602167804094741);
\draw[draw=none,fill=color0] (axis cs:-4.15919309401624,0) rectangle (axis cs:-3.67610526716651,0.0120433560818948);
\draw[draw=none,fill=color0] (axis cs:-3.67610526716651,0) rectangle (axis cs:-3.19301744031678,0.00802890405459655);
\draw[draw=none,fill=color0] (axis cs:-3.19301744031678,0) rectangle (axis cs:-2.70992961346705,0.0100361300682457);
\draw[draw=none,fill=color0] (axis cs:-2.70992961346705,0) rectangle (axis cs:-2.22684178661732,0.0100361300682457);
\draw[draw=none,fill=color0] (axis cs:-2.22684178661732,0) rectangle (axis cs:-1.7437539597676,0.0220794861501405);
\draw[draw=none,fill=color0] (axis cs:-1.7437539597676,0) rectangle (axis cs:-1.26066613291787,0.0301083902047371);
\draw[draw=none,fill=color0] (axis cs:-1.26066613291787,0) rectangle (axis cs:-0.777578306068142,0.058209554395825);
\draw[draw=none,fill=color0] (axis cs:-0.777578306068142,0) rectangle (axis cs:-0.294490479218414,0.0742673625050181);
\draw[draw=none,fill=color0] (axis cs:-0.294490479218414,0) rectangle (axis cs:0.188597347631315,0.0822962665596147);
\draw[draw=none,fill=color0] (axis cs:0.188597347631315,0) rectangle (axis cs:0.671685174481041,0.152549177037334);
\draw[draw=none,fill=color0] (axis cs:0.671685174481041,0) rectangle (axis cs:1.15477300133077,0.200722601364914);
\draw[draw=none,fill=color0] (axis cs:1.15477300133077,0) rectangle (axis cs:1.6378608281805,0.228823765556002);
\draw[draw=none,fill=color0] (axis cs:1.6378608281805,0) rectangle (axis cs:2.12094865503023,0.242874347651546);
\draw[draw=none,fill=color0] (axis cs:2.12094865503023,0) rectangle (axis cs:2.60403648187996,0.283018867924528);
\draw[draw=none,fill=color0] (axis cs:2.60403648187996,0) rectangle (axis cs:3.08712430872968,0.337213970293054);
\draw[draw=none,fill=color0] (axis cs:3.08712430872968,0) rectangle (axis cs:3.57021213557941,0.36932958651144);
\draw[draw=none,fill=color0] (axis cs:3.57021213557941,0) rectangle (axis cs:4.05329996242914,0.431553592934562);
\draw[draw=none,fill=color0] (axis cs:4.05329996242914,0) rectangle (axis cs:4.53638778927887,0.427539140907264);
\draw[draw=none,fill=color0] (axis cs:4.53638778927887,0) rectangle (axis cs:5.0194756161286,0.523885989562421);
\draw[draw=none,fill=color0] (axis cs:5.0194756161286,0) rectangle (axis cs:5.50256344297832,0.52589321557607);
\draw[draw=none,fill=color0] (axis cs:5.50256344297832,0) rectangle (axis cs:5.98565126982805,0.64030509835407);
\draw[draw=none,fill=color0] (axis cs:5.98565126982805,0) rectangle (axis cs:6.46873909667778,0.788839823364104);
\draw[draw=none,fill=color0] (axis cs:6.46873909667778,0) rectangle (axis cs:6.95182692352751,0.772782015254911);
\draw[draw=none,fill=color0] (axis cs:6.95182692352751,0) rectangle (axis cs:7.43491475037724,1.00160578081091);
\draw[draw=none,fill=color0] (axis cs:7.43491475037724,0) rectangle (axis cs:7.91800257722696,1.21035728623043);
\draw[draw=none,fill=color0] (axis cs:7.91800257722696,0) rectangle (axis cs:8.40109040407669,1.52749899638701);
\draw[draw=none,fill=color0] (axis cs:8.40109040407669,0) rectangle (axis cs:8.88417823092642,1.88077077478927);
\draw[draw=none,fill=color0] (axis cs:8.88417823092642,0) rectangle (axis cs:9.36726605777615,2.38458450421518);
\draw[draw=none,fill=color0] (axis cs:9.36726605777615,0) rectangle (axis cs:9.85035388462587,2.90847049377757);
\draw[draw=none,fill=color0] (axis cs:9.85035388462587,0) rectangle (axis cs:10.3334417114756,3.08109193095138);
\draw[draw=none,fill=color0] (axis cs:10.3334417114756,0) rectangle (axis cs:10.8165295383253,3.32597350461656);
\draw[draw=none,fill=color0] (axis cs:10.8165295383253,0) rectangle (axis cs:11.2996173651751,3.16138097149734);
\draw[draw=none,fill=color0] (axis cs:11.2996173651751,0) rectangle (axis cs:11.7827051920248,3.20152549177032);
\draw[draw=none,fill=color0] (axis cs:11.7827051920248,0) rectangle (axis cs:12.2657930188745,3.05499799277395);
\draw[draw=none,fill=color0] (axis cs:12.2657930188745,0) rectangle (axis cs:12.7488808457242,2.93657165796866);
\draw[draw=none,fill=color0] (axis cs:12.7488808457242,0) rectangle (axis cs:13.231968672574,3.06101967081489);
\draw[draw=none,fill=color0] (axis cs:13.231968672574,0) rectangle (axis cs:13.7150564994237,3.048976314733);
\draw[draw=none,fill=color0] (axis cs:13.7150564994237,0) rectangle (axis cs:14.1981443262734,3.00682456844637);
\draw[draw=none,fill=color0] (axis cs:14.1981443262734,0) rectangle (axis cs:14.6812321531232,3.02890405459651);
\draw[draw=none,fill=color0] (axis cs:14.6812321531232,0) rectangle (axis cs:15.1643199799729,3.01485347250097);
\draw[draw=none,fill=color0] (axis cs:15.1643199799729,0) rectangle (axis cs:15.6474078068226,2.91248494580487);
\draw[draw=none,fill=color0] (axis cs:15.6474078068226,0) rectangle (axis cs:16.1304956336723,3.07707747892409);
\draw[draw=none,fill=color0] (axis cs:16.1304956336723,0) rectangle (axis cs:16.6135834605221,3.13327980730626);
\draw[draw=none,fill=color0] (axis cs:16.6135834605221,0) rectangle (axis cs:17.0966712873718,3.16940987555194);
\draw[draw=none,fill=color0] (axis cs:17.0966712873718,0) rectangle (axis cs:17.5797591142215,2.9425933360096);
\draw[draw=none,fill=color0] (axis cs:17.5797591142215,0) rectangle (axis cs:18.0628469410712,3.10919309514247);
\draw[draw=none,fill=color0] (axis cs:18.0628469410712,0) rectangle (axis cs:18.545934767921,2.99478121236448);
\draw[draw=none,fill=color0] (axis cs:18.545934767921,0) rectangle (axis cs:19.0290225947707,2.83018867924526);
\draw[draw=none,fill=color0] (axis cs:19.0290225947707,0) rectangle (axis cs:19.5121104216204,2.85427539140905);
\draw[draw=none,fill=color0] (axis cs:19.5121104216204,0) rectangle (axis cs:19.9951982484702,2.78402248093133);
\draw[draw=none,fill=color0] (axis cs:19.9951982484702,0) rectangle (axis cs:20.4782860753199,2.6475311120032);
\draw[draw=none,fill=color0] (axis cs:20.4782860753199,0) rectangle (axis cs:20.9613739021696,2.38859895624248);
\draw[draw=none,fill=color0] (axis cs:20.9613739021696,0) rectangle (axis cs:21.4444617290193,2.26214371738259);
\draw[draw=none,fill=color0] (axis cs:21.4444617290193,0) rectangle (axis cs:21.9275495558691,2.19991971095947);
\draw[draw=none,fill=color0] (axis cs:21.9275495558691,0) rectangle (axis cs:22.4106373827188,1.86872741870737);
\draw[draw=none,fill=color0] (axis cs:22.4106373827188,0) rectangle (axis cs:22.8937252095685,1.55359293456445);
\draw[draw=none,fill=color0] (axis cs:22.8937252095685,0) rectangle (axis cs:23.3768130364183,1.37294259333602);
\draw[draw=none,fill=color0] (axis cs:23.3768130364183,0) rectangle (axis cs:23.859900863268,1.10999598554797);
\draw[draw=none,fill=color0] (axis cs:23.859900863268,0) rectangle (axis cs:24.3429886901177,0.927338418305892);
\draw[draw=none,fill=color0] (axis cs:24.3429886901177,0) rectangle (axis cs:24.8260765169674,0.70654355680449);
\draw[draw=none,fill=color0] (axis cs:24.8260765169674,0) rectangle (axis cs:25.3091643438172,0.52589321557607);
\draw[draw=none,fill=color0] (axis cs:25.3091643438172,0) rectangle (axis cs:25.7922521706669,0.493777599357685);
\draw[draw=none,fill=color0] (axis cs:25.7922521706669,0) rectangle (axis cs:26.2753399975166,0.283018867924528);
\draw[draw=none,fill=color0] (axis cs:26.2753399975166,0) rectangle (axis cs:26.7584278243664,0.260939381774388);
\draw[draw=none,fill=color0] (axis cs:26.7584278243664,0) rectangle (axis cs:27.2415156512161,0.146527498996387);
\draw[draw=none,fill=color0] (axis cs:27.2415156512161,0) rectangle (axis cs:27.7246034780658,0.134484142914492);
\draw[draw=none,fill=color0] (axis cs:27.7246034780658,0) rectangle (axis cs:28.2076913049155,0.0682456844640707);
\draw[draw=none,fill=color0] (axis cs:28.2076913049155,0) rectangle (axis cs:28.6907791317653,0.0461661983139302);
\draw[draw=none,fill=color0] (axis cs:28.6907791317653,0) rectangle (axis cs:29.173866958615,0.0381372942593336);
\draw[draw=none,fill=color0] (axis cs:29.173866958615,0) rectangle (axis cs:29.6569547854647,0.0200722601364914);
\draw[draw=none,fill=color0] (axis cs:29.6569547854647,0) rectangle (axis cs:30.1400426123144,0.00602167804094741);
\draw[draw=none,fill=color0] (axis cs:30.1400426123144,0) rectangle (axis cs:30.6231304391642,0.00602167804094741);
\draw[draw=none,fill=color0] (axis cs:30.6231304391642,0) rectangle (axis cs:31.1062182660139,0.00401445202729827);
\draw[draw=none,fill=color0] (axis cs:31.1062182660139,0) rectangle (axis cs:31.5893060928636,0.00200722601364914);

\addlegendimage{ybar, ybar legend, draw=none, fill=color1, line width=0pt};
\addlegendentry{DetNet $x_n=1$}
\draw[draw=none,fill=color1] (axis cs:-20,0) rectangle (axis cs:-19.6000003814697,0.0341228432953358);
\draw[draw=none,fill=color1] (axis cs:-19.6000022888184,0) rectangle (axis cs:-19.2000026702881,0);
\draw[draw=none,fill=color1] (axis cs:-19.2000007629395,0) rectangle (axis cs:-18.7999992370605,0);
\draw[draw=none,fill=color1] (axis cs:-18.7999992370605,0) rectangle (axis cs:-18.3999996185303,0);
\draw[draw=none,fill=color1] (axis cs:-18.4000015258789,0) rectangle (axis cs:-18.0000019073486,0);
\draw[draw=none,fill=color1] (axis cs:-18,0) rectangle (axis cs:-17.6000003814697,0);
\draw[draw=none,fill=color1] (axis cs:-17.6000022888184,0) rectangle (axis cs:-17.2000026702881,0);
\draw[draw=none,fill=color1] (axis cs:-17.2000007629395,0) rectangle (axis cs:-16.7999992370605,0);
\draw[draw=none,fill=color1] (axis cs:-16.7999992370605,0) rectangle (axis cs:-16.3999996185303,0);
\draw[draw=none,fill=color1] (axis cs:-16.4000015258789,0) rectangle (axis cs:-16.0000019073486,0);
\draw[draw=none,fill=color1] (axis cs:-16,0) rectangle (axis cs:-15.6000003814697,0);
\draw[draw=none,fill=color1] (axis cs:-15.6000003814697,0) rectangle (axis cs:-15.1999988555908,0);
\draw[draw=none,fill=color1] (axis cs:-15.1999998092651,0) rectangle (axis cs:-14.8000001907349,0);
\draw[draw=none,fill=color1] (axis cs:-14.8000011444092,0) rectangle (axis cs:-14.3999996185303,0);
\draw[draw=none,fill=color1] (axis cs:-14.3999996185303,0) rectangle (axis cs:-14,0);
\draw[draw=none,fill=color1] (axis cs:-14,0) rectangle (axis cs:-13.6000003814697,0);
\draw[draw=none,fill=color1] (axis cs:-13.6000003814697,0) rectangle (axis cs:-13.1999988555908,0);
\draw[draw=none,fill=color1] (axis cs:-13.1999998092651,0) rectangle (axis cs:-12.8000001907349,0);
\draw[draw=none,fill=color1] (axis cs:-12.8000011444092,0) rectangle (axis cs:-12.3999996185303,0);
\draw[draw=none,fill=color1] (axis cs:-12.3999996185303,0) rectangle (axis cs:-12,0);
\draw[draw=none,fill=color1] (axis cs:-12,0) rectangle (axis cs:-11.6000003814697,0);
\draw[draw=none,fill=color1] (axis cs:-11.6000003814697,0) rectangle (axis cs:-11.1999988555908,0);
\draw[draw=none,fill=color1] (axis cs:-11.1999998092651,0) rectangle (axis cs:-10.8000001907349,0);
\draw[draw=none,fill=color1] (axis cs:-10.8000011444092,0) rectangle (axis cs:-10.3999996185303,0);
\draw[draw=none,fill=color1] (axis cs:-10.3999996185303,0) rectangle (axis cs:-10,0);
\draw[draw=none,fill=color1] (axis cs:-10,0) rectangle (axis cs:-9.60000038146973,0);
\draw[draw=none,fill=color1] (axis cs:-9.60000038146973,0) rectangle (axis cs:-9.19999885559082,0);
\draw[draw=none,fill=color1] (axis cs:-9.19999980926514,0) rectangle (axis cs:-8.80000019073486,0);
\draw[draw=none,fill=color1] (axis cs:-8.80000114440918,0) rectangle (axis cs:-8.39999961853027,0);
\draw[draw=none,fill=color1] (axis cs:-8.39999961853027,0) rectangle (axis cs:-8,0);
\draw[draw=none,fill=color1] (axis cs:-8,0) rectangle (axis cs:-7.60000038146973,0);
\draw[draw=none,fill=color1] (axis cs:-7.59999942779541,0) rectangle (axis cs:-7.19999980926514,0);
\draw[draw=none,fill=color1] (axis cs:-7.19999980926514,0) rectangle (axis cs:-6.80000019073486,0);
\draw[draw=none,fill=color1] (axis cs:-6.80000019073486,0) rectangle (axis cs:-6.40000057220459,0);
\draw[draw=none,fill=color1] (axis cs:-6.39999961853027,0) rectangle (axis cs:-6,0.0020072259940207);
\draw[draw=none,fill=color1] (axis cs:-6,0) rectangle (axis cs:-5.60000038146973,0);
\draw[draw=none,fill=color1] (axis cs:-5.59999942779541,0) rectangle (axis cs:-5.19999980926514,0);
\draw[draw=none,fill=color1] (axis cs:-5.19999980926514,0) rectangle (axis cs:-4.80000019073486,0);
\draw[draw=none,fill=color1] (axis cs:-4.80000019073486,0) rectangle (axis cs:-4.40000057220459,0.0120433559641242);
\draw[draw=none,fill=color1] (axis cs:-4.39999961853027,0) rectangle (axis cs:-4,0.0040144519880414);
\draw[draw=none,fill=color1] (axis cs:-3.99999976158142,0) rectangle (axis cs:-3.60000014305115,0.0140505824238062);
\draw[draw=none,fill=color1] (axis cs:-3.59999990463257,0) rectangle (axis cs:-3.20000028610229,0.0100361295044422);
\draw[draw=none,fill=color1] (axis cs:-3.19999980926514,0) rectangle (axis cs:-2.80000019073486,0.022079486399889);
\draw[draw=none,fill=color1] (axis cs:-2.79999971389771,0) rectangle (axis cs:-2.40000009536743,0.0120433559641242);
\draw[draw=none,fill=color1] (axis cs:-2.39999985694885,0) rectangle (axis cs:-2.00000023841858,0.0120433559641242);
\draw[draw=none,fill=color1] (axis cs:-1.99999976158142,0) rectangle (axis cs:-1.60000014305115,0.0060216779820621);
\draw[draw=none,fill=color1] (axis cs:-1.59999990463257,0) rectangle (axis cs:-1.20000028610229,0.0301083903759718);
\draw[draw=none,fill=color1] (axis cs:-1.19999980926514,0) rectangle (axis cs:-0.800000190734863,0.0341228432953358);
\draw[draw=none,fill=color1] (axis cs:-0.799999833106995,0) rectangle (axis cs:-0.400000214576721,0.0381372943520546);
\draw[draw=none,fill=color1] (axis cs:-0.399999797344208,0) rectangle (axis cs:-1.78813934326172e-07,0.0421517454087734);
\draw[draw=none,fill=color1] (axis cs:1.9371509552002e-07,0) rectangle (axis cs:0.399999797344208,0.0521878749132156);
\draw[draw=none,fill=color1] (axis cs:0.400000214576721,0) rectangle (axis cs:0.799999833106995,0.0381372943520546);
\draw[draw=none,fill=color1] (axis cs:0.800000190734863,0) rectangle (axis cs:1.19999980926514,0.0321156159043312);
\draw[draw=none,fill=color1] (axis cs:1.20000028610229,0) rectangle (axis cs:1.59999990463257,0.0521878749132156);
\draw[draw=none,fill=color1] (axis cs:1.60000014305115,0) rectangle (axis cs:1.99999976158142,0.0662384554743767);
\draw[draw=none,fill=color1] (axis cs:2.00000023841858,0) rectangle (axis cs:2.39999985694885,0.0562023296952248);
\draw[draw=none,fill=color1] (axis cs:2.40000009536743,0) rectangle (axis cs:2.79999971389771,0.170614212751389);
\draw[draw=none,fill=color1] (axis cs:2.80000019073486,0) rectangle (axis cs:3.19999980926514,0.206744283437729);
\draw[draw=none,fill=color1] (axis cs:3.20000028610229,0) rectangle (axis cs:3.59999990463257,0.301083892583847);
\draw[draw=none,fill=color1] (axis cs:3.60000014305115,0) rectangle (axis cs:3.99999976158142,0.357286214828491);
\draw[draw=none,fill=color1] (axis cs:4,0) rectangle (axis cs:4.39999961853027,0.389401853084564);
\draw[draw=none,fill=color1] (axis cs:4.40000057220459,0) rectangle (axis cs:4.80000019073486,0.30710557103157);
\draw[draw=none,fill=color1] (axis cs:4.80000019073486,0) rectangle (axis cs:5.19999980926514,0.258932143449783);
\draw[draw=none,fill=color1] (axis cs:5.19999980926514,0) rectangle (axis cs:5.59999942779541,0.186672016978264);
\draw[draw=none,fill=color1] (axis cs:5.60000038146973,0) rectangle (axis cs:6,0.138498589396477);
\draw[draw=none,fill=color1] (axis cs:6,0) rectangle (axis cs:6.39999961853027,0.0702529102563858);
\draw[draw=none,fill=color1] (axis cs:6.40000057220459,0) rectangle (axis cs:6.80000019073486,0.0461661964654922);
\draw[draw=none,fill=color1] (axis cs:6.80000019073486,0) rectangle (axis cs:7.19999980926514,0.0240867119282484);
\draw[draw=none,fill=color1] (axis cs:7.19999980926514,0) rectangle (axis cs:7.59999942779541,0.018065033480525);
\draw[draw=none,fill=color1] (axis cs:7.60000038146973,0) rectangle (axis cs:8,0.0160578079521656);
\draw[draw=none,fill=color1] (axis cs:8,0) rectangle (axis cs:8.39999961853027,0.0160578079521656);
\draw[draw=none,fill=color1] (axis cs:8.39999961853027,0) rectangle (axis cs:8.80000114440918,0);
\draw[draw=none,fill=color1] (axis cs:8.80000019073486,0) rectangle (axis cs:9.19999980926514,0.0020072259940207);
\draw[draw=none,fill=color1] (axis cs:9.19999885559082,0) rectangle (axis cs:9.60000038146973,0.0060216779820621);
\draw[draw=none,fill=color1] (axis cs:9.60000038146973,0) rectangle (axis cs:10,0);
\draw[draw=none,fill=color1] (axis cs:10,0) rectangle (axis cs:10.3999996185303,0.0040144519880414);
\draw[draw=none,fill=color1] (axis cs:10.3999996185303,0) rectangle (axis cs:10.8000011444092,0);
\draw[draw=none,fill=color1] (axis cs:10.8000001907349,0) rectangle (axis cs:11.1999998092651,0);
\draw[draw=none,fill=color1] (axis cs:11.1999988555908,0) rectangle (axis cs:11.6000003814697,0);
\draw[draw=none,fill=color1] (axis cs:11.6000003814697,0) rectangle (axis cs:12,0);
\draw[draw=none,fill=color1] (axis cs:12,0) rectangle (axis cs:12.3999996185303,0.0020072259940207);
\draw[draw=none,fill=color1] (axis cs:12.3999996185303,0) rectangle (axis cs:12.8000011444092,0);
\draw[draw=none,fill=color1] (axis cs:12.8000001907349,0) rectangle (axis cs:13.1999998092651,0);
\draw[draw=none,fill=color1] (axis cs:13.1999988555908,0) rectangle (axis cs:13.6000003814697,0);
\draw[draw=none,fill=color1] (axis cs:13.6000003814697,0) rectangle (axis cs:14,0);
\draw[draw=none,fill=color1] (axis cs:14,0) rectangle (axis cs:14.3999996185303,0);
\draw[draw=none,fill=color1] (axis cs:14.3999996185303,0) rectangle (axis cs:14.8000011444092,0);
\draw[draw=none,fill=color1] (axis cs:14.8000001907349,0) rectangle (axis cs:15.1999998092651,0);
\draw[draw=none,fill=color1] (axis cs:15.1999988555908,0) rectangle (axis cs:15.6000003814697,0);
\draw[draw=none,fill=color1] (axis cs:15.6000003814697,0) rectangle (axis cs:16,0);
\draw[draw=none,fill=color1] (axis cs:16,0) rectangle (axis cs:16.3999996185303,0);
\draw[draw=none,fill=color1] (axis cs:16.3999977111816,0) rectangle (axis cs:16.7999973297119,0);
\draw[draw=none,fill=color1] (axis cs:16.7999992370605,0) rectangle (axis cs:17.2000007629395,0);
\draw[draw=none,fill=color1] (axis cs:17.2000007629395,0) rectangle (axis cs:17.6000003814697,0);
\draw[draw=none,fill=color1] (axis cs:17.5999984741211,0) rectangle (axis cs:17.9999980926514,0);
\draw[draw=none,fill=color1] (axis cs:18,0) rectangle (axis cs:18.3999996185303,0);
\draw[draw=none,fill=color1] (axis cs:18.3999977111816,0) rectangle (axis cs:18.7999973297119,0);
\draw[draw=none,fill=color1] (axis cs:18.7999992370605,0) rectangle (axis cs:19.2000007629395,0);
\draw[draw=none,fill=color1] (axis cs:19.2000007629395,0) rectangle (axis cs:19.6000003814697,0);
\draw[draw=none,fill=color1] (axis cs:39.5999984741211 - 0.5 + 0.2, 0) rectangle (axis cs:39.9999980926514 + 0.5 + 0.2, 96.9088745117188 - 93);

\addlegendimage{ybar,ybar legend,draw=none,fill=color2, line width=0pt};
\addlegendentry{CMDNet $x_n=-1$}
\draw[draw=none,fill=color2] (axis cs:-30.5352777108035,0) rectangle (axis cs:-30.0525188425273,0.00996412913511359);
\draw[draw=none,fill=color2] (axis cs:-30.0525188425273,0) rectangle (axis cs:-29.5697599742511,0.00797130330809087);
\draw[draw=none,fill=color2] (axis cs:-29.5697599742511,0) rectangle (axis cs:-29.0870011059748,0.0199282582702272);
\draw[draw=none,fill=color2] (axis cs:-29.0870011059748,0) rectangle (axis cs:-28.6042422376986,0.0338780390593862);
\draw[draw=none,fill=color2] (axis cs:-28.6042422376986,0) rectangle (axis cs:-28.1214833694224,0.049820645675568);
\draw[draw=none,fill=color2] (axis cs:-28.1214833694224,0) rectangle (axis cs:-27.6387245011462,0.0876843363889996);
\draw[draw=none,fill=color2] (axis cs:-27.6387245011462,0) rectangle (axis cs:-27.1559656328699,0.127540852929454);
\draw[draw=none,fill=color2] (axis cs:-27.1559656328699,0) rectangle (axis cs:-26.6732067645937,0.167397369469909);
\draw[draw=none,fill=color2] (axis cs:-26.6732067645937,0) rectangle (axis cs:-26.1904478963175,0.245117576723795);
\draw[draw=none,fill=color2] (axis cs:-26.1904478963175,0) rectangle (axis cs:-25.7076890280412,0.384615384615384);
\draw[draw=none,fill=color2] (axis cs:-25.7076890280412,0) rectangle (axis cs:-25.224930159765,0.476285372658429);
\draw[draw=none,fill=color2] (axis cs:-25.224930159765,0) rectangle (axis cs:-24.7421712914888,0.595854922279794);
\draw[draw=none,fill=color2] (axis cs:-24.7421712914888,0) rectangle (axis cs:-24.2594124232126,0.775209246711842);
\draw[draw=none,fill=color2] (axis cs:-24.2594124232126,0) rectangle (axis cs:-23.7766535549363,0.974491829414118);
\draw[draw=none,fill=color2] (axis cs:-23.7766535549363,0) rectangle (axis cs:-23.2938946866601,1.09804703068953);
\draw[draw=none,fill=color2] (axis cs:-23.2938946866601,0) rectangle (axis cs:-22.8111358183839,1.38899960143485);
\draw[draw=none,fill=color2] (axis cs:-22.8111358183839,0) rectangle (axis cs:-22.3283769501077,1.56038262255881);
\draw[draw=none,fill=color2] (axis cs:-22.3283769501077,0) rectangle (axis cs:-21.8456180818314,1.79952172180154);
\draw[draw=none,fill=color2] (axis cs:-21.8456180818314,0) rectangle (axis cs:-21.3628592135552,2.05659625348748);
\draw[draw=none,fill=color2] (axis cs:-21.3628592135552,0) rectangle (axis cs:-20.880100345279,2.37544838581112);
\draw[draw=none,fill=color2] (axis cs:-20.880100345279,0) rectangle (axis cs:-20.3973414770028,2.52690314866485);
\draw[draw=none,fill=color2] (axis cs:-20.3973414770028,0) rectangle (axis cs:-19.9145826087265,2.55280988441614);
\draw[draw=none,fill=color2] (axis cs:-19.9145826087265,0) rectangle (axis cs:-19.4318237404503,2.72618573136712);
\draw[draw=none,fill=color2] (axis cs:-19.4318237404503,0) rectangle (axis cs:-18.9490648721741,2.87963332004788);
\draw[draw=none,fill=color2] (axis cs:-18.9490648721741,0) rectangle (axis cs:-18.4663060038979,2.80988441610208);
\draw[draw=none,fill=color2] (axis cs:-18.4663060038978,0) rectangle (axis cs:-17.9835471356216,2.9952172180152);
\draw[draw=none,fill=color2] (axis cs:-17.9835471356216,0) rectangle (axis cs:-17.5007882673454,3.0569948186529);
\draw[draw=none,fill=color2] (axis cs:-17.5007882673454,0) rectangle (axis cs:-17.0180293990692,2.97130330809092);
\draw[draw=none,fill=color2] (axis cs:-17.0180293990692,0) rectangle (axis cs:-16.5352705307929,3.14866480669595);
\draw[draw=none,fill=color2] (axis cs:-16.5352705307929,0) rectangle (axis cs:-16.0525116625167,3.06297329613397);
\draw[draw=none,fill=color2] (axis cs:-16.0525116625167,0) rectangle (axis cs:-15.5697527942405,3.02710243124756);
\draw[draw=none,fill=color2] (axis cs:-15.5697527942405,0) rectangle (axis cs:-15.0869939259643,3.0569948186529);
\draw[draw=none,fill=color2] (axis cs:-15.0869939259643,0) rectangle (axis cs:-14.604235057688,3.09087285771229);
\draw[draw=none,fill=color2] (axis cs:-14.604235057688,0) rectangle (axis cs:-14.1214761894118,3.20645675567961);
\draw[draw=none,fill=color2] (axis cs:-14.1214761894118,0) rectangle (axis cs:-13.6387173211356,3.12873654842572);
\draw[draw=none,fill=color2] (axis cs:-13.6387173211356,0) rectangle (axis cs:-13.1559584528594,2.87963332004788);
\draw[draw=none,fill=color2] (axis cs:-13.1559584528594,0) rectangle (axis cs:-12.6731995845831,3.09087285771229);
\draw[draw=none,fill=color2] (axis cs:-12.6731995845831,0) rectangle (axis cs:-12.1904407163069,3.04105221203672);
\draw[draw=none,fill=color2] (axis cs:-12.1904407163069,0) rectangle (axis cs:-11.7076818480307,3.13471502590679);
\draw[draw=none,fill=color2] (axis cs:-11.7076818480307,0) rectangle (axis cs:-11.2249229797545,3.17457154244724);
\draw[draw=none,fill=color2] (axis cs:-11.2249229797545,0) rectangle (axis cs:-10.7421641114782,3.25627740135518);
\draw[draw=none,fill=color2] (axis cs:-10.7421641114782,0) rectangle (axis cs:-10.259405243202,3.18652849740938);
\draw[draw=none,fill=color2] (axis cs:-10.259405243202,0) rectangle (axis cs:-9.77664637492578,3.12873654842572);
\draw[draw=none,fill=color2] (axis cs:-9.77664637492578,0) rectangle (axis cs:-9.29388750664955,2.61857313670789);
\draw[draw=none,fill=color2] (axis cs:-9.29388750664955,0) rectangle (axis cs:-8.81112863837333,2.2160223196493);
\draw[draw=none,fill=color2] (axis cs:-8.81112863837333,0) rectangle (axis cs:-8.3283697700971,1.87524910322841);
\draw[draw=none,fill=color2] (axis cs:-8.3283697700971,0) rectangle (axis cs:-7.84561090182087,1.5683539258669);
\draw[draw=none,fill=color2] (axis cs:-7.84561090182087,0) rectangle (axis cs:-7.36285203354465,1.19370267038662);
\draw[draw=none,fill=color2] (axis cs:-7.36285203354465,0) rectangle (axis cs:-6.88009316526842,0.956556396970913);
\draw[draw=none,fill=color2] (axis cs:-6.88009316526842,0) rectangle (axis cs:-6.3973342969922,0.823037066560388);
\draw[draw=none,fill=color2] (axis cs:-6.3973342969922,0) rectangle (axis cs:-5.91457542871597,0.719410123555205);
\draw[draw=none,fill=color2] (axis cs:-5.91457542871597,0) rectangle (axis cs:-5.43181656043974,0.605819051414908);
\draw[draw=none,fill=color2] (axis cs:-5.43181656043974,0) rectangle (axis cs:-4.94905769216352,0.514149063371861);
\draw[draw=none,fill=color2] (axis cs:-4.94905769216352,0) rectangle (axis cs:-4.46629882388729,0.476285372658429);
\draw[draw=none,fill=color2] (axis cs:-4.46629882388729,0) rectangle (axis cs:-3.98353995561106,0.452371462734156);
\draw[draw=none,fill=color2] (axis cs:-3.98353995561106,0) rectangle (axis cs:-3.50078108733484,0.424471901155839);
\draw[draw=none,fill=color2] (axis cs:-3.50078108733484,0) rectangle (axis cs:-3.01802221905861,0.384615384615384);
\draw[draw=none,fill=color2] (axis cs:-3.01802221905861,0) rectangle (axis cs:-2.53526335078239,0.324830609804703);
\draw[draw=none,fill=color2] (axis cs:-2.53526335078239,0) rectangle (axis cs:-2.05250448250616,0.31486648066959);
\draw[draw=none,fill=color2] (axis cs:-2.05250448250616,0) rectangle (axis cs:-1.56974561422993,0.259067357512954);
\draw[draw=none,fill=color2] (axis cs:-1.56974561422993,0) rectangle (axis cs:-1.08698674595371,0.223196492626545);
\draw[draw=none,fill=color2] (axis cs:-1.08698674595371,0) rectangle (axis cs:-0.604227877677481,0.171383021123954);
\draw[draw=none,fill=color2] (axis cs:-0.604227877677481,0) rectangle (axis cs:-0.121469009401252,0.141490633718613);
\draw[draw=none,fill=color2] (axis cs:-0.121469009401252,0) rectangle (axis cs:0.361289858874972,0.0956556396970905);
\draw[draw=none,fill=color2] (axis cs:0.361289858874972,0) rectangle (axis cs:0.844048727151197,0.0677560781187724);
\draw[draw=none,fill=color2] (axis cs:0.844048727151197,0) rectangle (axis cs:1.32680759542743,0.0538062973296134);
\draw[draw=none,fill=color2] (axis cs:1.32680759542743,0) rectangle (axis cs:1.80956646370365,0.0418493423674771);
\draw[draw=none,fill=color2] (axis cs:1.80956646370365,0) rectangle (axis cs:2.29232533197988,0.0119569549621363);
\draw[draw=none,fill=color2] (axis cs:2.29232533197988,0) rectangle (axis cs:2.77508420025611,0.0119569549621363);
\draw[draw=none,fill=color2] (axis cs:2.77508420025611,0) rectangle (axis cs:3.25784306853233,0.00996412913511359);
\draw[draw=none,fill=color2] (axis cs:3.25784306853233,0) rectangle (axis cs:3.74060193680856,0.0119569549621363);
\draw[draw=none,fill=color2] (axis cs:3.74060193680856,0) rectangle (axis cs:4.22336080508478,0.00996412913511359);
\draw[draw=none,fill=color2] (axis cs:4.22336080508478,0) rectangle (axis cs:4.70611967336101,0.00797130330809087);
\draw[draw=none,fill=color2] (axis cs:4.70611967336101,0) rectangle (axis cs:5.18887854163723,0.00398565165404544);
\draw[draw=none,fill=color2] (axis cs:5.18887854163723,0) rectangle (axis cs:5.67163740991346,0.00597847748106816);
\draw[draw=none,fill=color2] (axis cs:5.67163740991346,0) rectangle (axis cs:6.15439627818969,0.00398565165404544);
\draw[draw=none,fill=color2] (axis cs:6.15439627818969,0) rectangle (axis cs:6.63715514646591,0);
\draw[draw=none,fill=color2] (axis cs:6.63715514646591,0) rectangle (axis cs:7.11991401474214,0.00199282582702272);
\draw[draw=none,fill=color2] (axis cs:7.11991401474214,0) rectangle (axis cs:7.60267288301837,0.00797130330809087);
\draw[draw=none,fill=color2] (axis cs:7.60267288301837,0) rectangle (axis cs:8.0854317512946,0);
\draw[draw=none,fill=color2] (axis cs:8.0854317512946,0) rectangle (axis cs:8.56819061957082,0);
\draw[draw=none,fill=color2] (axis cs:8.56819061957082,0) rectangle (axis cs:9.05094948784705,0);
\draw[draw=none,fill=color2] (axis cs:9.05094948784705,0) rectangle (axis cs:9.53370835612327,0);
\draw[draw=none,fill=color2] (axis cs:9.53370835612327,0) rectangle (axis cs:10.0164672243995,0.00398565165404544);
\draw[draw=none,fill=color2] (axis cs:10.0164672243995,0) rectangle (axis cs:10.4992260926757,0.00597847748106816);
\draw[draw=none,fill=color2] (axis cs:10.4992260926757,0) rectangle (axis cs:10.981984960952,0);
\draw[draw=none,fill=color2] (axis cs:10.981984960952,0) rectangle (axis cs:11.4647438292282,0.00199282582702272);
\draw[draw=none,fill=color2] (axis cs:11.4647438292282,0) rectangle (axis cs:11.9475026975044,0.00199282582702272);
\draw[draw=none,fill=color2] (axis cs:11.9475026975044,0) rectangle (axis cs:12.4302615657806,0.00199282582702272);
\draw[draw=none,fill=color2] (axis cs:12.4302615657806,0) rectangle (axis cs:12.9130204340569,0.00199282582702272);
\draw[draw=none,fill=color2] (axis cs:12.9130204340569,0) rectangle (axis cs:13.3957793023331,0);
\draw[draw=none,fill=color2] (axis cs:13.3957793023331,0) rectangle (axis cs:13.8785381706093,0);
\draw[draw=none,fill=color2] (axis cs:13.8785381706093,0) rectangle (axis cs:14.3612970388855,0);
\draw[draw=none,fill=color2] (axis cs:14.3612970388855,0) rectangle (axis cs:14.8440559071618,0);
\draw[draw=none,fill=color2] (axis cs:14.8440559071618,0) rectangle (axis cs:15.326814775438,0);
\draw[draw=none,fill=color2] (axis cs:15.326814775438,0) rectangle (axis cs:15.8095736437142,0.00199282582702272);
\draw[draw=none,fill=color2] (axis cs:15.8095736437142,0) rectangle (axis cs:16.2923325119904,0);
\draw[draw=none,fill=color2] (axis cs:16.2923325119904,0) rectangle (axis cs:16.7750913802667,0);
\draw[draw=none,fill=color2] (axis cs:16.7750913802667,0) rectangle (axis cs:17.2578502485429,0);
\draw[draw=none,fill=color2] (axis cs:17.2578502485429,0) rectangle (axis cs:17.7406091168191,0.00199282582702272);

\addlegendimage{ybar, ybar legend, draw=none, fill=color3, line width=0pt};
\addlegendentry{DetNet $x_n=-1$}
\draw[draw=none,fill=color3] (axis cs:-40 - 0.5 - 0.2,0) rectangle (axis cs:-39.6000003814697 + 0.5 - 0.2,97.3854141235352- 93);
\draw[draw=none,fill=color3] (axis cs:-19.6000022888184,0) rectangle (axis cs:-19.2000026702881,0);
\draw[draw=none,fill=color3] (axis cs:-19.2000007629395,0) rectangle (axis cs:-18.7999992370605,0);
\draw[draw=none,fill=color3] (axis cs:-18.7999992370605,0) rectangle (axis cs:-18.3999996185303,0);
\draw[draw=none,fill=color3] (axis cs:-18.4000015258789,0) rectangle (axis cs:-18.0000019073486,0);
\draw[draw=none,fill=color3] (axis cs:-18,0) rectangle (axis cs:-17.6000003814697,0);
\draw[draw=none,fill=color3] (axis cs:-17.6000022888184,0) rectangle (axis cs:-17.2000026702881,0);
\draw[draw=none,fill=color3] (axis cs:-17.2000007629395,0) rectangle (axis cs:-16.7999992370605,0);
\draw[draw=none,fill=color3] (axis cs:-16.7999992370605,0) rectangle (axis cs:-16.3999996185303,0);
\draw[draw=none,fill=color3] (axis cs:-16.4000015258789,0) rectangle (axis cs:-16.0000019073486,0.001992825884372);
\draw[draw=none,fill=color3] (axis cs:-16,0) rectangle (axis cs:-15.6000003814697,0);
\draw[draw=none,fill=color3] (axis cs:-15.6000003814697,0) rectangle (axis cs:-15.1999988555908,0);
\draw[draw=none,fill=color3] (axis cs:-15.1999998092651,0) rectangle (axis cs:-14.8000001907349,0);
\draw[draw=none,fill=color3] (axis cs:-14.8000011444092,0) rectangle (axis cs:-14.3999996185303,0);
\draw[draw=none,fill=color3] (axis cs:-14.3999996185303,0) rectangle (axis cs:-14,0);
\draw[draw=none,fill=color3] (axis cs:-14,0) rectangle (axis cs:-13.6000003814697,0);
\draw[draw=none,fill=color3] (axis cs:-13.6000003814697,0) rectangle (axis cs:-13.1999988555908,0);
\draw[draw=none,fill=color3] (axis cs:-13.1999998092651,0) rectangle (axis cs:-12.8000001907349,0);
\draw[draw=none,fill=color3] (axis cs:-12.8000011444092,0) rectangle (axis cs:-12.3999996185303,0);
\draw[draw=none,fill=color3] (axis cs:-12.3999996185303,0) rectangle (axis cs:-12,0);
\draw[draw=none,fill=color3] (axis cs:-12,0) rectangle (axis cs:-11.6000003814697,0);
\draw[draw=none,fill=color3] (axis cs:-11.6000003814697,0) rectangle (axis cs:-11.1999988555908,0);
\draw[draw=none,fill=color3] (axis cs:-11.1999998092651,0) rectangle (axis cs:-10.8000001907349,0.001992825884372);
\draw[draw=none,fill=color3] (axis cs:-10.8000011444092,0) rectangle (axis cs:-10.3999996185303,0);
\draw[draw=none,fill=color3] (axis cs:-10.3999996185303,0) rectangle (axis cs:-10,0);
\draw[draw=none,fill=color3] (axis cs:-10,0) rectangle (axis cs:-9.60000038146973,0.00398565176874399);
\draw[draw=none,fill=color3] (axis cs:-9.60000038146973,0) rectangle (axis cs:-9.19999885559082,0.00398565176874399);
\draw[draw=none,fill=color3] (axis cs:-9.19999980926514,0) rectangle (axis cs:-8.80000019073486,0.00398565176874399);
\draw[draw=none,fill=color3] (axis cs:-8.80000114440918,0) rectangle (axis cs:-8.39999961853027,0.00597847765311599);
\draw[draw=none,fill=color3] (axis cs:-8.39999961853027,0) rectangle (axis cs:-8,0.00398565176874399);
\draw[draw=none,fill=color3] (axis cs:-8,0) rectangle (axis cs:-7.60000038146973,0.011956955306232);
\draw[draw=none,fill=color3] (axis cs:-7.59999942779541,0) rectangle (axis cs:-7.19999980926514,0.011956955306232);
\draw[draw=none,fill=color3] (axis cs:-7.19999980926514,0) rectangle (axis cs:-6.80000019073486,0.0298923887312412);
\draw[draw=none,fill=color3] (axis cs:-6.80000019073486,0) rectangle (axis cs:-6.40000057220459,0.0378636904060841);
\draw[draw=none,fill=color3] (axis cs:-6.39999961853027,0) rectangle (axis cs:-6,0.0418493449687958);
\draw[draw=none,fill=color3] (axis cs:-6,0) rectangle (axis cs:-5.60000038146973,0.085691511631012);
\draw[draw=none,fill=color3] (axis cs:-5.59999942779541,0) rectangle (axis cs:-5.19999980926514,0.113591074943542);
\draw[draw=none,fill=color3] (axis cs:-5.19999980926514,0) rectangle (axis cs:-4.80000019073486,0.145476296544075);
\draw[draw=none,fill=color3] (axis cs:-4.80000019073486,0) rectangle (axis cs:-4.40000057220459,0.161418899893761);
\draw[draw=none,fill=color3] (axis cs:-4.39999961853027,0) rectangle (axis cs:-4,0.209246724843979);
\draw[draw=none,fill=color3] (axis cs:-3.99999976158142,0) rectangle (axis cs:-3.60000014305115,0.277002811431885);
\draw[draw=none,fill=color3] (axis cs:-3.59999990463257,0) rectangle (axis cs:-3.20000028610229,0.255081713199615);
\draw[draw=none,fill=color3] (axis cs:-3.19999980926514,0) rectangle (axis cs:-2.80000019073486,0.199282586574554);
\draw[draw=none,fill=color3] (axis cs:-2.79999971389771,0) rectangle (axis cs:-2.40000009536743,0.211239546537399);
\draw[draw=none,fill=color3] (axis cs:-2.39999985694885,0) rectangle (axis cs:-2.00000023841858,0.0936628133058548);
\draw[draw=none,fill=color3] (axis cs:-1.99999976158142,0) rectangle (axis cs:-1.60000014305115,0.0797130316495895);
\draw[draw=none,fill=color3] (axis cs:-1.59999990463257,0) rectangle (axis cs:-1.20000028610229,0.0936628133058548);
\draw[draw=none,fill=color3] (axis cs:-1.19999980926514,0) rectangle (axis cs:-0.800000190734863,0.0757273808121681);
\draw[draw=none,fill=color3] (axis cs:-0.799999833106995,0) rectangle (axis cs:-0.400000214576721,0.0518134720623493);
\draw[draw=none,fill=color3] (axis cs:-0.399999797344208,0) rectangle (axis cs:-1.78813934326172e-07,0.05380629748106);
\draw[draw=none,fill=color3] (axis cs:1.9371509552002e-07,0) rectangle (axis cs:0.399999797344208,0.0358708649873734);
\draw[draw=none,fill=color3] (axis cs:0.400000214576721,0) rectangle (axis cs:0.799999833106995,0.0418493449687958);
\draw[draw=none,fill=color3] (axis cs:0.800000190734863,0) rectangle (axis cs:1.19999980926514,0.0418493449687958);
\draw[draw=none,fill=color3] (axis cs:1.20000028610229,0) rectangle (axis cs:1.59999990463257,0.0338780395686626);
\draw[draw=none,fill=color3] (axis cs:1.60000014305115,0) rectangle (axis cs:1.99999976158142,0.0338780395686626);
\draw[draw=none,fill=color3] (axis cs:2.00000023841858,0) rectangle (axis cs:2.39999985694885,0.00398565176874399);
\draw[draw=none,fill=color3] (axis cs:2.40000009536743,0) rectangle (axis cs:2.79999971389771,0.0318852141499519);
\draw[draw=none,fill=color3] (axis cs:2.80000019073486,0) rectangle (axis cs:3.19999980926514,0.0259067360311747);
\draw[draw=none,fill=color3] (axis cs:3.20000028610229,0) rectangle (axis cs:3.59999990463257,0.0179354324936867);
\draw[draw=none,fill=color3] (axis cs:3.60000014305115,0) rectangle (axis cs:3.99999976158142,0.015942607074976);
\draw[draw=none,fill=color3] (axis cs:4,0) rectangle (axis cs:4.39999961853027,0.00797130353748798);
\draw[draw=none,fill=color3] (axis cs:4.40000057220459,0) rectangle (axis cs:4.80000019073486,0.00797130353748798);
\draw[draw=none,fill=color3] (axis cs:4.80000019073486,0) rectangle (axis cs:5.19999980926514,0.00797130353748798);
\draw[draw=none,fill=color3] (axis cs:5.19999980926514,0) rectangle (axis cs:5.59999942779541,0.00398565176874399);
\draw[draw=none,fill=color3] (axis cs:5.60000038146973,0) rectangle (axis cs:6,0.00597847765311599);
\draw[draw=none,fill=color3] (axis cs:6,0) rectangle (axis cs:6.39999961853027,0.001992825884372);
\draw[draw=none,fill=color3] (axis cs:6.40000057220459,0) rectangle (axis cs:6.80000019073486,0);
\draw[draw=none,fill=color3] (axis cs:6.80000019073486,0) rectangle (axis cs:7.19999980926514,0);
\draw[draw=none,fill=color3] (axis cs:7.19999980926514,0) rectangle (axis cs:7.59999942779541,0);
\draw[draw=none,fill=color3] (axis cs:7.60000038146973,0) rectangle (axis cs:8,0);
\draw[draw=none,fill=color3] (axis cs:8,0) rectangle (axis cs:8.39999961853027,0);
\draw[draw=none,fill=color3] (axis cs:8.39999961853027,0) rectangle (axis cs:8.80000114440918,0);
\draw[draw=none,fill=color3] (axis cs:8.80000019073486,0) rectangle (axis cs:9.19999980926514,0);
\draw[draw=none,fill=color3] (axis cs:9.19999885559082,0) rectangle (axis cs:9.60000038146973,0);
\draw[draw=none,fill=color3] (axis cs:9.60000038146973,0) rectangle (axis cs:10,0);
\draw[draw=none,fill=color3] (axis cs:10,0) rectangle (axis cs:10.3999996185303,0);
\draw[draw=none,fill=color3] (axis cs:10.3999996185303,0) rectangle (axis cs:10.8000011444092,0);
\draw[draw=none,fill=color3] (axis cs:10.8000001907349,0) rectangle (axis cs:11.1999998092651,0);
\draw[draw=none,fill=color3] (axis cs:11.1999988555908,0) rectangle (axis cs:11.6000003814697,0);
\draw[draw=none,fill=color3] (axis cs:11.6000003814697,0) rectangle (axis cs:12,0);
\draw[draw=none,fill=color3] (axis cs:12,0) rectangle (axis cs:12.3999996185303,0);
\draw[draw=none,fill=color3] (axis cs:12.3999996185303,0) rectangle (axis cs:12.8000011444092,0);
\draw[draw=none,fill=color3] (axis cs:12.8000001907349,0) rectangle (axis cs:13.1999998092651,0);
\draw[draw=none,fill=color3] (axis cs:13.1999988555908,0) rectangle (axis cs:13.6000003814697,0);
\draw[draw=none,fill=color3] (axis cs:13.6000003814697,0) rectangle (axis cs:14,0);
\draw[draw=none,fill=color3] (axis cs:14,0) rectangle (axis cs:14.3999996185303,0);
\draw[draw=none,fill=color3] (axis cs:14.3999996185303,0) rectangle (axis cs:14.8000011444092,0);
\draw[draw=none,fill=color3] (axis cs:14.8000001907349,0) rectangle (axis cs:15.1999998092651,0);
\draw[draw=none,fill=color3] (axis cs:15.1999988555908,0) rectangle (axis cs:15.6000003814697,0);
\draw[draw=none,fill=color3] (axis cs:15.6000003814697,0) rectangle (axis cs:16,0);
\draw[draw=none,fill=color3] (axis cs:16,0) rectangle (axis cs:16.3999996185303,0);
\draw[draw=none,fill=color3] (axis cs:16.3999977111816,0) rectangle (axis cs:16.7999973297119,0);
\draw[draw=none,fill=color3] (axis cs:16.7999992370605,0) rectangle (axis cs:17.2000007629395,0);
\draw[draw=none,fill=color3] (axis cs:17.2000007629395,0) rectangle (axis cs:17.6000003814697,0);
\draw[draw=none,fill=color3] (axis cs:17.5999984741211,0) rectangle (axis cs:17.9999980926514,0);
\draw[draw=none,fill=color3] (axis cs:18,0) rectangle (axis cs:18.3999996185303,0);
\draw[draw=none,fill=color3] (axis cs:18.3999977111816,0) rectangle (axis cs:18.7999973297119,0);
\draw[draw=none,fill=color3] (axis cs:18.7999992370605,0) rectangle (axis cs:19.2000007629395,0);
\draw[draw=none,fill=color3] (axis cs:19.2000007629395,0) rectangle (axis cs:19.6000003814697,0);
\draw[draw=none,fill=color3] (axis cs:19.5999984741211,0) rectangle (axis cs:19.9999980926514,0.0298923887312412);
\end{axis}

\begin{axis}[
width=12.0cm,
height=7cm, 
scale=0.7,
xmin=-41, xmax=41,
ymin=0, ymax=3.5,
xtick = \empty,
ytick = \empty,
tick align=outside,
tick pos=left,
x tick style = {draw = none},
y tick style = {draw = none},
extra x ticks={-35, 35}, extra x tick style={draw = white, tick label style={xshift=0cm, yshift=0.55cm}}, extra x tick labels = {\color{black}{\fontsize{17.42}{18}$\parallel$}, \color{black}{\fontsize{17.42}{18}$\parallel$}},
extra y ticks={2.75}, extra y tick style={{tick label style={xshift=0.55cm, yshift=0cm}}}, extra y tick labels = {\color{black}{\rotatebox{90}{\fontsize{17.42}{18}$\parallel$}}},
tick pos=left,
]
\addplot[draw=none] {x};
\end{axis}

\begin{axis}[
width=12.0cm,
height=7cm, 
scale=0.7,
xmin=-41, xmax=41,
ymin=0, ymax=3.5,
xtick = \empty,
ymin=0, ymax=4.5,
ytick = \empty,
extra x ticks={0, -20, 20, -10, 10, -30, 30, -40, 40}, extra x tick labels = { , , , , , , , , },
extra y ticks={0, 1, 2, 3, 4},  extra y tick labels = { , , , , },
tick pos=left,
]
\addplot[draw=none] {x};
\end{axis}

\end{tikzpicture}

%% file: TikZ/complexity_64x64_resized.tikz
\begin{tikzpicture}[scale=1]
\pgfplotsset{
	every axis plot/.style={line width=1.4pt},
	every axis/.style={grid style={line width=0.7pt, dashed}, 
					 },
	every outer x axis line/.style={line width=1.4pt},
	every outer y axis line/.style={line width=1.4pt},
	every tick/.style={black,
					line width=0.7pt,
					},
	label style={font=\fontsize{8}{9}\selectfont},
	every y tick label/.style={font=\fontsize{8}{9}\color{black}},
	every x tick label/.style={font=\fontsize{8}{9}\color{black}},
	every extra x tick/.style={grid style={solid, violet, line width=2.8pt},
							x tick label style={/pgf/number format/.cd,precision=10}
							},
}
\begin{axis}[
width=12.5cm,
height=8.7cm, 
scale=0.7,
separate axis lines,
log basis y={10},
tick pos=left,
xmajorgrids,
xmin=-0.84, xmax=7.84,
xminorgrids,
xtick = \empty,
ylabel={$\#$ of Multiplications},
ymajorgrids,
ymin=1000, ymax=2000000000,
yminorgrids,
ymode=log,
]
\draw[draw=none,fill=\clmmse!50!black] (axis cs:1.6,0) rectangle (axis cs:2.4,1146880);
\draw[draw=none,fill=\clamp!50!black] (axis cs:2.6,0) rectangle (axis cs:3.4,630787);
\draw[draw=none,fill=\clcmd!50!black] (axis cs:3.6,0) rectangle (axis cs:4.4,598144);
\draw[draw=none,fill=\cldetnet!50!black] (axis cs:4.6,0) rectangle (axis cs:5.4,41455616); 
\draw[draw=none,fill=\cloamp!50!black] (axis cs:5.6,0) rectangle (axis cs:6.4,73674816);
\draw[draw=none,fill=\clsd] (axis cs:-0.4,0) rectangle (axis cs:0.4,1340384720.53254); 
\draw[draw=none,fill=black] (axis cs:0.6,0) rectangle (axis cs:1.4,4096);
\draw[draw=none,fill=\clmmse!50!white] (axis cs:1.6,0) rectangle (axis cs:2.4,355616); 
\draw[draw=none,fill=\clamp] (axis cs:2.6,0) rectangle (axis cs:3.4,134160);
\draw[draw=none,fill=\clcmd] (axis cs:3.6,0) rectangle (axis cs:4.4,138305);
\draw[draw=none,fill=\cldetnet] (axis cs:4.6,0) rectangle (axis cs:5.4,1976320); 
\draw[draw=none,fill=\cloamp] (axis cs:5.6,0) rectangle (axis cs:6.4,18418704);
\draw[draw=none,fill=\clsdr!50!black] (axis cs:6.6,0) rectangle (axis cs:7.4,309047739.70833); 
\draw[draw=none,fill=\clsdr] (axis cs:6.6,0) rectangle (axis cs:7.4,28973225.5976559 * 0.5); 
\end{axis}

\begin{axis}[
width=12.5cm,
height=8.7cm, 
scale=0.7,
separate axis lines,
xmin=-0.84, xmax=7.84,
xtick = \empty,
ytick = \empty,
tick pos=left,
axis y line = none,
extra x ticks={0, 1, 2, 3, 4, 5, 6, 7}, extra x tick labels = {SD $10$dB, MF, MMSE, AMP, CMD, DetNet, OAMP, SDR},
xticklabel style = {font = \fontsize{7}{8}\selectfont},
]
\addplot[draw=none] {x};
\end{axis}

\end{tikzpicture}